\documentclass{ws-ijmpd}
\usepackage{epsfig}

\newcommand{\beq}{\begin{equation}}
\newcommand{\eeq}{\end{equation}}
\newcommand{\be}{\begin{eqnarray}}
\newcommand{\ee}{\end{eqnarray}}

\def\l{\ell}

\def\apj{ApJ}

\def\apjs{ApJS}

\def\mnras{MNRAS}

\newcommand{\nbi}{{Niels Bohr Institute, Blegdamsvej 17,
DK-2100 Copenhagen, Denmark}}
\newcommand{\nbia}{{Niels Bohr International Academy,
Blegdamsvej 17, DK-2100 Copenhagen, Denmark}}
\newcommand{\sao}{{Special Astrophysical Observatory, Nizhnij Arkhyz,
Karachaj-Cherkesia, 369167, Russia}}

\newcommand{\asc}{{Astro Space Center of Lebedev Physical Institute,
Profsoyuznaya 84/32, Moscow, Russia}}
\newcommand{\imperial}{{Imperial College, London, United Kingdom}}
\newcommand{\iam}{{Keldysh Institute of Applied Math, Russian Academy of
Science, Moscow, 125047, Russia}}
\newcommand{\iaat}{{Institute of Astronomy and Astrophysics, Academia Sinica,
P.O.Box 23-141, Taipei 10617, Taiwan, Republic of China}}

\begin{document}

\title{The Gauss-Legendre Sky Pixelization for the CMB polarization
(GLESP-pol). Errors due to pixelization of the CMB sky}

\author{Andrei G.~Doroshkevich}
\address{\asc, dorr@asc.rssi.ru}
\author{Oleg V. Verkhodanov}
\address{\sao, vo@sao.ru}
\author{Pavel D. Naselsky}
\address{\nbi, naselsky@nbi.dk}
\author{Jaiseung  Kim}
\address{\nbi, jkim@nbi.dk}
\author{Dmitry I. Novikov}
\address{\imperial,\\ \asc, d.novikov@imperial.ac.uk}
\author{Viktor I.~Turchaninov}
\address{\iam}
\author{Igor D.~Novikov}
\address{\asc,\\ \nbi, \\ \nbia, novikov@asc.rssi.ru}
\author{Lung-Yih~Chiang}
\address{\iaat, lychiang@asiaa.sinica.edu.tw}
\author{Martin~Hansen}
\address{\nbi}

\date{Accepted 2009 ???? ???; Received 2009 ???? ???}

\maketitle

\begin{abstract}
We present developing of method of the numerical analysis of polarization
in the Gauss--Legendre Sky Pixelization (GLESP) scheme for the
CMB maps. This incorporation of the polarization transforms in
the pixelization scheme GLESP completes the creation of our new
method for the numerical analysis of CMB maps. The comparison
of GLESP and HEALPix calculations is done.
\end{abstract}

\section{Introduction}
The analysis of the anisotropy of CMB temperature and 
polarization is one of the most effective methods for the 
extraction of the cosmological information and for tests of 
cosmology and fundamental physics.

The measurements 
of the anisotropy of CMB temperature performed by the WMAP 
mission allow us to establish parameters of the cosmological 
model of the Universe with unprecedented precision
Refs.~\refcite{hinshaw2007},
\refcite{hinshaw2009},          % (Hinshaw et al. 2007, 2009; %++
\refcite{spergel2007},          % Spergel et al. 2007;        %+
\refcite{nolta2009}.            % Nolta et al. 2009).         %+

Increasing sensitivity and angular resolution of the CMB data, including
recently available  WMAP 5 year data
Ref.~\refcite{hinshaw2009} ,       % (Hinshaw et al. 2009),            %+
ACBAR
Ref.~\refcite{acbar}               % (Reichardt et al. 2008),   %+
QUaD
Refs.~\refcite{quad1},\refcite{quad2},%(Pryke etal.2009,Hiderks etal.2009)%++
stimulates significant
development and increasing
predictability of the corresponding software
(see, for instance,
CAMB
Ref.~\refcite{camb},                % (Challinor and Lewis 2005), %+
COSMOMC
Ref.~\refcite{cosmomc},             % (Lewis and Bridle 2002), %+
RICO
Ref.~\refcite{rico},                %  (Fendt et al. 2008)         %+
etc.)

During the next decade, after the PLANCK experiment, an investigation of
the CMB polarization (including the B-mode) will be at the focus of
the CMB science. Planning the
CMBpol mission
Ref.~\refcite{CMBpol},              % (Baumann et al. 2008),     %+
B-pol mission
Ref.~\refcite{Bpol},                % (de Bernardis et al. 2008) %+
etc. requires significant
improvement in estimation of the errors of the signal.
Partially an uncertainties due to pixelization of the CMB sky and
their propogation
to the CMB power and the map can be potential sources of the error.
For planning high resolution the CMB experiments the most frequently used the
HEALPix package
Ref.~\refcite{hpix}               % (G\'orski et al. 1999, 2005)   %++
as well as GLESP
Ref.~\refcite{glesp},             % (Doroshkevich et al. 2005),   %+
ECP
Ref.~\refcite{ecp},               % (Muciaccia, Natoli and Vittorio 1997),     %+
and some others, needs to be tested in order
to provide an exact information about the error bars of
the convolution of the $T,E,B$-maps to the corresponding coefficients of
 the spherical harmonics decomposition not only for the power spectrum,
but for the real and imaginary part of the coefficients as well.

The information about the multipole structure of the CMB signal is vital
for the low multipoles ($\ell=2,3..$), since a  lot of theoretical
predictions about the properties of the cosmological model are related
directly to the global morphology of the signal.
As an example, we would like to mention widely discussed the Bianchi $VII_h$
anisotropic cosmological model
Refs.~\refcite{eriksenasym},         % Eriksen et al. 2004a;
\refcite{jaffebianchi},        % (Jaffe et al. 2005;
\refcite{bridges}              % Bridges et al. 2007), %+++
which can mimic the anisotropy of the CMB power at the range of
multipoles $\ell \le 20$, and
the Cold Spot
Ref.~\refcite{cruz}                 % (Cruz et al. 2006)    %+
as well.
The whole sky decomposition is vital for testing the alignment and
planarity of the CMB anisotropy multipoles at $2\le\ell\le 5$,
discussed in
Ref.~\refcite{lilc}                 % Eriksen et al. (2004b).  %+
The exact information about the phases
of the CMB signal is very useful for investigation of statistical
anisotropy and non-Gaussianity of the CMB
% (Chiang et al. 2003; Naselsky et al. 2003--2006). %++++
Refs.~\refcite{nong},
\refcite{pcor1},
\refcite{pcor2},
\refcite{cmbook}.

  This Paper is devoted to presentation of  a new package
GLESP-pol (see Appendix), and  investigation in details the errors of
the standard transition ``map to $a_{\l,m}$'' and vice verse
for the most frequently used
HEALPix\,2.11 and  newly released the GLESP-pol package.
We would like to point out that both these packages
reveal some peculiarities of the reconstruction of the coefficients
of decomposition, especially for
polarization. The major part of the error belongs to the $\ell, m=0$,
and $\ell, m=2$ modes, when the simplest
variants of decomposition were used.
For the HEALPix\,2.11, it is ``zero iteration'' key which blocks
the correction of the $a_{\l,m}$ taken from the map.
For the GLESP-pol package, the maxima of the $a_{\l,m}$ errors
correspond to the GLESP\,1.0 pixelization
% (Doroshkevich et al, 2005a,b).  %++
Refs.~\refcite{glesp},
\refcite{glespa}.
However, all these problems can be successfully resolved by implementation
of iterations for the HEALPix\,2.11
(the key ``iterative analysis '' 3 or 4 iterations) and
the GLESP-pol pixelization for the polarization.

The outline of the paper is the follows.
In Section 2, we  discuss the difference between the HEALPix and the
GLESP-pol scheme of the CMB sky pixelization,
focusing on the polarization of the CMB. Section 3 is devoted
to investigation of the errors of
the ``map$\rightarrow a_{\l,m}\rightarrow$ map'' transition for
the CMB temperature anisotropy
for the  HEALPix  and the GLESP-pol. In Section 4, we discuss
the same issue  for the Q,U  Stokes parameters and E and B-modes of
polarization.
In Appendix the basic relations and description of the GLESP-pol package
are presented.

\section{Basic definitions}

The temperature and polarization CMB anisotropy can be described
in terms of the Stokes parameters $T,Q, U$ through spherical harmonics
decomposition $Y_{\l,m}(\theta,\phi)$, and spin $\pm2$ spherical
harmonics${}_{\pm 2}Y_{\l,m}(\theta,\phi)$
\begin{eqnarray}
 T(\theta,\phi)=\sum_\l\sum_m a_{\l,m}Y_{\l,m}(\theta,\phi),\nonumber\\
Q(\theta,\phi)\pm iU(\theta,\phi)=
   \sum_\l\sum_m {}_{\pm 2}a_{\l,m}{}_{\pm 2}Y_{\l,m}(\theta,\phi)
\label{eq1}
\end{eqnarray}
Here $\theta,\phi$ are the polar and azimuthal angles of the polar
system of coordinates, $a_{\l,m}$ stands for the temperature anisotropy
and  the spin  coefficients ${}_{\pm 2}a_{\l,m}$
can be decomposed into E and B modes of polarization
(see Appendix for details):
\begin{eqnarray}
{}_{\pm 2}a_{\l,m}=-(a^E_{\l,m}\pm ia^B_{\l,m}),\hspace {0.5cm}a^{E,B}_{\l,m}=
	(-1)^m(a^{E,B})^{*}_{\l,-m}
\label{eq2}
\end{eqnarray}

The conversion of the $T,Q,U$ signals to corresponding
$a_{\l,m}$ and ${}_{\pm 2}a_{\l,m}$ coefficients is given by following
integrals:
\begin{eqnarray}
 {}_{0,\pm 2}a_{\l,m}=\nonumber\\
\int_{-1}^1dx\int_0^{2\pi}d\phi\left(T(x,\phi),Q(x,\phi),U(x,\phi)\right)
    {}_{0,\pm 2}Y^{*}_{\l,m}(x,\phi),\nonumber\\
\label{eq3}
\end{eqnarray}
where index $0$ marks the temperature anisotropy, and $x=\cos\theta$.
As it seen from Eq(\ref{eq3}), the mathematical basis of any schemes of
the pixelization of the CMB sky is very simple. We need to estimate
the integrals in Eq(\ref{eq3}) with very high accuracy taking into
account the properties of discrete representation of the signal on
the sphere. However, the modern CMB experiments
normally  deal with the incomplete sky due to peculiarities of design,
scan strategy or implementation of different sort of mask. In this
case the scheme of the pixelization of the sky become even more important,
since we need to use the pixel domain for estimation of the power spectrum
and investigation of the
statistical properties of the CMB signal without implementation of
the  ${}_{0,\pm 2}a_{\l,m}$-coefficients.
For these purpose the basic idea of the HEALPix package (equal area
isolatitude pixelization) is very useful and more advanced in comparison
to other pixelization.
However, for the whole sky analysis of the $T,Q,U$ signals
${}_{0,\pm 2}a_{\l,m}$-domain seems to be more optimal,
in terms of the CPU timing, as from the scientific point of view.
This is why in this paper we
propose the GLESP-pol pixelization as complementary approach to
the HEALPix package.
%(see
%{\tt http://www.apc.univ-paris7.fr/$\sim$radek/s2hat/docs/S2HATdocs.html}).

 For numerical evaluation of the integral
Eq(\ref{eq3}), we use the Gaussian quadratures
Ref.~\refcite{glesp}.         %(Doroshkevich et al. 2005a).
This approach was proposed by Gauss in 1814,
and developed later by Christoffel in 1877. As the integral
over $x$ in  Eq. (\ref{eq3}) is an integral over a
polynomial of $x$ we can use the following equality
Ref.~\refcite{nr}:            % (Press  et al. 1992):
\begin{equation}
\int^1_{-1}dx \Delta T(x,\phi) Y^{*}_{\ell m}(x,\phi)=
\sum^N_{j=1}w_j\Delta T(x_j,\phi) Y^{*}_{\ell m}(x_j,\phi)\,.
\label{eq4}
\end{equation}
where both $\Delta T(x_j,\phi)Y^{*}_{\ell m}(x_j,\phi)$ and
the proper Gaussian quadrature weighting functions, $w_j=
w(x_j)$, are taken at points $x_j$ which are the net of roots
of the Legendre polynomial
\begin{equation}
P_N(x_j)=0\,.
\label{root}
\end{equation}
Here $N$ is the maximal rank of the polynomial under
consideration.

It is well known that the equation $P_N(x_j)=0$
has $N$ number of zeros in interval $-1\le x\le 1$. For the
Gaussian--Legendre method Eq(\ref{eq4}), the weighting
coefficients are
% (Press et al. 1992)     %+
\begin{equation}
w_j= {2\over 1-x^2_j} [P_N^{'}(x_j)]^{-2}\,,
\label{eq6}
\end{equation}
where ${'}$ denotes a derivative. They can be calculated together
with the set of $x_j$ with the `{\tt gauleg}' code
Ref.~\refcite{nr}. % (Press et al. 1992, Sec. 4.5).

The old GLESP (version 1.0) pixelization scheme
Ref.~\refcite{glespa}  % (Doroshkevich et al. 2005a)
was defined as follows:
\begin{itemize}
%\item
\item
     In the polar direction $x=\cos\theta$, we define
     $x_j, j=1,2,\ldots,N$, as
     the net of roots
     of Eq. (\ref{root}).
\item
     Each root $x_j$ determines the position of a ring with
   $N_{\phi}^j$ pixel centers with $\phi$--coordinates $\phi_i$.
\item
     All the pixels have nearly equal area.
\item
     Each pixel has weight $w_j$
    (see Eq (\ref{eq6})).
\end{itemize}

\begin{figure}[!th]
\hbox{\hspace*{0.01cm}
%\begingroup
\centerline{\includegraphics[width=0.45\linewidth]{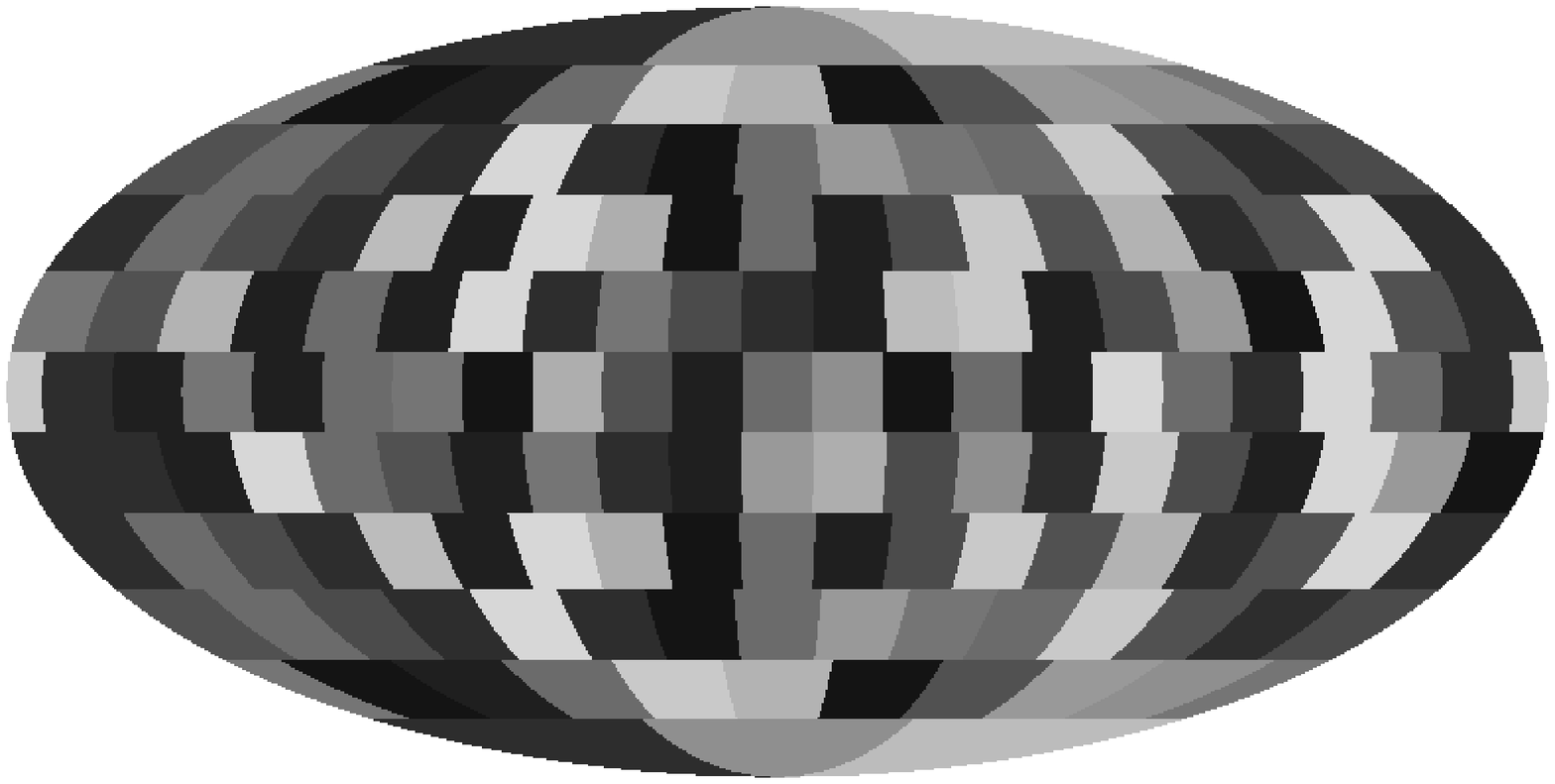}%}}
\includegraphics[width=0.28\linewidth]{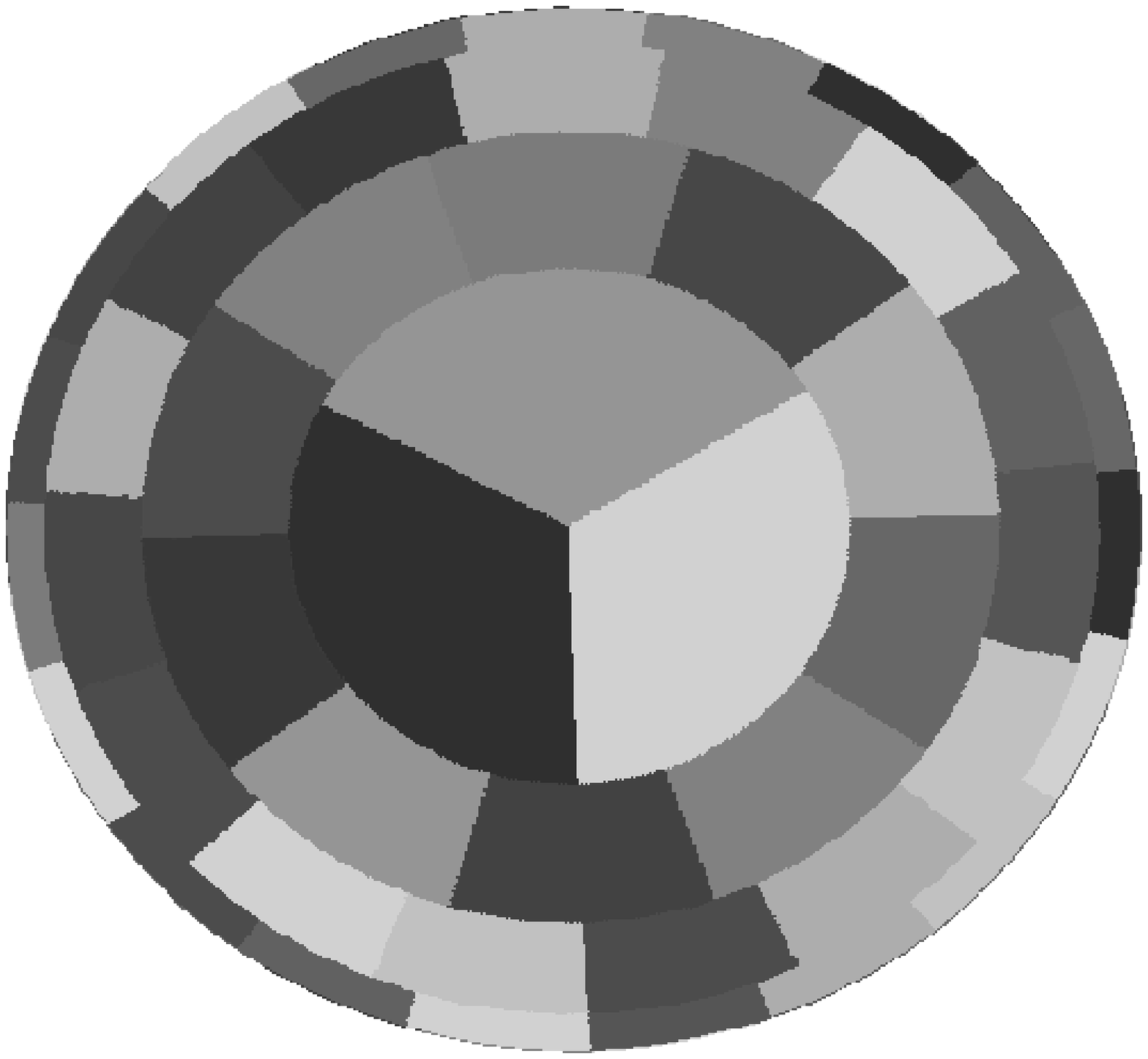}}}
\hbox{\hspace*{0.01cm}
\centerline{\includegraphics[width=0.45\linewidth]{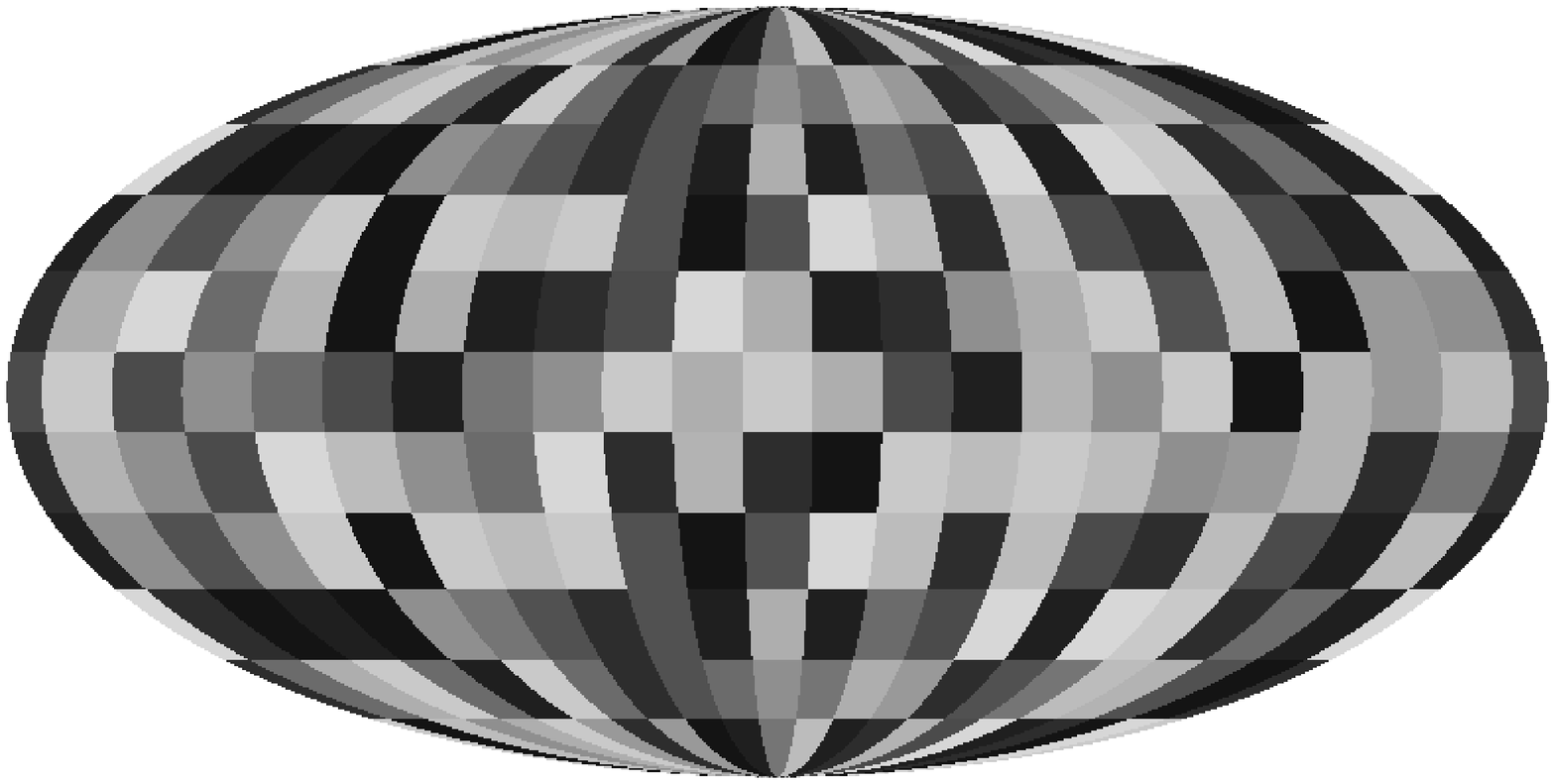}%}}
\includegraphics[width=0.28\linewidth]{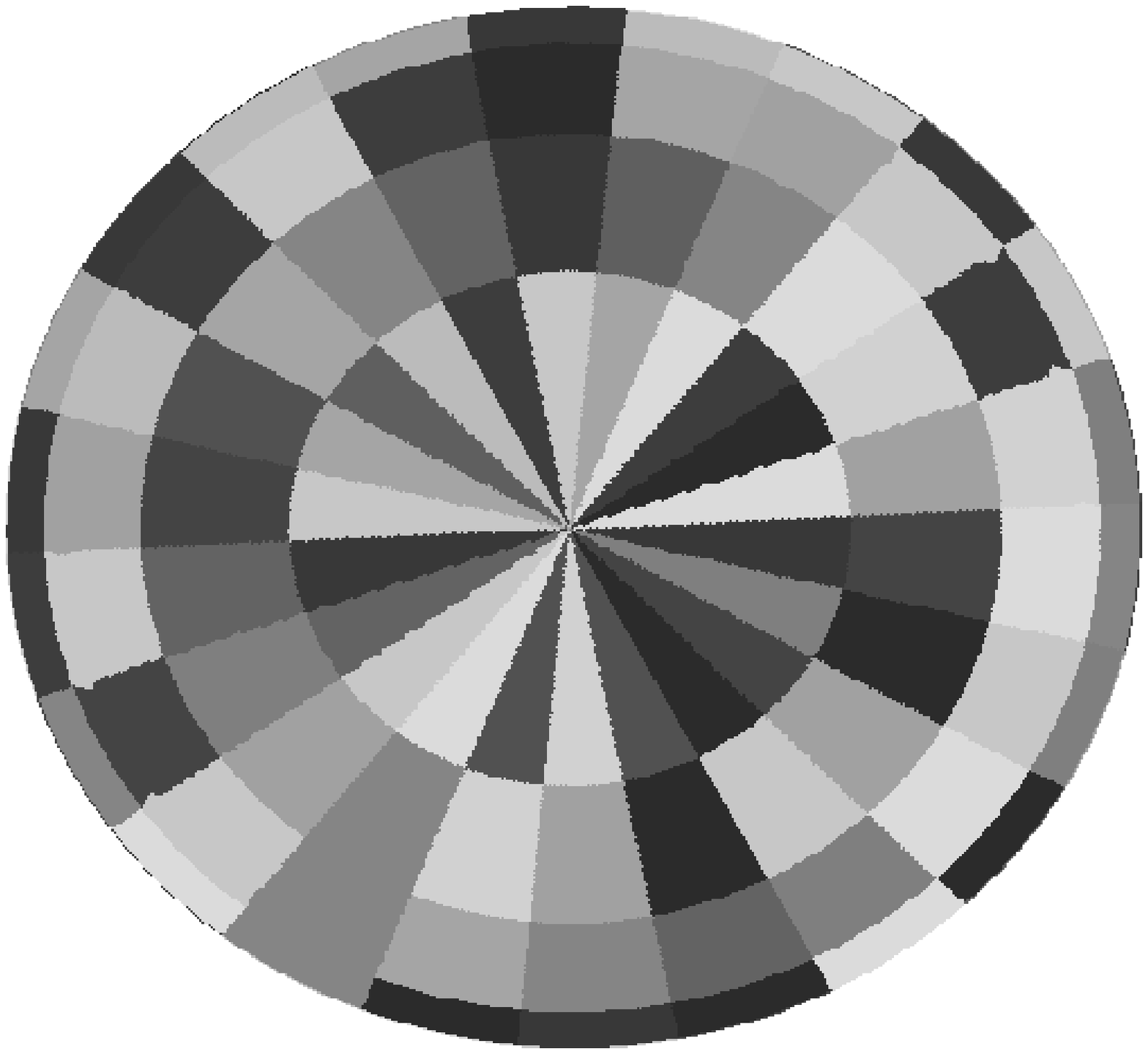}}}
\hbox{\hspace*{0.01cm}
\centerline{\includegraphics[width=0.45\linewidth]{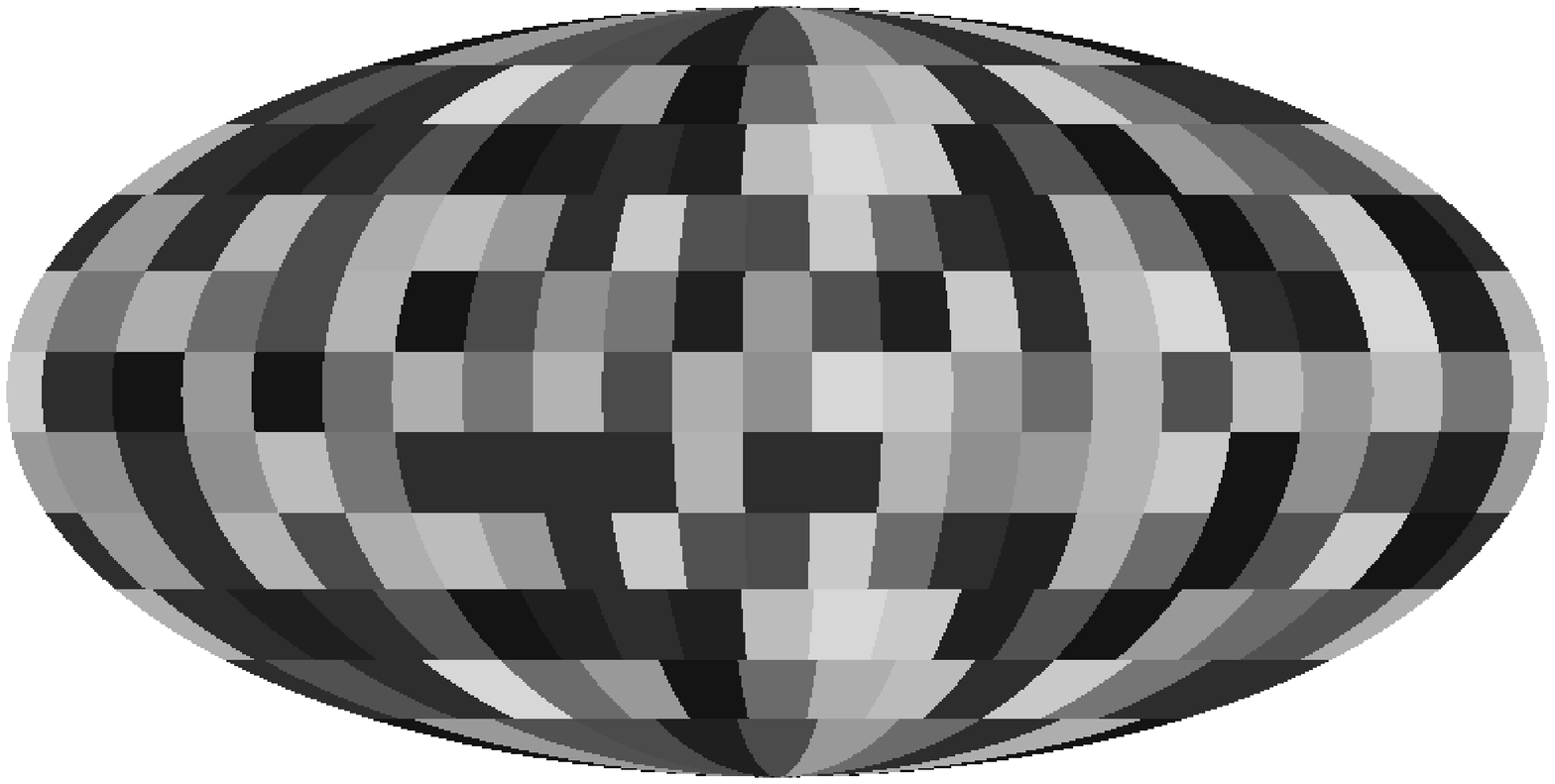}%}}
\includegraphics[width=0.28\linewidth]{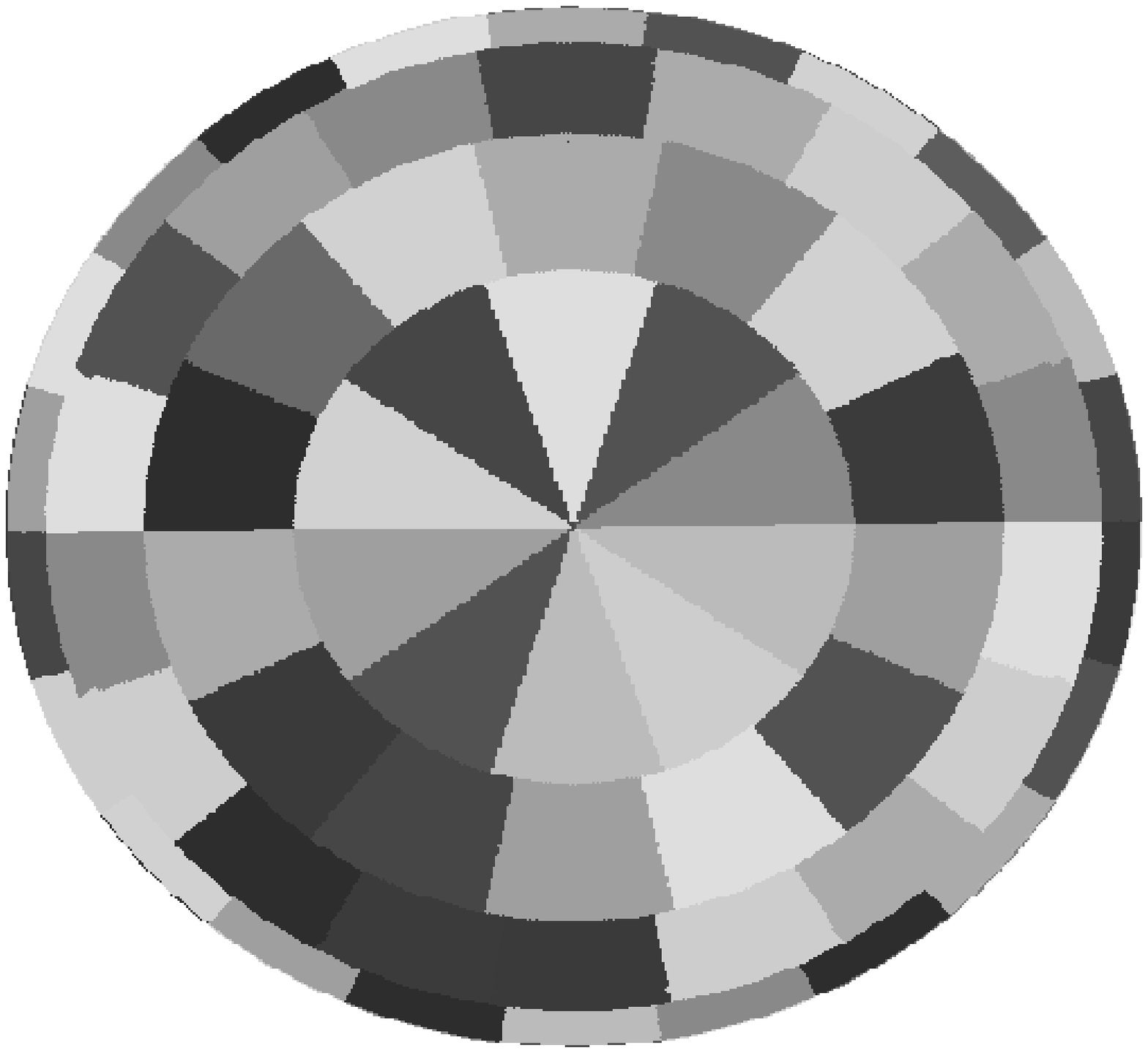}}}
\hbox{\hspace*{0.01cm}
\centerline{\includegraphics[width=0.45\linewidth]{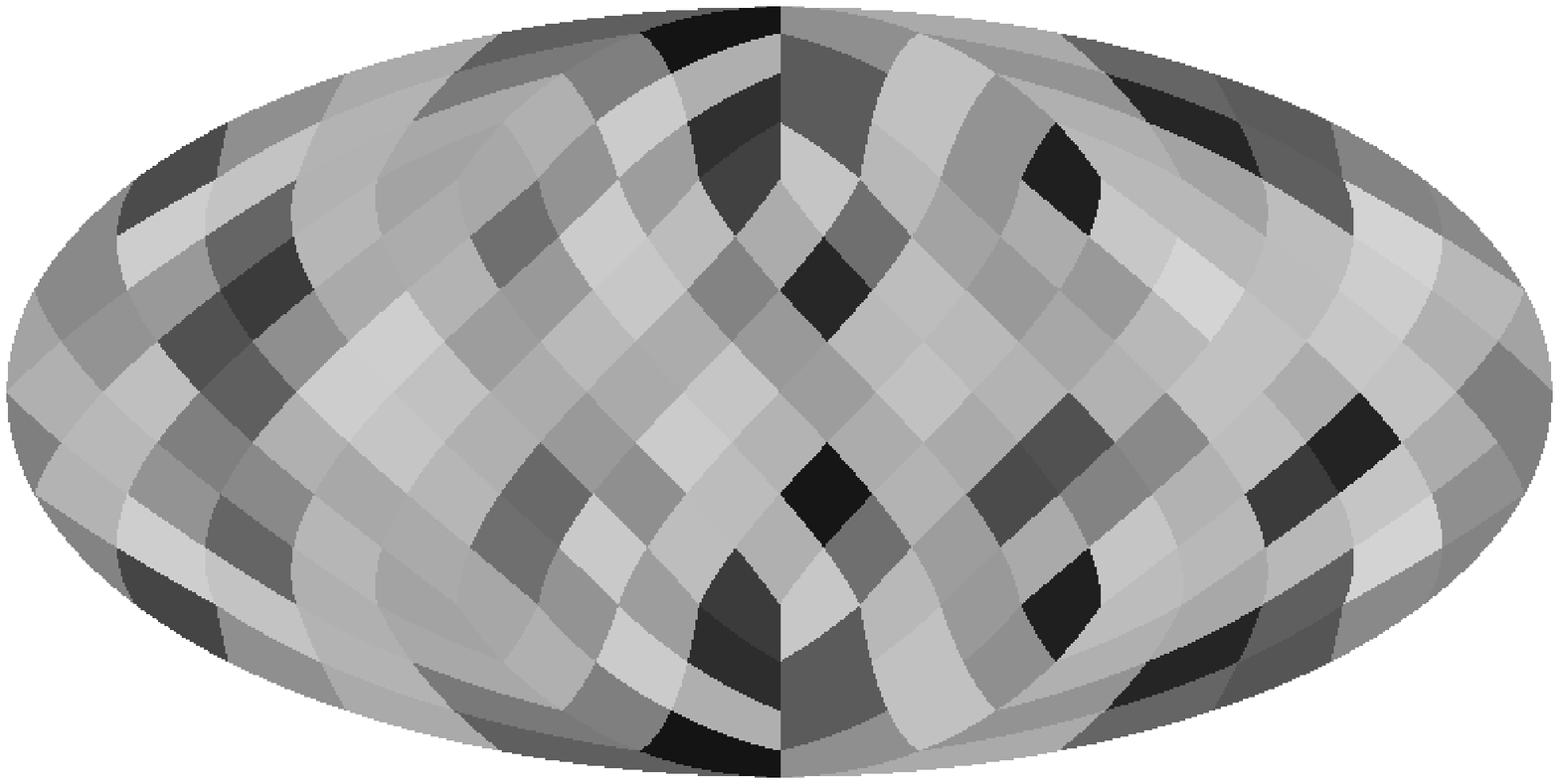}%}}
\includegraphics[width=0.28\linewidth]{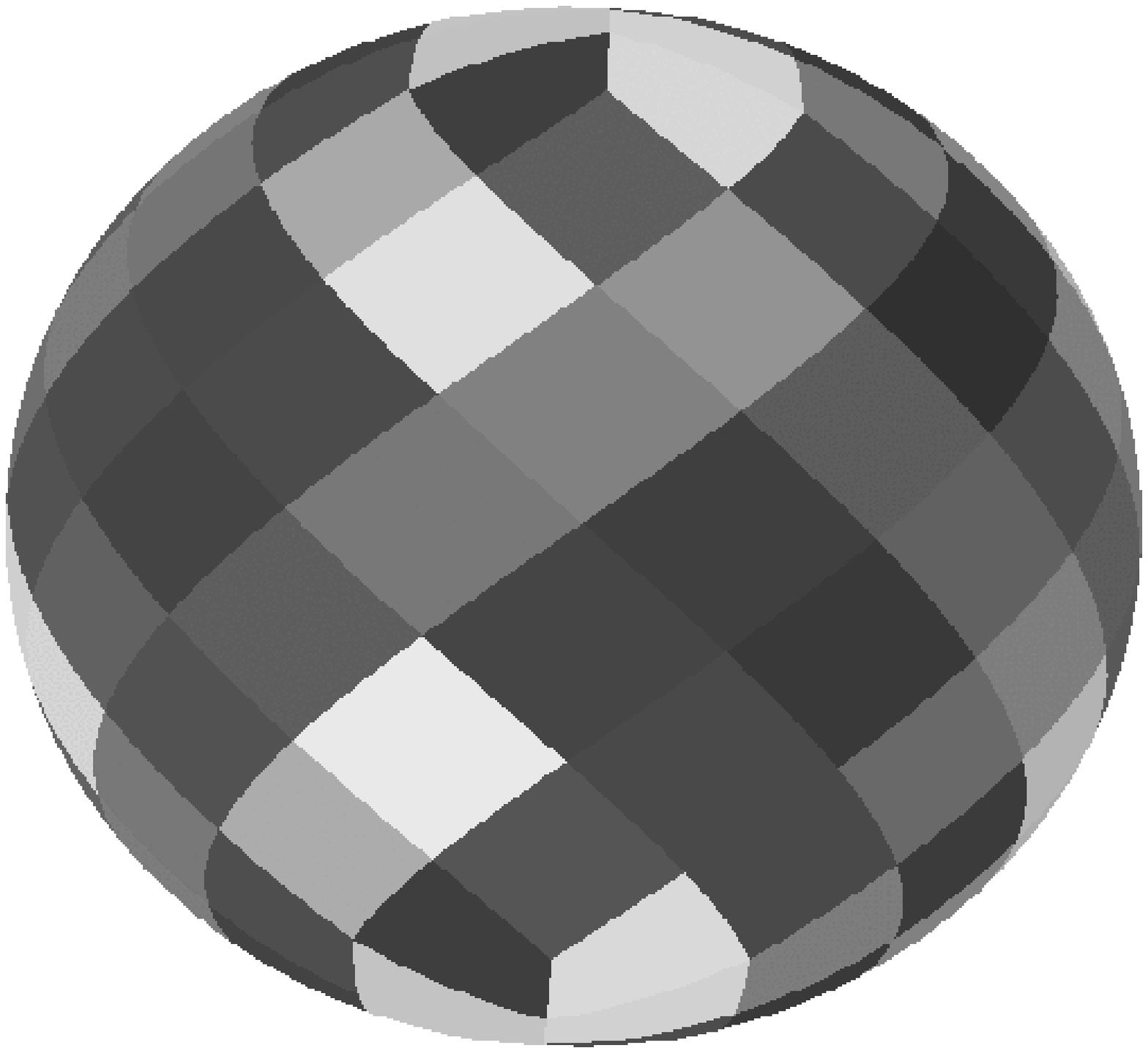}}}
\caption{Left column is the  Molldweide projection of pixelization grids
(from top to bottom):
(a) the standard pixelization grid of the GLESP\,1.0
% described in
% \refcite{glesp},  % Doroshkevich et al. (2005),
(b) the GLESP-pol rectangular grid with the same number of pixels
per each ring
(the so called case `grN'),
(c) the GLESP-pol grid with an pixel number increment 4 (case `grS')
starting
from 10 pixels near poles but not greater than a given resolution in the
equator ring,
(d) the HEALPix grid.
The right column shows the corresponding pixelization in the vicinity of
the polar cups.
 }
\label{fig1}
\end{figure}

This scheme for the temperature anisotropy was realized in the publicly
available code presented
in {\tt www.glesp.nbi.dk} and below we call it ``GLESP\,1.0''
denoting as `grA' (grid of equal areas).
The new code GLESP-pol
is based on the same definition of the roots of the Legendre polynomials,
but different  definition of the
pixel area (see the item 3 from the top):
\begin{itemize}
\item
     All the rings have the same number of pixels
     (case `grN' --- grid of equal number of pixels in ring)
     by default. That is a cylindric projection of the sphere.
     We have also checked a special case when
     the number pixels of ring has
     an increment 4 starting from 10 pixels near the poles
     (case `grS' --- special grid). The old scheme (GLESP\,1.0)
     when all the pixels have nearly equal area (grA) is also accessible.
\end{itemize}
In fig.\ref{fig1}, we show the differences in pixelization of
the GLESP\,1.0, the GLESP-pol (grN), the GLESP-pol (grS) and the HEALPix.
One can see, in the central part of the map the properties of the GLESP-pol
(grN,grS) are the same as for the GLESP\,1.0,
but in the vicinity of the polar cups they are significantly different
due to over-pixelization. This modification is vital for estimation of
the ${}_{0,\pm 2}a_{\l,m}$-coefficients due to specific
behaviour of ${}_{\pm 2}Y_{\l,m}(\theta,\phi)$-spherical harmonics
(see Appendix for details).

\section{Errors ``$a_{\ell,m}\rightarrow$ map$\rightarrow a_{\ell,m}$''
transition}

This section is devoted to estimation of the error of the $a_{\l,m}$
reconstruction by implementation different regimes of the HEALPix\,2.11
and the GLESP-pol (grN and grS) packages.
The error bars were defined in the following way. Let us take
the coefficients $a^{T,E,B}_{\l,m}$  for temperature anisotropy T and E and B
modes of polarization for $\Lambda$CDM concordance model by implementation
of the Monte Carlo simulation of the random Gaussian
signal\footnote{We have used the Gaussian signal for
simplicity.
However, only for the Gaussian signal the power spectrum
$C(\l)=(2\ell+1)^{-1}\sum_{m=-\ell}^{\ell}|a^{T,E,B}_{\l,m}|^2$
is the unique characteristic, which determines all the statistical
properties.}.
For the $C_{BB}$--spectrum, we have taken a value equal to $10^{-11}$ just
to escape zero in the test calculations.

Then, by implementation of the HEALPix\,2.11 and the GLESP-pol
packages we used these $a^{T,E,B}_{\l,m}$ coefficients
to create the map of the signal, keeping for both packages the same number
of pixels. Thus, the HEALPix map $M^i_H$, and the CLESP-pol map
$ M^p_G$ are defined as
\begin{eqnarray}
M^i_H = \textbf{H}a^{T,E,B}_{\l,m},\hspace{0.5cm}
	 M^p_G=\textbf{G}a^{T,E,B}_{\l,m}
\label{eq7}
\end{eqnarray}
where $\textbf{H}$ and $\textbf{G}$ are the HEALPix and
the GLESP-pol operators for ``$a_{\l,m}$ to map'' transition,
the index $i$ marks the number of iteration (0 as default, or 1-4),
the index $p=1$ marks the GLESP-pol grN pixelization, and $p=2$ corresponds
to the grS pixelization. Let us define the corresponding transition
``map to $a_{\l,m}$'' as  $\textbf{H}^{-1}$ for the HEALPix and
$\textbf{G}^{-1}$ for the GLESP-pol:
\begin{eqnarray}
b^{T,E,B}_{\l,m}=
   \textbf{H}^{-1}M^i_H=\textbf{H}^{-1}\textbf{H}a^{T,E,B}_{\l,m},\nonumber\\
c^{T,E,B}_{\l,m}= \textbf{G}^{-1}M^p_G=
   \textbf{G}^{-1}\textbf{G}a^{T,E,B}_{\l,m}
\label{eq8}
\end{eqnarray}
where $b^{T,E,B}_{\l,m}$ and $c^{T,E,B}_{\l,m}$ are now the reconstructed
coefficients for the HEALPix and GLESP-pol correspondingly. For idealistic
case, when $a_{\ell,m}\rightarrow$ map$\rightarrow a_{\ell,m}$'' transition
has no error bars, the reconstructed $b^{T,E,B}_{\l,m}$ and $c^{T,E,B}_{\l,m}$
have to be identical to the input coefficients $a^{T,E,B}_{\l,m}$,
and $\textbf{H}^{-1}\textbf{H}=\textbf{H}\textbf{H}^{-1}=\textbf{I}$,
$\textbf{G}^{-1}\textbf{G}=\textbf{G}\textbf{G}^{-1}=\textbf{I}$,
where $\textbf{I}$ is just a unit matrix.
In reality, neither the HEALPix, nor the  GLESP-pol packages have
non-zero error of the reconstruction due to
window functions of the pixels and computational errors for
the spherical harmonics. That means that corresponding absolute errors
for the `` $a_{\ell,m}\rightarrow$ map$\rightarrow a_{\ell,m}$''
transition can be defined as follows:
\begin{eqnarray}
{R^i_H}_{\l,m}=
  \Re e (b^{T,E,B}_{\l,m})-\Re e (a^{T,E,B}_{\l,m}),\hspace{0.5cm}%\nonumber\\
{I^i_H}_{\l,m}=
  \Im m (b^{T,E,B}_{\l,m})-\Im m (a^{T,E,B}_{\l,m}),\nonumber\\
{R^p_G}_{\l,m}=
  \Re e (c^{T,E,B}_{\l,m})-\Re e (a^{T,E,B}_{\l,m}),\hspace{0.5cm}%\nonumber\\
{I^p_G}_{\l,m}=
  \Im m (c^{T,E,B}_{\l,m})-\Im m (a^{T,E,B}_{\l,m}),\nonumber\\
\label{eq9}
\end{eqnarray}
where $\Re e$ and $\Im m$ stand for the real and imaginary parts
of the coefficients.
Thus, the relative error is given by
\begin{eqnarray}
{r^i_H}_{\l,m}=\frac{{R^i_H}_{\l,m}}{\Re e (a^{T,E,B}_{\l,m})},\hspace{0.5cm}
{y^i_H}_{\l,m}=\frac{{I^i_H}_{\l,m}}{\Im m (a^{T,E,B}_{\l,m})},\nonumber\\
{r^p_G}_{\l,m}=\frac{{R^p_G}_{\l,m}}{\Re e (a^{T,E,B}_{\l,m})},\hspace{0.5cm}
{y^p_G}_{\l,m}=\frac{{I^p_G}_{\l,m}}{\Im m (a^{T,E,B}_{\l,m})},\nonumber\\
\label{eq10}
\end{eqnarray}

Note that defined in Eq(\ref{eq10}) relative errors are related
to the error of the power spectrum
\begin{eqnarray}
\frac{ \Delta C_{\ell}}{C_{\ell}}=
  \frac{\sum_m(|g_{\l,m}|^2-|a_{\l,m}|^2)}{\sum_m|a_{\l,m}|^2}=
  \frac{\sum_m|a_{\l,m}|^2(\delta_{\l,m}+\delta^*_{\l,m})}{\sum_m|a_{\l,m}|^2}
\label{eq11}
\end{eqnarray}
where $g_{\l,m}$ and $a_{\l,m}$ denote the reconstructed  and
input multipole coefficients, and $g_{\l,m}=a_{\l,m}(1+\delta_{\l,m})$.
Taking into account that $\delta_{\l,m}+\delta^*_{\l,m}=2\Re e(\delta_{\l,m})$
and $\Re e(\delta_{\l,m})=
    [(\Re ea_{\l,m})^2r_{\l,m}+(\Im ma_{\l,m})^2y_{\l,m}]/|a_{\l,m}|^2$,
where $r_{\l,m}$ and $y_{\l,m}$ denote relative errors for real and
imaginary parts from Eq(\ref{eq10}), we gets:
\begin{eqnarray}
 \frac{ \Delta C_{\ell}}{C_{\ell}}=
   2\frac{\sum_m[(\Re ea_{\l,m})^2r_{\l,m}+
     (\Im ma_{\l,m})^2y_{\l,m}]}{\sum_m|a_{\l,m}|^2}
\label{eq12}
\end{eqnarray}
From Eq(\ref{eq12}) clearly seen that the error of reconstruction of
real and imaginary part of each $\ell,m$ coefficient propagates to
the error of the power spectrum through weighting coefficients
\begin{eqnarray}
w^r_{\l,m}=\frac{(\Re ea_{\l,m})^2}{\sum_m|a_{\l,m}|^2},
    \hspace{0.5cm}w^y_{\l,m}=\frac{(\Im m a_{\l,m})^2}{\sum_m|a_{\l,m}|^2}
\label{eq13}
\end{eqnarray}
and formally depends on the power spectrum and the morphology of
the input signal. This is why in addition to
the input random Gaussian CMB signal we will discuss the errors of
reconstruction of the multipoles and the power spectrum for
very asymmetric maps, like WMAP\,5 Q and V bands.

\subsection{Errors for HEALPix and GLESP-pol temperature anisotropy}

\begin{figure}[!th]
 \hbox{\hspace*{0.01cm}
%\begingroup
\centerline{\includegraphics[width=0.3\linewidth]{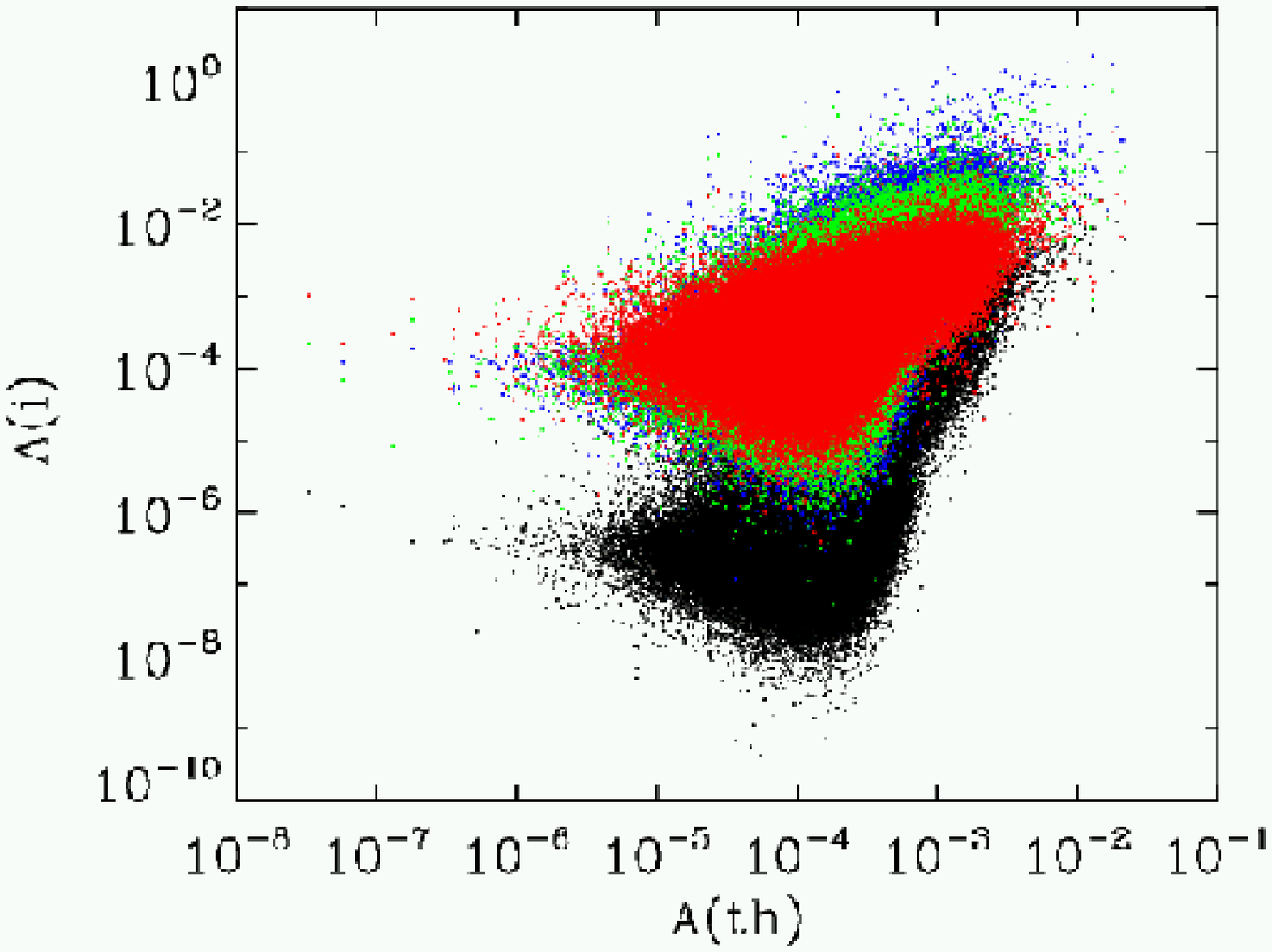}%}}
\includegraphics[width=0.3\linewidth]{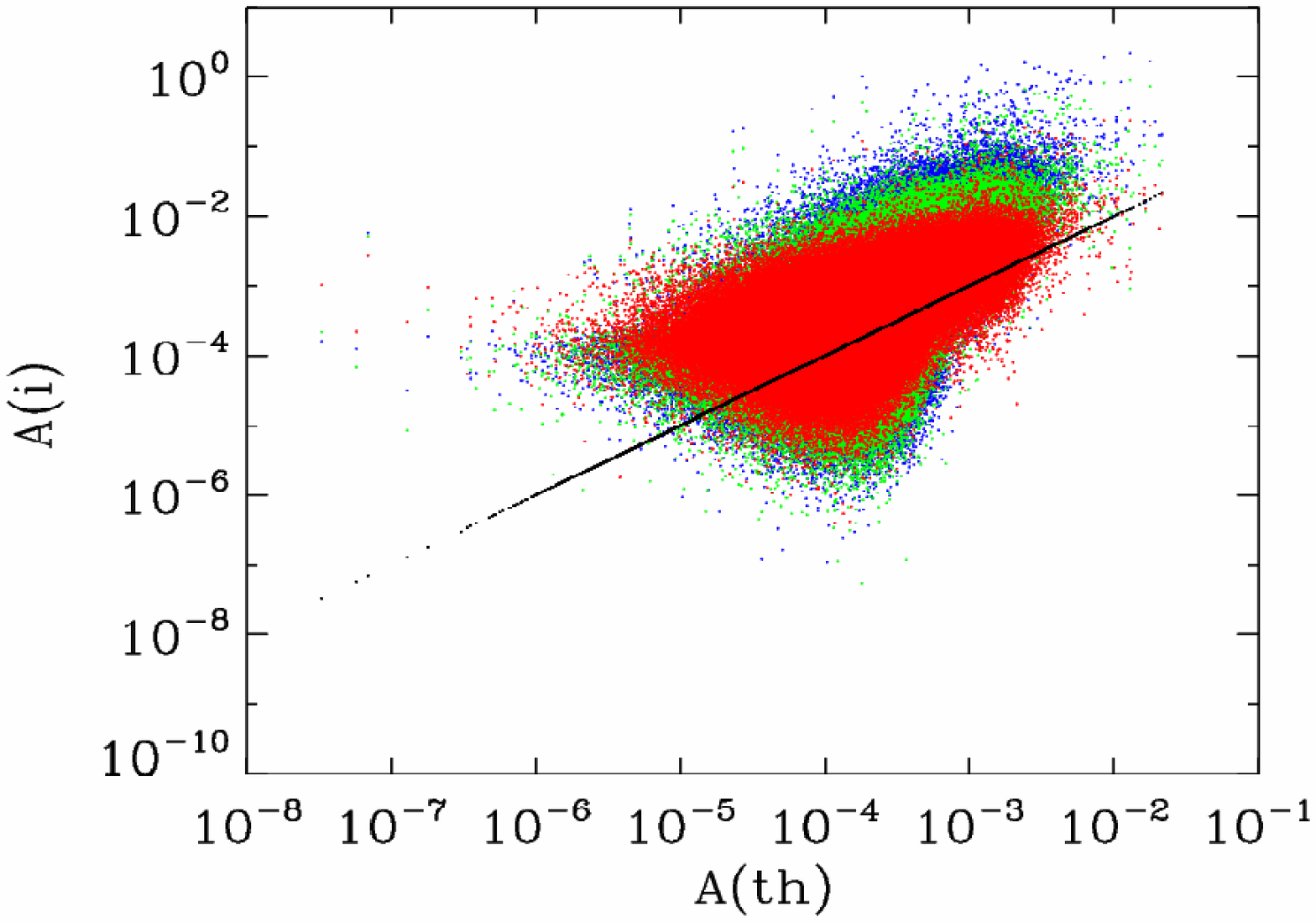}%}}
\includegraphics[width=0.28\linewidth]{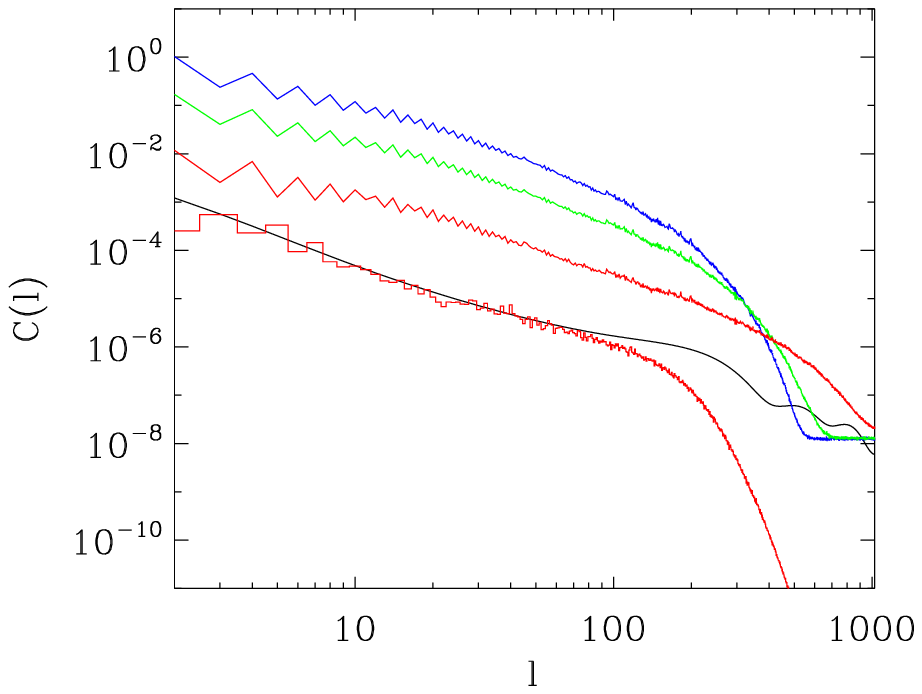}}}
\caption{Left. The $A(i)$ versus $A(th)$ diagram for the WILC5
(the black dots), K-band (the blue dots),
KA-band (the green dots), V-band (the red dots).
Both $A(i)$ and $A(th)$ are in mK.
Middle. The same as the left plot, but for replacement of the WILC5 by
$A(th)$. Right. The power spectrum $C(\l) mK^2$ for the WILC5
(the red bottom line), for random realization (the black line),
the K band ( the solid blue line), the KA-band (the green solid line),
and for V band (the red solid line).
} 
\label{f1}
\end{figure}

We will start our analysis from estimation of the dynamical range of
variations of the temperature anisotropy for Nside=1024.
For that we will use
the input signal, which corresponds to random Gaussian CMB, and Internal
Linear Combination Map (WILC5) from the LAMBDA archive
({\tt http://lambda.gsfc.nasa.gov/})
and corresponding K, Ka, Q, V and W total channels maps. In Fig\ref{f1}
we plot the diagram  $A(i)$ versus $A(th)$, where$A(i)=string(|a_{\l,m})|$,
where
the operator ''string'' transforms the $|a_{\l,m}|$ coefficients to
one dimensional string $|a_{1,0}|,|a_{1,1}|...$
for K--W bands, and $A(th)$ is the string for
the random Gaussian CMB signal.
From this diagram one can see that, for example, to estimate
the $a_{\l,m}$ coefficients by using
ILC method we should have, at least, the relative error better than
$10^{-3}-10^{-4}$ if the low frequency
K-band is included to the analysis. T
he accuracy of the $a_{\l,m}$ coefficients should be about 3--4 orders of
magnitude better if we are interested in different sort of coupling
between different multipoles (so called the non-Gaussianity tool).

\begin{figure}[!th]
\hbox{\hspace*{0.01cm}
%\begingroup
\centerline{\includegraphics[width=0.3\linewidth]{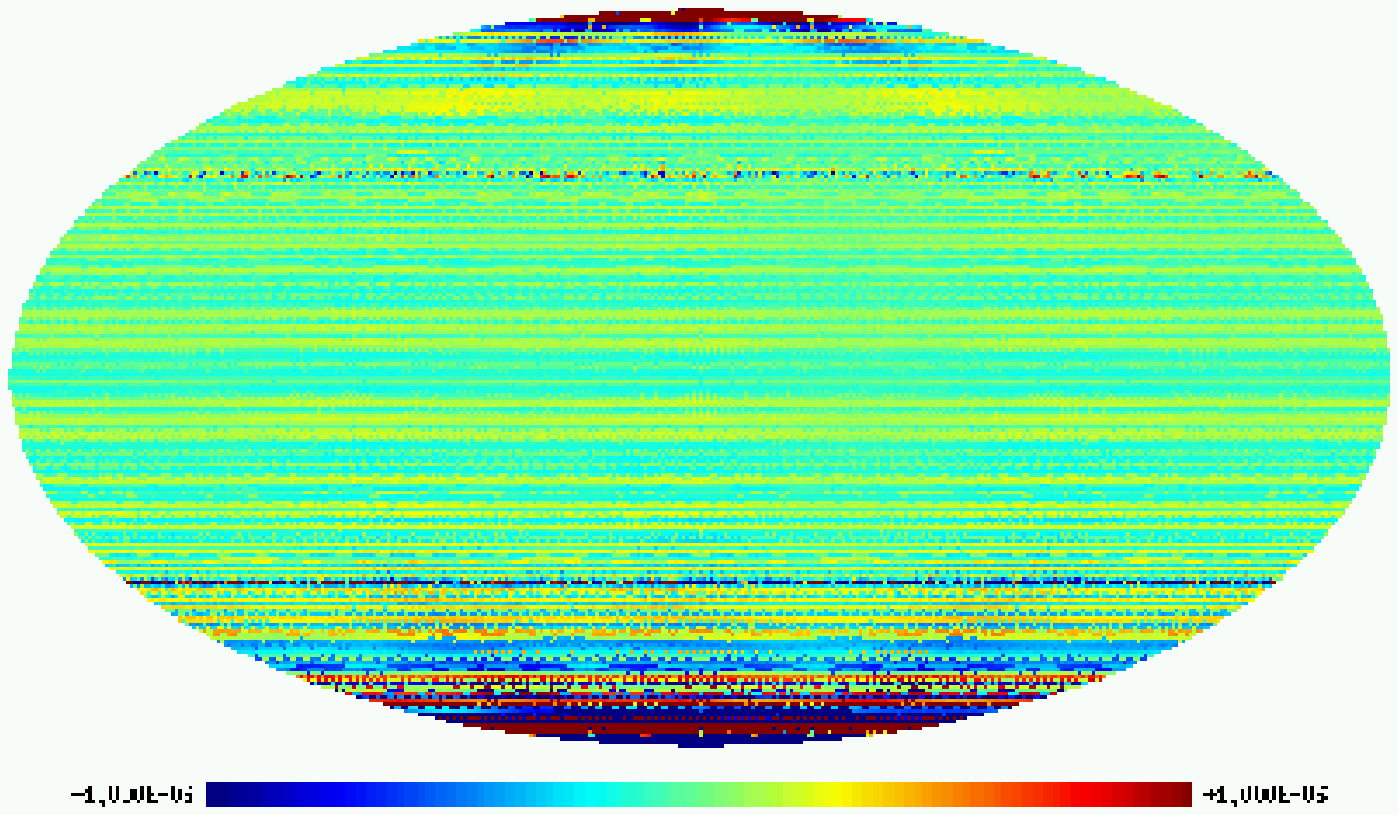}%}}
%\hbox{\hspace*{0.01cm}
\includegraphics[width=0.3\linewidth]{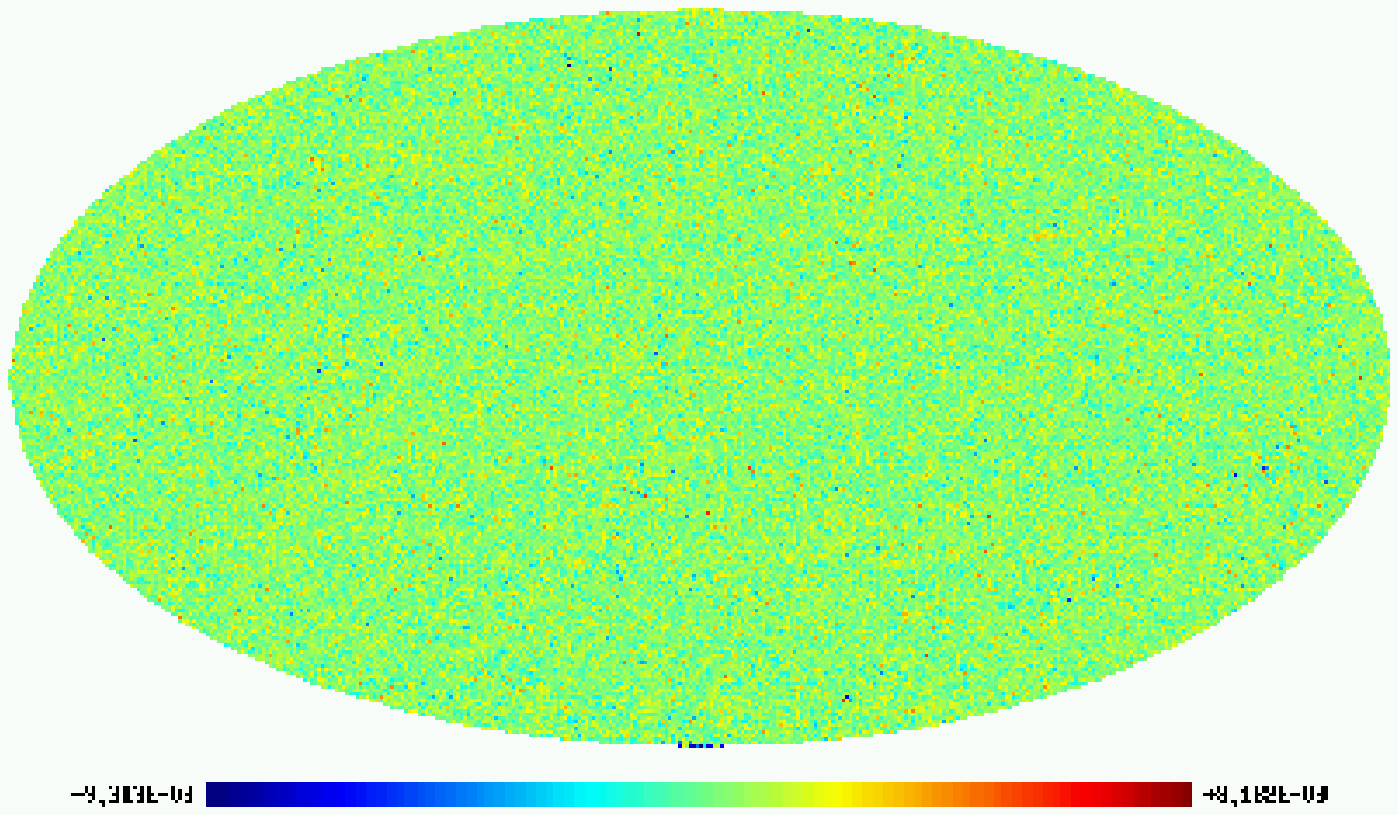}}}
\hbox{\hspace*{0.01cm}
\centerline{\includegraphics[width=0.3\linewidth]{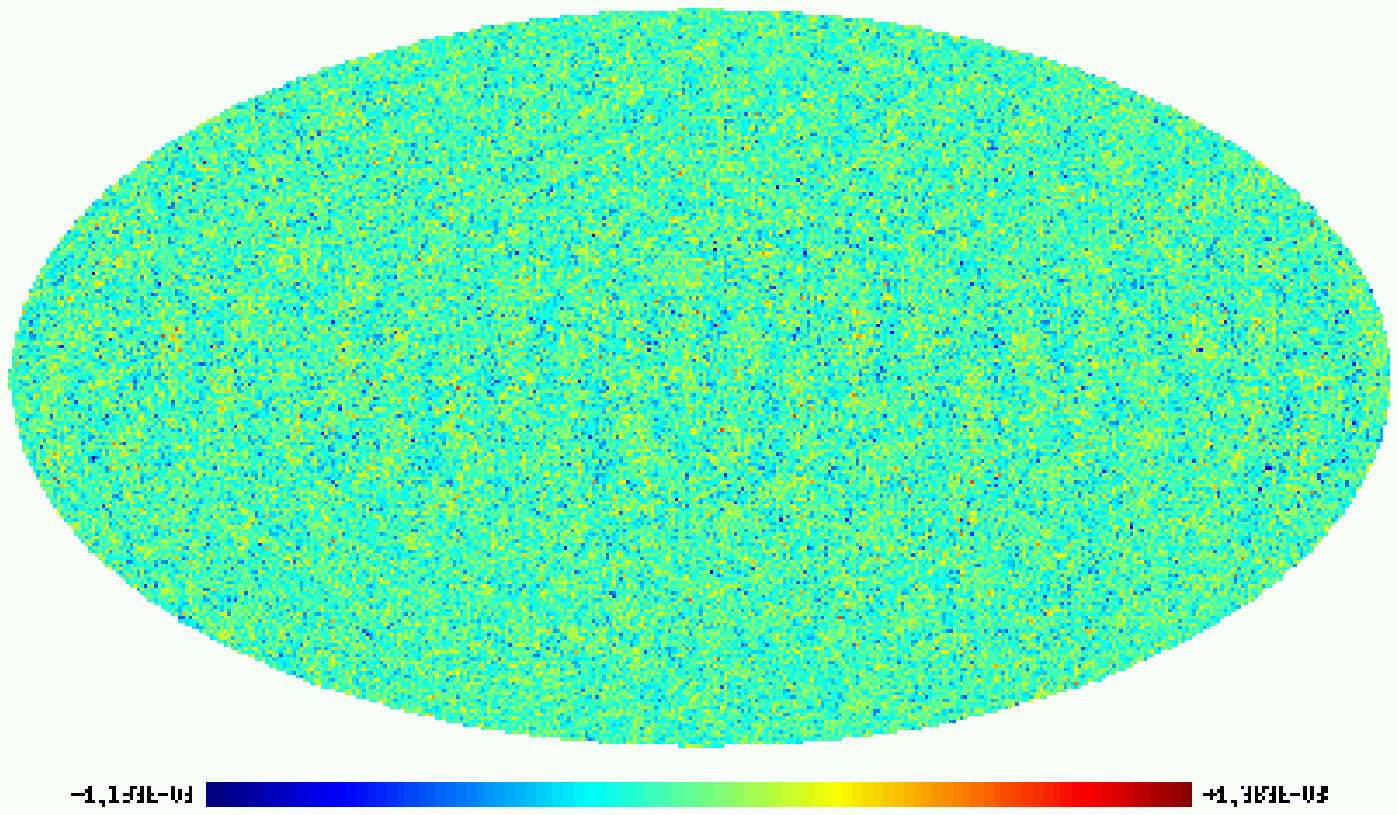}%}}
%\hbox{\hspace*{0.01cm}
\includegraphics[width=0.3\linewidth]{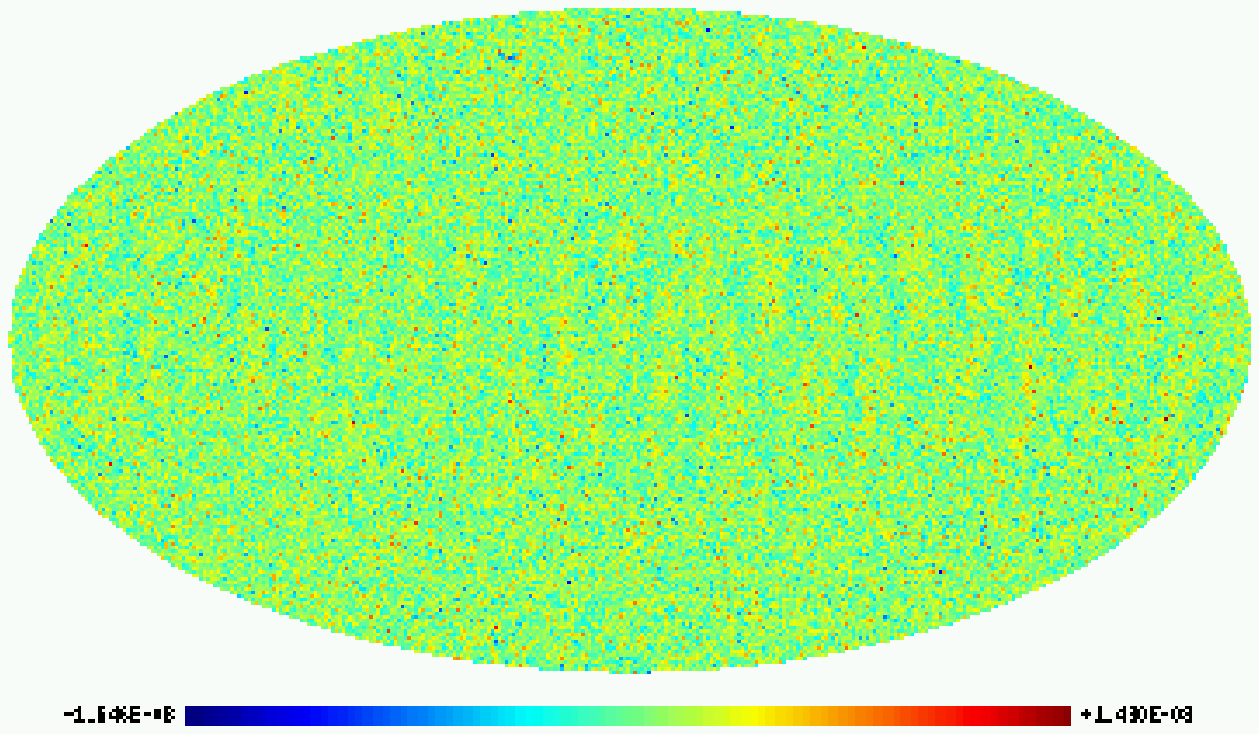}}}
\caption{ The differences of reconstructed and input maps for
the HEALPix\,2.11 (the top pair) and the GLESP-pol (the bottom pair).
Top left plot corresponds to ``zero iteration`` key
(the color scale is $-10^{-6},10^{-6}mK$), top right is for 4 iterations
(the color scale is $-10^{-8},10^{-8}mK$).
Bottom left-the GLESP-pol reconstruction for the grN mode
(the color scale is $-10^{-8},10^{-8}mK$).
Top right corresponds to the grS mode (the color scale is $-10^{-8},10^{-8}mK$).
The number of pixels for the
HEALPix\,2.11 and the GLESP-pol are practically the same.
No correction by the  window function of the pixels.
 }
\label{fig2}
\end{figure}

To estimate the errors of the $a_{\ell,m}\rightarrow$ map$\rightarrow
a_{\ell,m}$'' transition, let us start from
the analysis of the temperature anisotropy for the HEALPix\,2.11 and
the GLESP-pol packages.
In Fig.\ref{fig2}, we show the maps for $b^0_{\l,m}-a_{\l,m}$-signal,
where $b^0_{\l,m}$ obtained  by implementation of the HEALPix\,2.11
``zero iteration`` key
(top left map) and 4 iterations (top right map).

\begin{figure}[!th]
\hbox{\hspace*{0.01cm}
%\begingroup
\centerline{\includegraphics[width=0.3\linewidth]{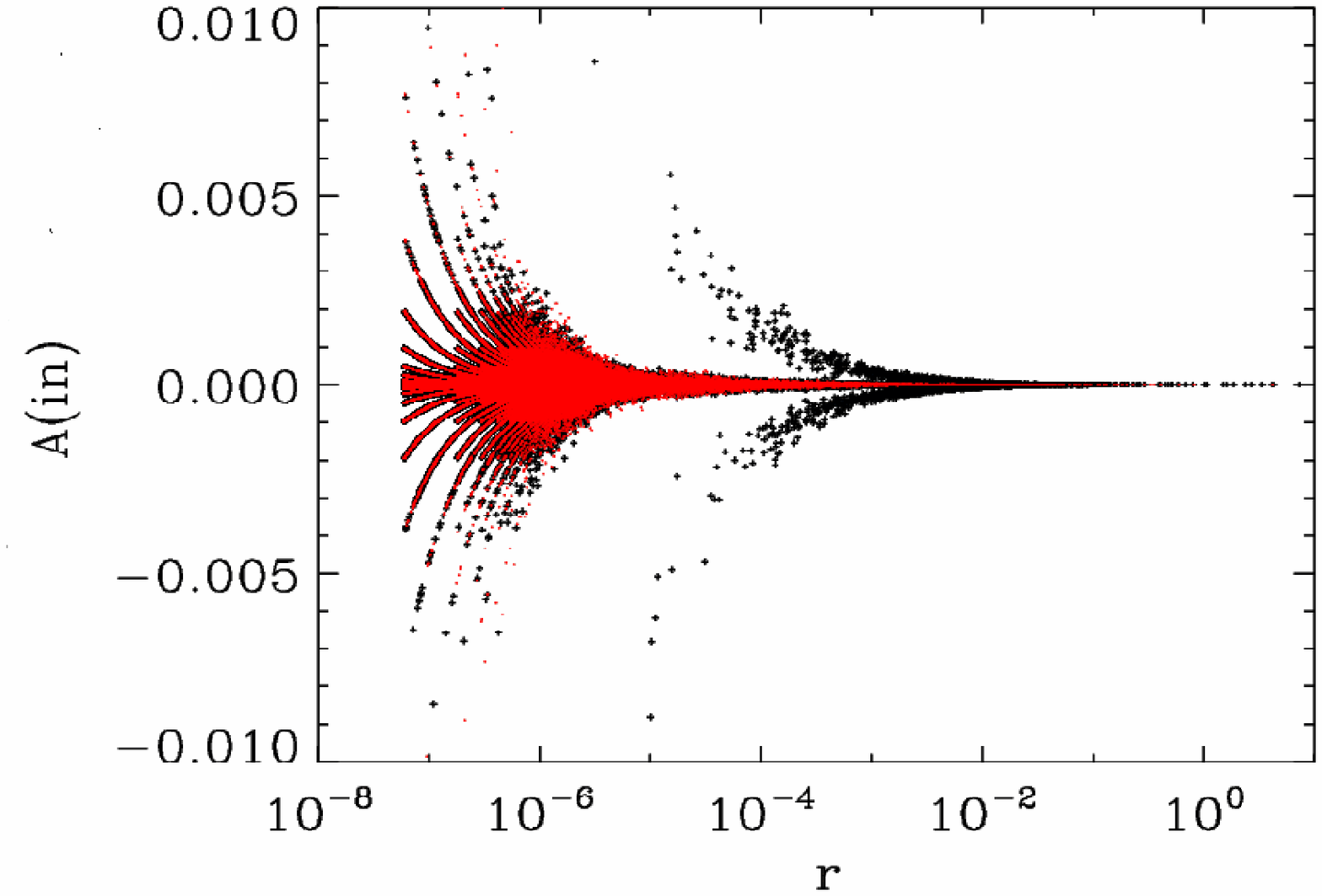}%}}
%\hbox{\hspace*{0.01cm}
\includegraphics[width=0.3\linewidth]{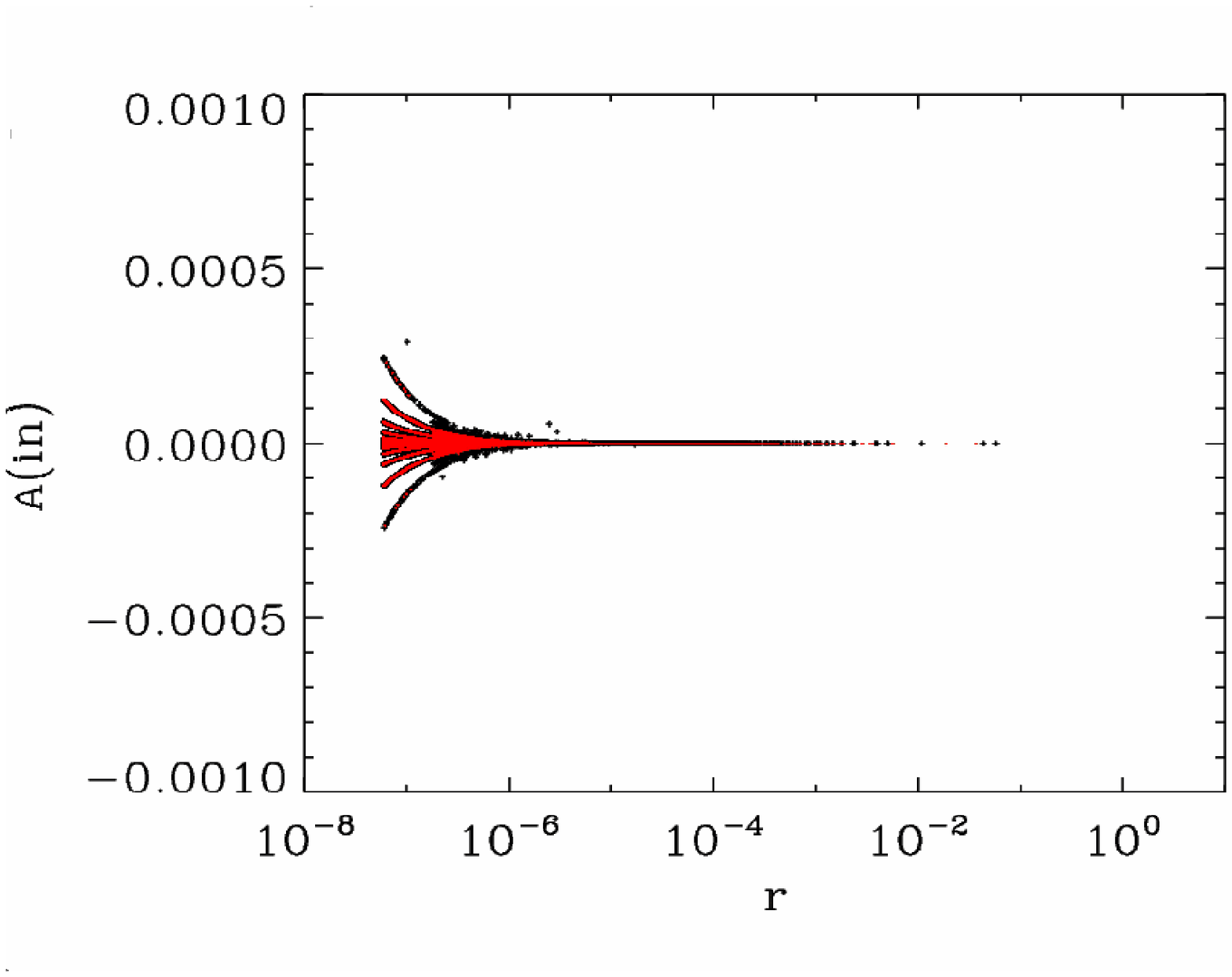}}}
\hbox{\hspace*{0.01cm}
\centerline{\includegraphics[width=0.3\linewidth]{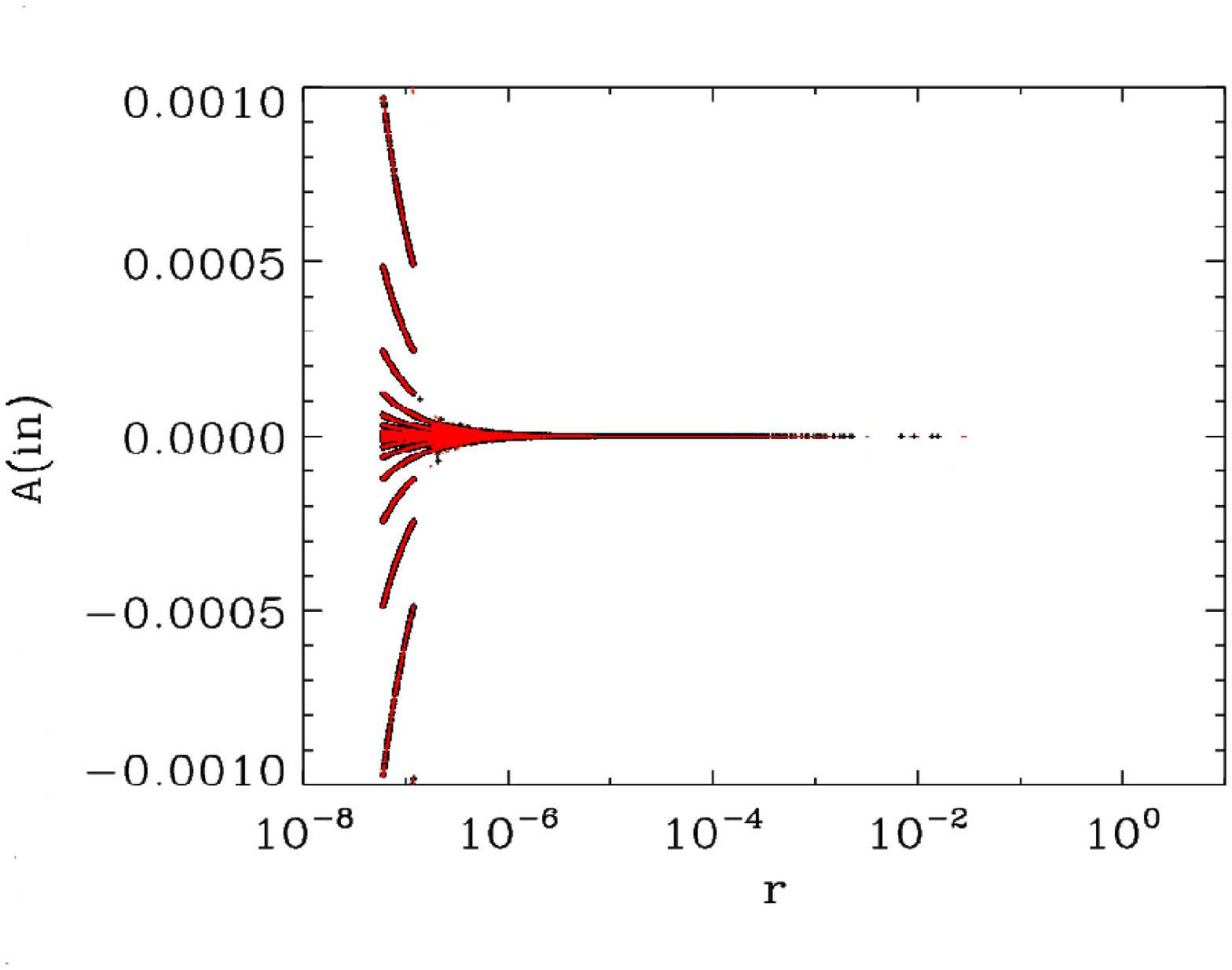}%}}
%\hbox{\hspace*{0.01cm}
\includegraphics[width=0.3\linewidth]{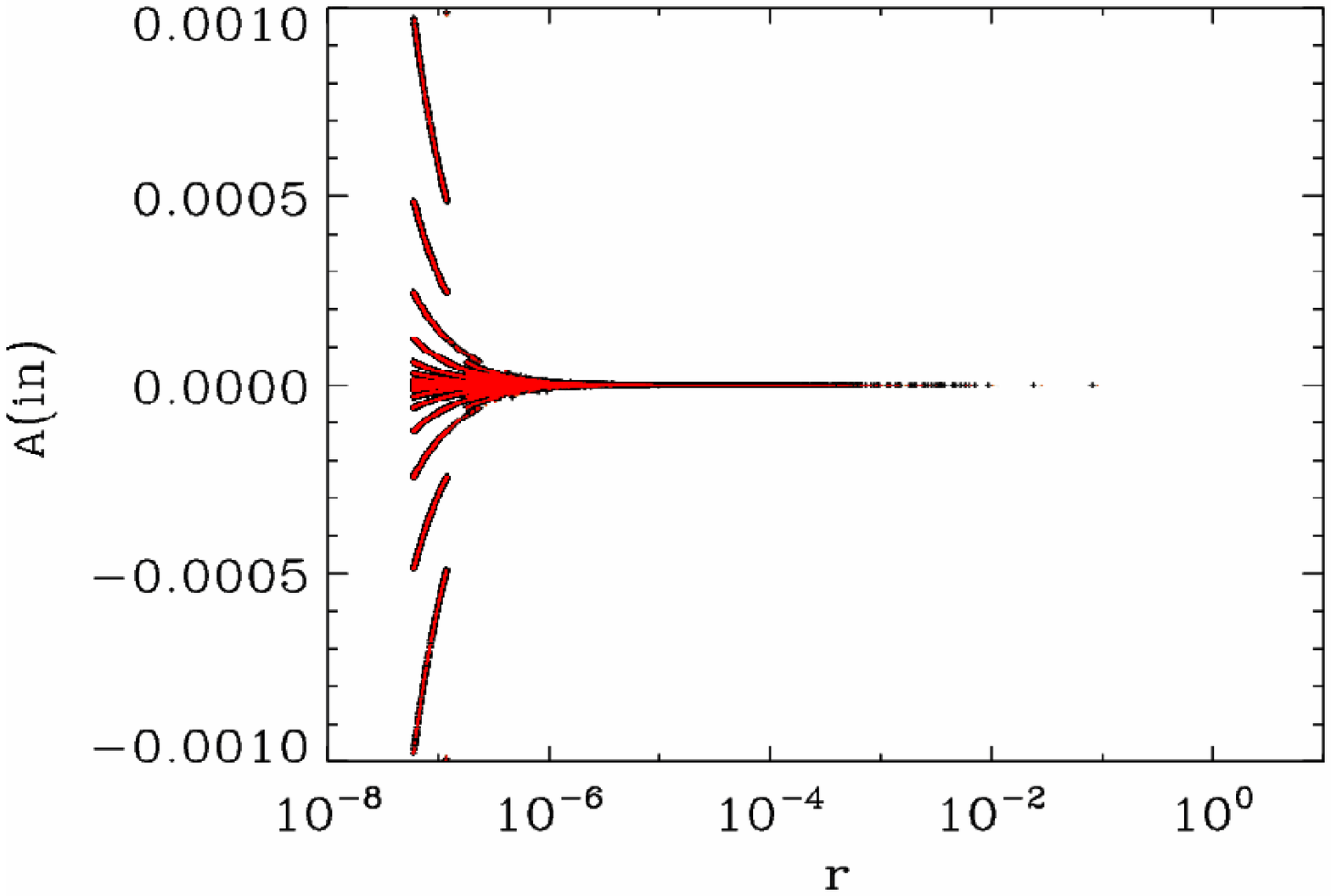}}}
\caption{ The errors of reconstructions  for the HEALPix\,2.11 (the top pair)
and the GLESP-pol (the bottom pair). Top left plot corresponds to
``zero iteration'' key, top right is for 4 iterations. Bottom left ---
the GLESP-pol reconstruction for the grN mode.
Top right corresponds to the grS mode. Black dots corresponds to
the real part of $\ell,m$ modes, the red dots show for the imaginary part.
 }
\label{fig3}
\end{figure}

One can see, that the top left map reveals all the peculiarities of
the pixelization, localized at the vicinity of the North and the South
poles.  The top left plot show the map of differences,
but after 4 iterations.
No visible large scale defects can be found.
The bottom left and right maps represent the GLESP-pol pixelization
without any iterations for the same number of pixels, as the top one.
In Fig.\ref{fig3}, we plot the real and imaginary parts of the absolute
errors from Eq(\ref{eq9}) for the maps, shown in fig.\ref{fig2}).

The definition of function $A(in)$ and $r$ in Fig.\ref{fig3} is
the following: $A(in)=string(a_{\l,m})$, where
the operator ''string'' transforms the $a_{\l,m}$ coefficients to
one dimensional string $a_{1,0},a_{1,1}...$, and $r=
|{r^0_H}_{\l,m}, {r^4_H}_{\l,m}, {y^0_H}_{\l,m}, {y^4_H}_{\l,m}|$
(see Eq.(\ref{eq10}).

Fig.\ref{fig3} clearly show that the HEALPix\,2.11 (no iteration)
reconstruction is different for the real and imaginary parts of the signal.
For real part (see Fig.\ref{fig3}, top left), the relative
error has the secondary zone, localized at $r>10^{-5}$, while for imaginary
part the major part of the points belongs to $r<10^{-5}$. Note that even
after implementation of 4 iterations for the HEALPix\,2.11 and
grN, grS modes of the GLESP-pol, there exists a very small number
($\sim 10$) of modes with error
1\%-10\%.
Thus,  the main conclusion is that for Nside=1024 the HEALPix iterations
significantly improve the global morphology of the map and the error of
reconstruction, which is practically the same the HEALPix and the GLESP-pol.

\begin{figure}[!th]
\hbox{\hspace*{0.01cm}
%\begingroup
\centerline{\includegraphics[width=0.3\linewidth]{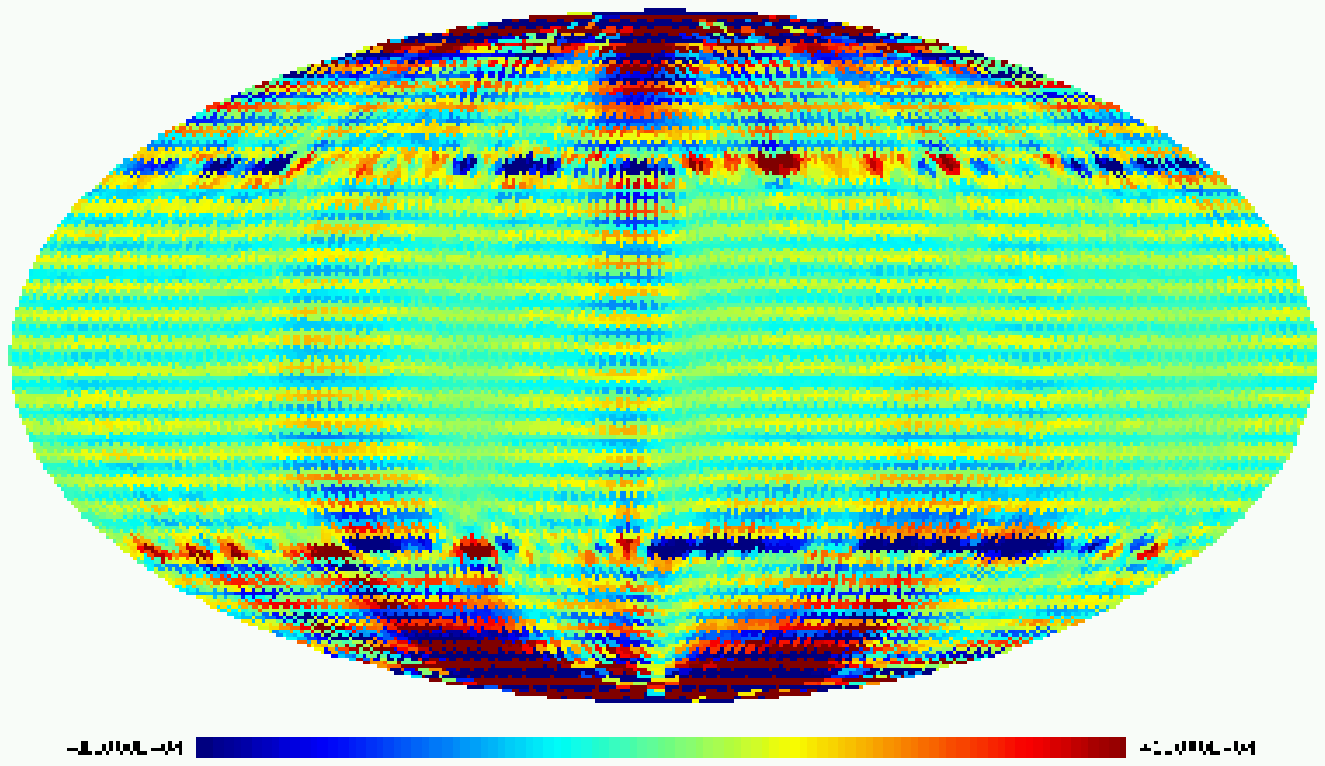}%}}
%\hbox{\hspace*{0.01cm}
\includegraphics[width=0.3\linewidth]{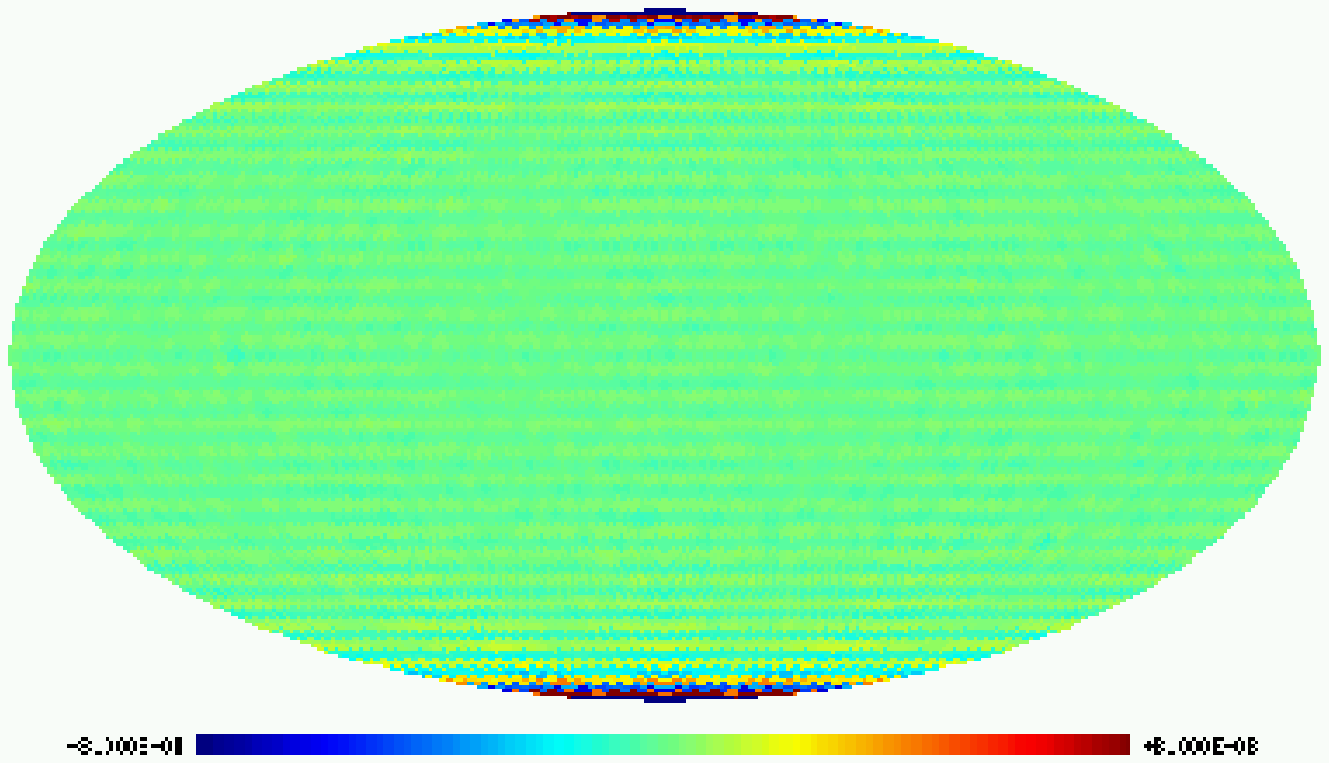}}}
\hbox{\hspace*{0.01cm}
\centerline{\includegraphics[width=0.3\linewidth]{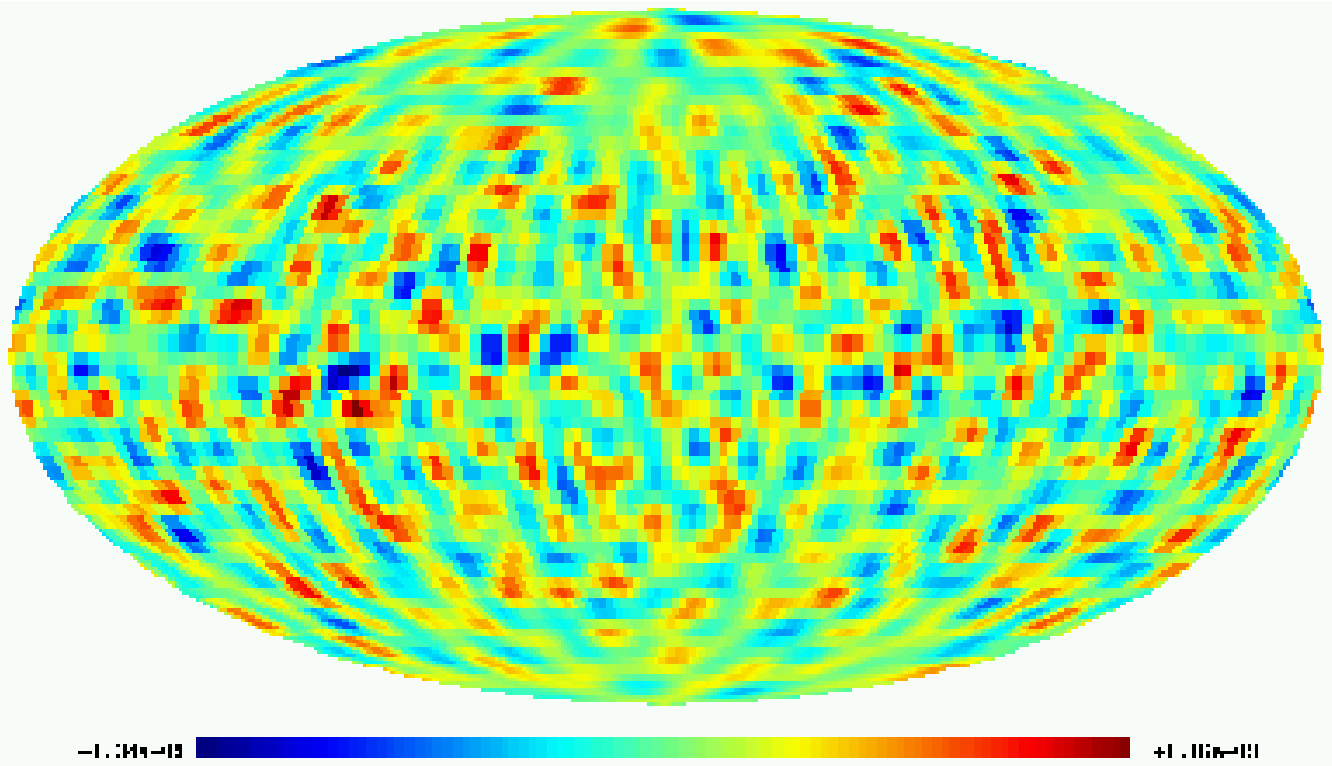}%}}
%\hbox{\hspace*{0.01cm}
\includegraphics[width=0.3\linewidth]{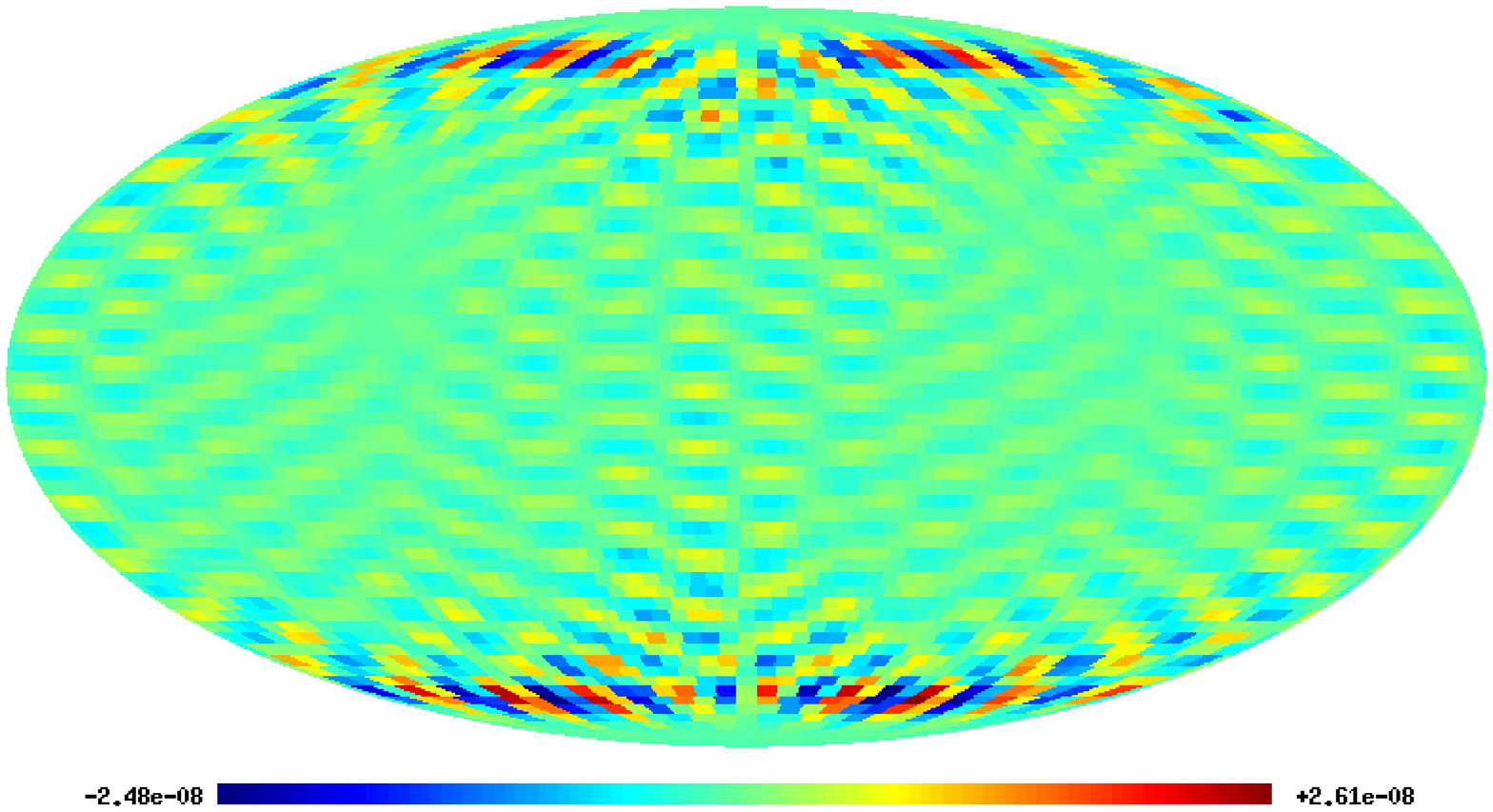}}}
\caption{ The differences of reconstructed and input maps for
the HEALPix\,2.11 for Nside=32 (the top pair) and the GLESP-pol
(the bottom pair). Top left plot corresponds to ``zero iteration''
key (the color scale is $-10^{-4},10^{-4}mK$),
top right is for 4 iterations (the color scale is
$-8\cdot10^{-8},8\cdot10^{-8}mK$).
Bottom left ---
the GLESP-pol reconstruction for the grN mode
(the color scale is $-10^{-8},10^{-8}mK$).
Bottom right corresponds to the grS mode (the color scale is
$-2.5\cdot10^{-8},2.5\cdot10^{-8}mK$). The number of pixels for the
HEALPix\,2.11 and the GLESP-pol are practically the same.
No correction by the  window function of the pixels.
 }
\label{fig4}
\end{figure}

\begin{figure}[!th]
\hbox{\hspace*{0.01cm}
%\begingroup
\centerline{\includegraphics[width=0.3\linewidth]{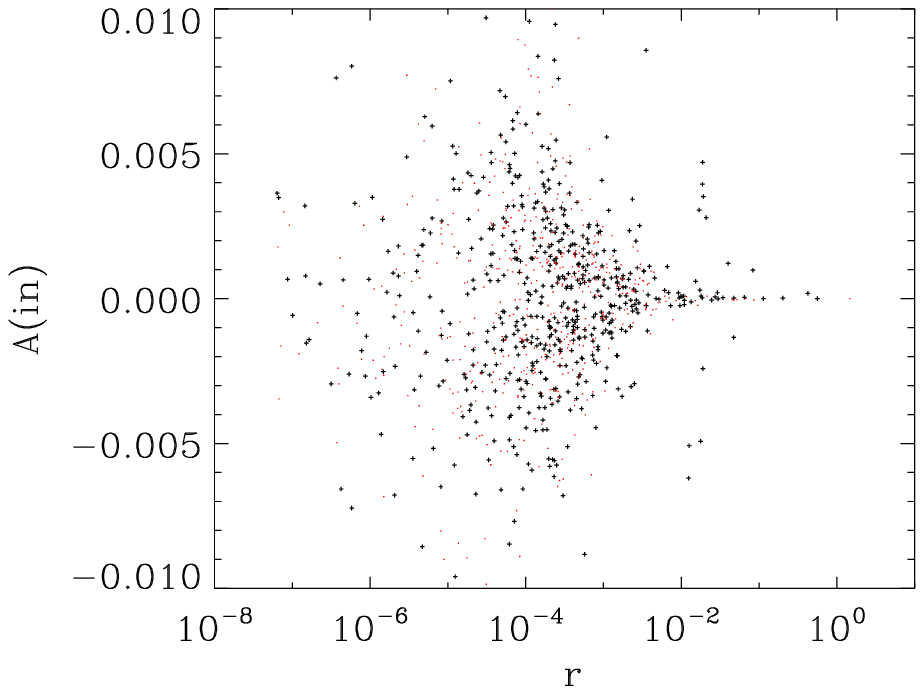}%}}
%\hbox{\hspace*{0.01cm}
\includegraphics[width=0.3\linewidth]{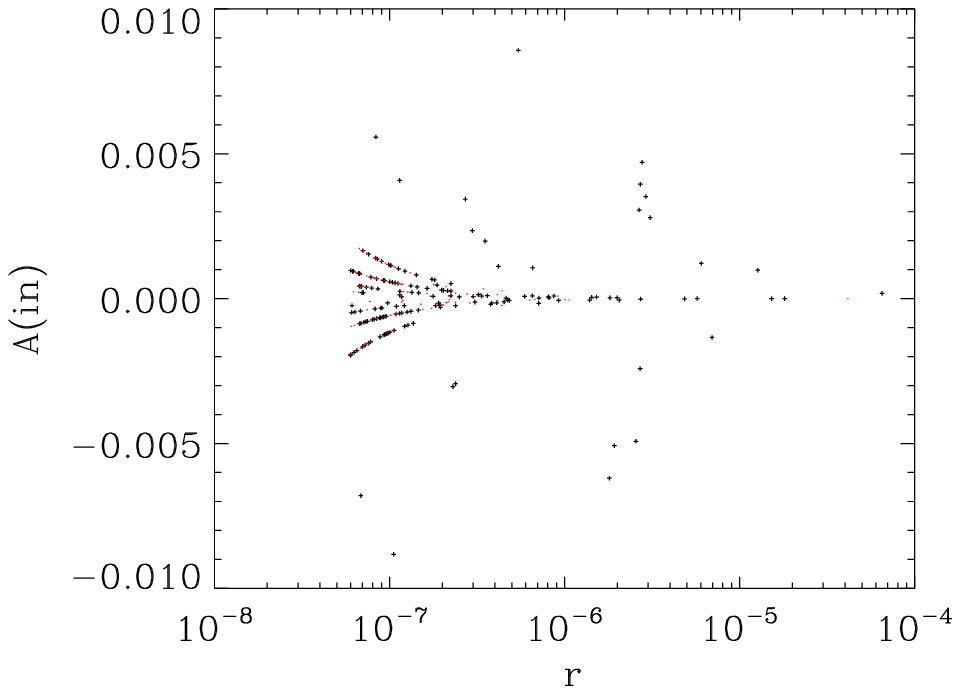}}}
\hbox{\hspace*{0.01cm}
\centerline{\includegraphics[width=0.3\linewidth]{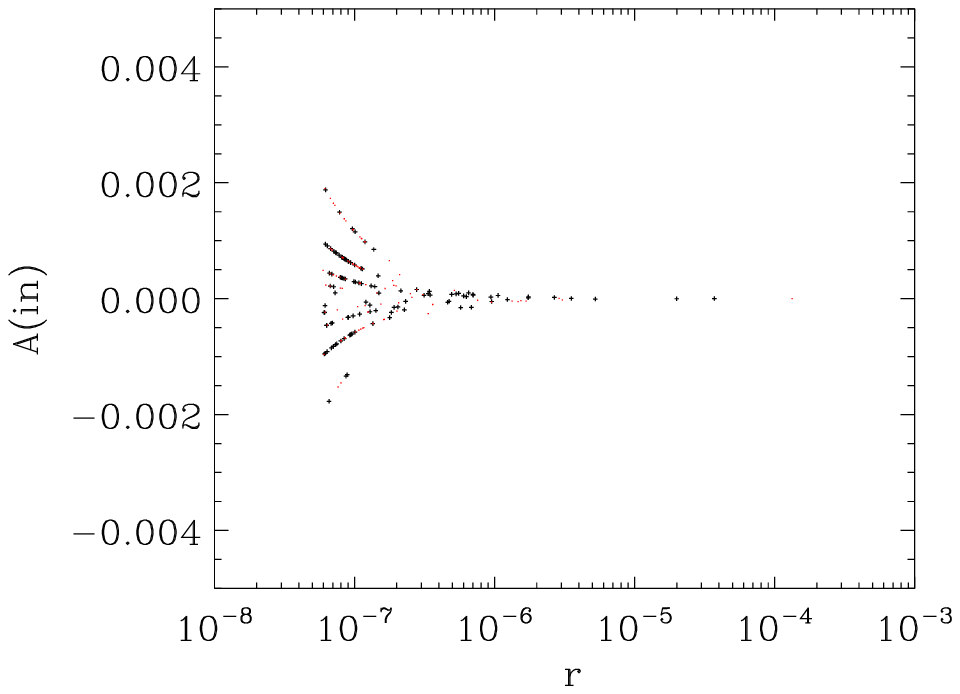}%}}
%\hbox{\hspace*{0.01cm}
\includegraphics[width=0.3\linewidth]{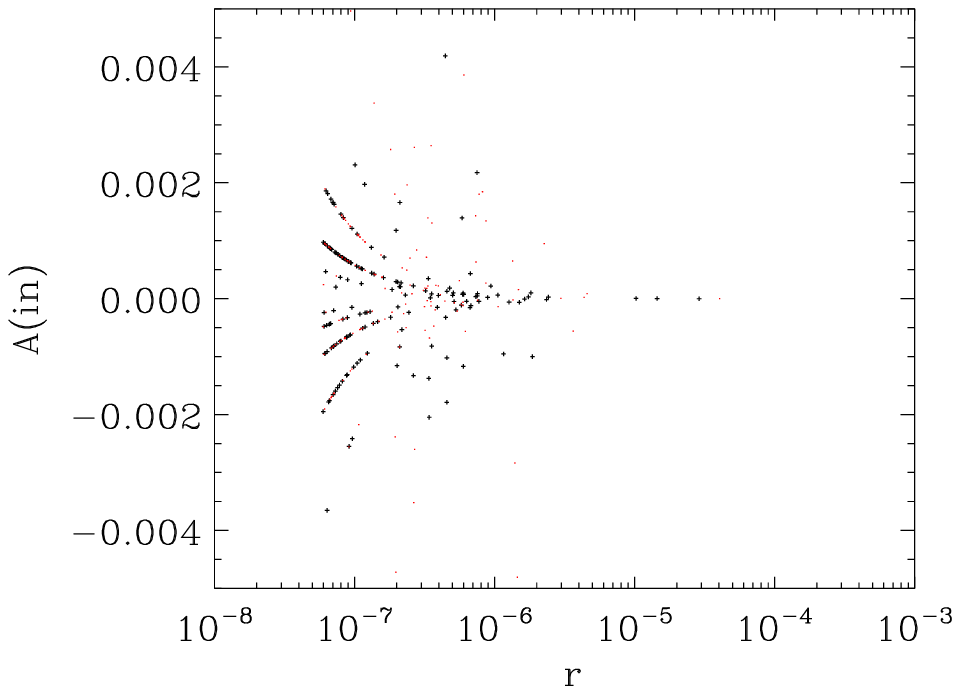}}}
\caption{ The errors of reconstructions  for the HEALPix\,2.11 (Nside=32)
(the top pair) and the GLESP-pol (the bottom pair).
Top left plot corresponds to ``zero iteration'' key, top right is
for 4 iterations. Bottom left ---
the GLESP-pol reconstruction for the grN mode. Top right corresponds
to the grS mode. Black dots corresponds to
the real part of $\ell,m$ modes, the red dots are for imaginary part.
 }
\label{fig5}
\end{figure}

The next question, which we would like to discuss, is how the error of
reconstruction depends on the number of pixels and their size.
Both these parameters are determined by the choice of Nside for the HEALPix,
or by
the maximal resolution of the map $\l_{max}$ for the GLESP-pol.
Answering this question in Fig.\ref{fig4} and
Fig.\ref{fig5}, we plot the corresponding maps of errors and $A(in)$
versus $r$ diagrams, similar to
Fig.\ref{fig2} and Fig.\ref{fig3}, but for less number of pixels.

It would be important to note that as for the high resolution map of
difference, shown in Fig.\ref{fig2}, as for the low resolution map
(see Fig.\ref{fig4}) for the HEALPix\,2.11 with zero iteration
the major component
of the error is related to $b_{\ell,m=0}$ mode.
These harmonics manifest themselves as horizontal
lines parallel to the Galactic plane. Two horizontal lines with
high amplitude signal along mark the HEALPix zones,
where the number of pixels for each equal latitude ring start to decrease,
when $\theta\rightarrow 0$ (the North pole), or $\theta\rightarrow  \pi$
(the South pole). However, after 4 iterations all these peculiarities of
the  map of errors were significantly suppressed, except two zones
around the North and the South cups. For the GLESP-pol package with
$\l_{max}=32$ the minimal level of errors for the reconstructed map
is given by the grN pixelization, when $r\ll 10^{-7}$ for major part of
the pixels, and $r\ll 10^{-5}$ for the grS pixelization.
It would be important to note that the accuracy of reconstruction of real and
imaginary parts of the coefficients of expansion are different for the
HEALPix and the GLESP-pol.
As it is follows from Fig\ref{fig3} and Fig\ref{fig5},
the HEALPix recovers the imaginary part significantly better than
a real one even for zero iteration mode. The GLESP-pol reconstructs real and
imaginary parts with nearly equal errors.

\section {The GLESP-pol and the HEALPix polarization.}

\begin{figure}[!th]
\hbox{\hspace*{0.01cm}
%\begingroup
\centerline{\includegraphics[width=0.3\linewidth]{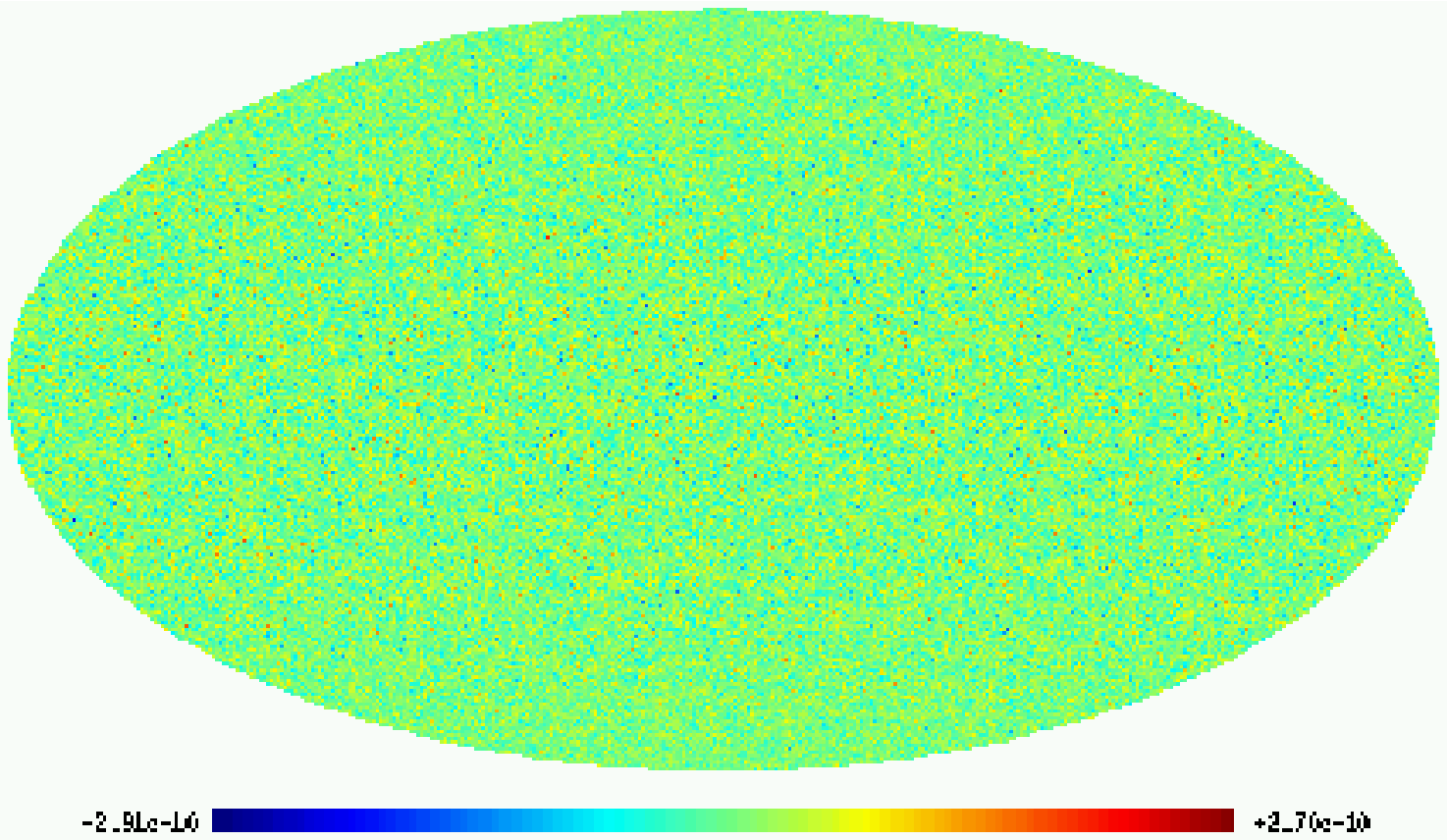}%}}
%\hbox{\hspace*{0.01cm}
\includegraphics[width=0.3\linewidth]{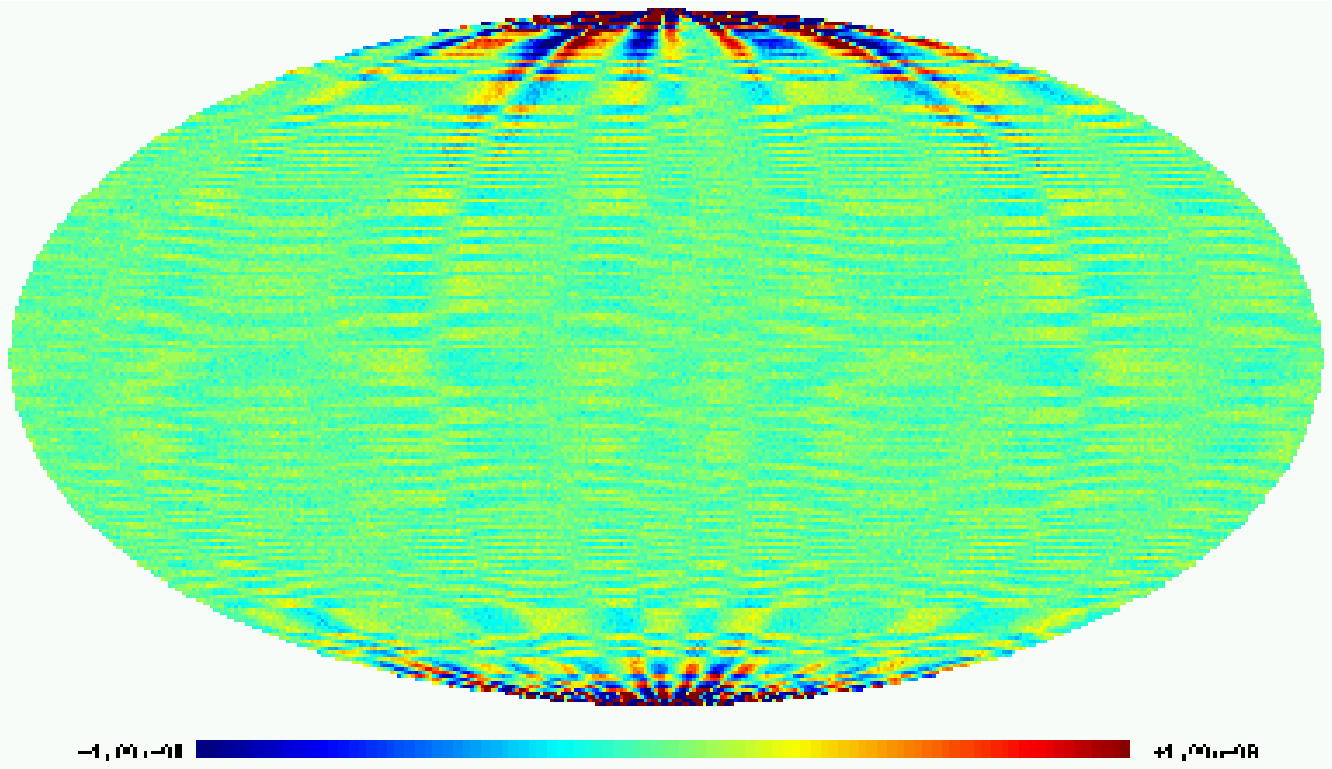}}}
\hbox{\hspace*{0.01cm}
\centerline{\includegraphics[width=0.3\linewidth]{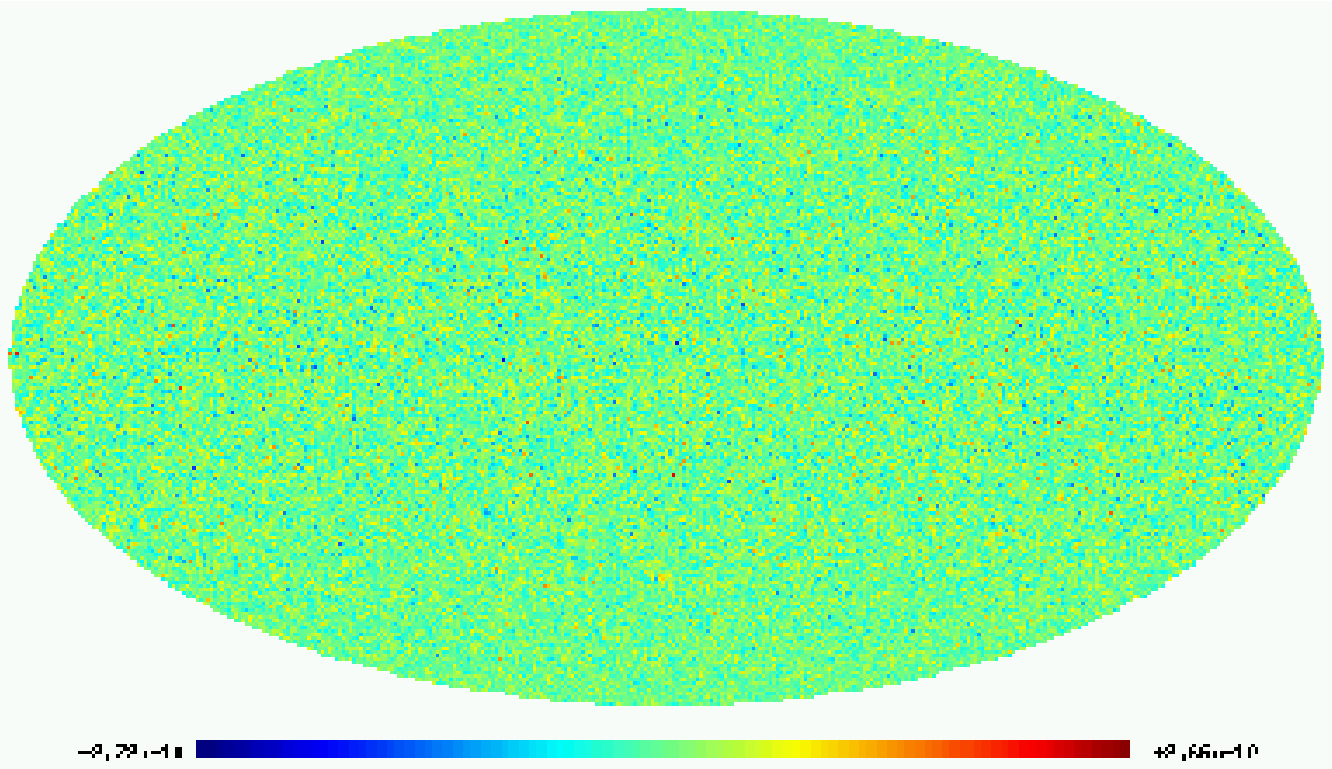}%}}
%\hbox{\hspace*{0.01cm}
\includegraphics[width=0.3\linewidth]{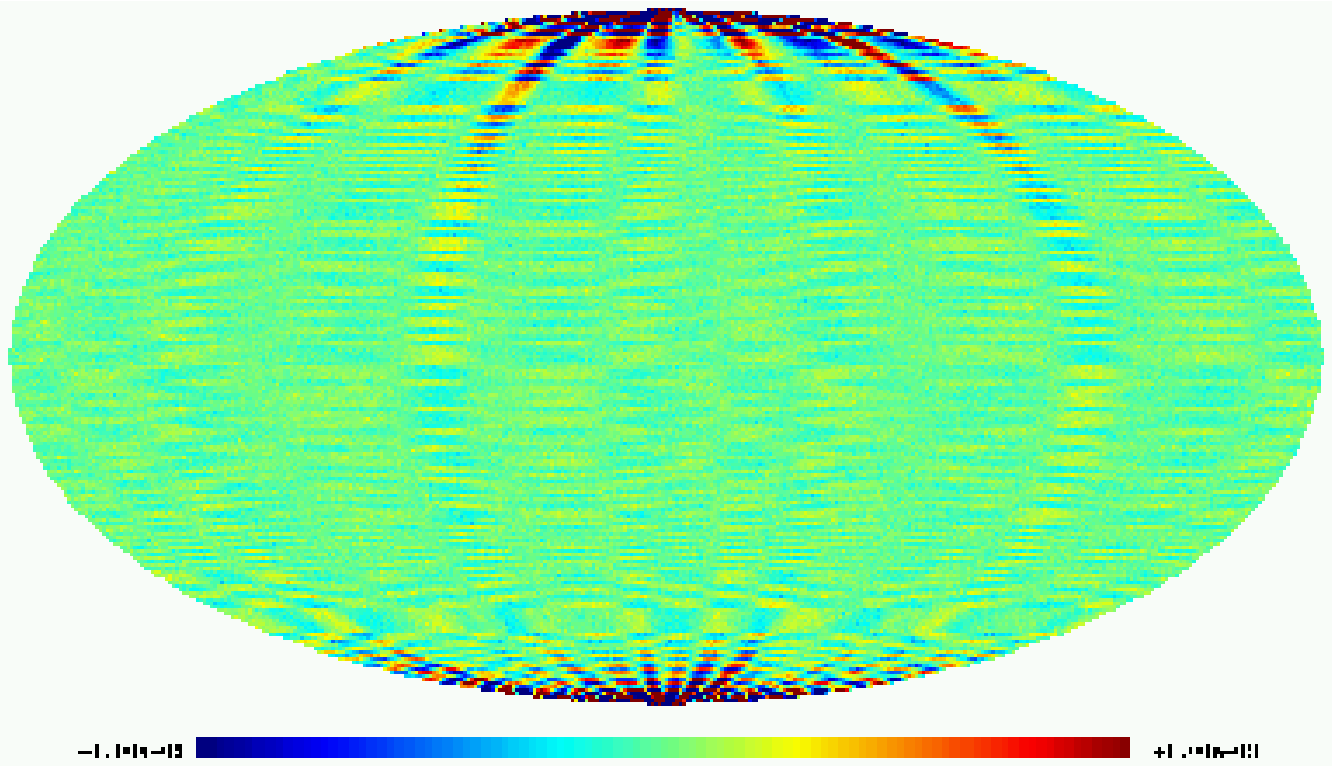}}}
\caption{ The differences of reconstructed and input maps for the GLESP-pol
for $\ell_{max}=1500$ (the top pair is for Q the Stokes parameter) and
the  bottom pair is for U). Top left plot corresponds to the grN
pixelization (the color scale is $-3\cdot10^{-10},3\cdot10^{-10}mK$),
top right is for the grS pixelization (
the color scale is $-10^{-9},10^{-9}mK$).
Bottom left ---
the GLESP-pol reconstruction of Q for grN mode
(the color scale is $-2.5\cdot10^{-10},2.5\cdot10^{-10}mK$).
Bottom right corresponds to the grS mode (the color
scale is $-10^{-9},10^{-9}mK$).
 }
\label{fig6}
\end{figure}

As we have mentioned already in Introduction, the GLESP-pol pixelization
was designed to assess the problem of accurate reconstruction of
the coefficients of spin $\pm2$ spherical harmonics decomposition
(see Appendix for details).  Since the spin $\pm2$ spherical harmonics
have a peculiarity, from the computational point of view, behaviour in the
vicinity of the polar cups, the differences in the GLESP-pol polarization
and the HEALPix  produce different error of reconstruction and become more
visible especially for polarization.
In Fig.\ref{fig6}, we plot the map of differences between input and output
signal, reconstructed by
the GLESP-pol (grN and grS pixelization).

\begin{figure}[!th]
\hbox{\hspace*{0.01cm}
%\begingroup
\centerline{\includegraphics[width=0.3\linewidth]{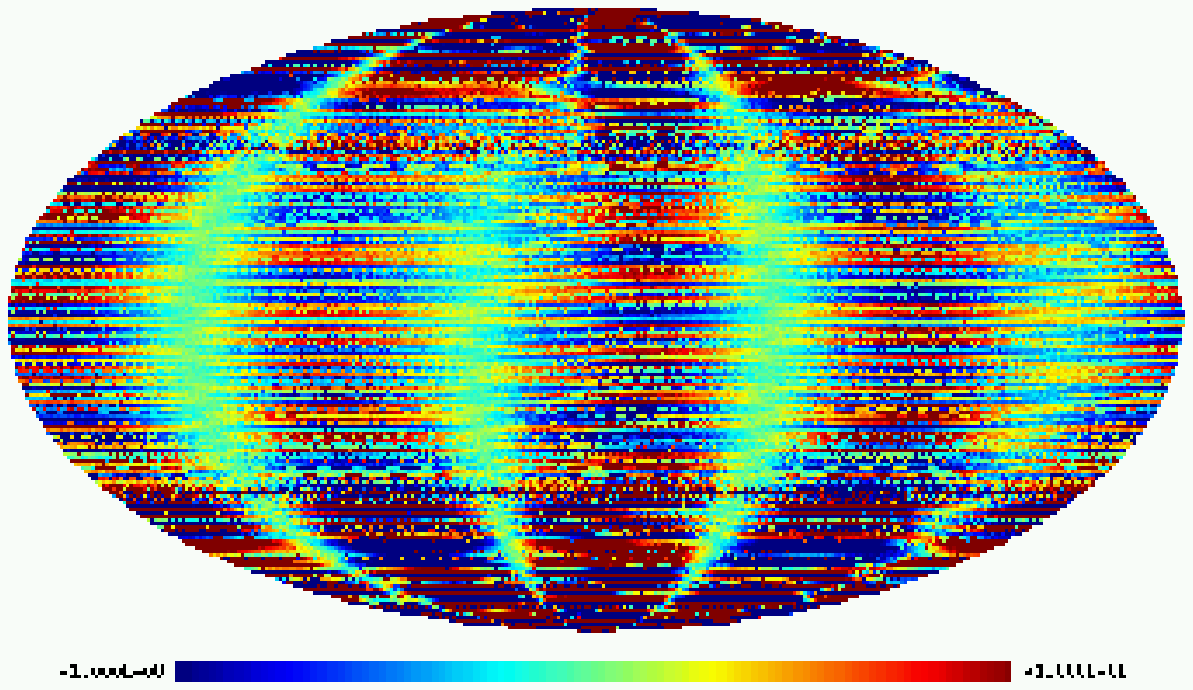}%}}
%\hbox{\hspace*{0.01cm}
\includegraphics[width=0.3\linewidth]{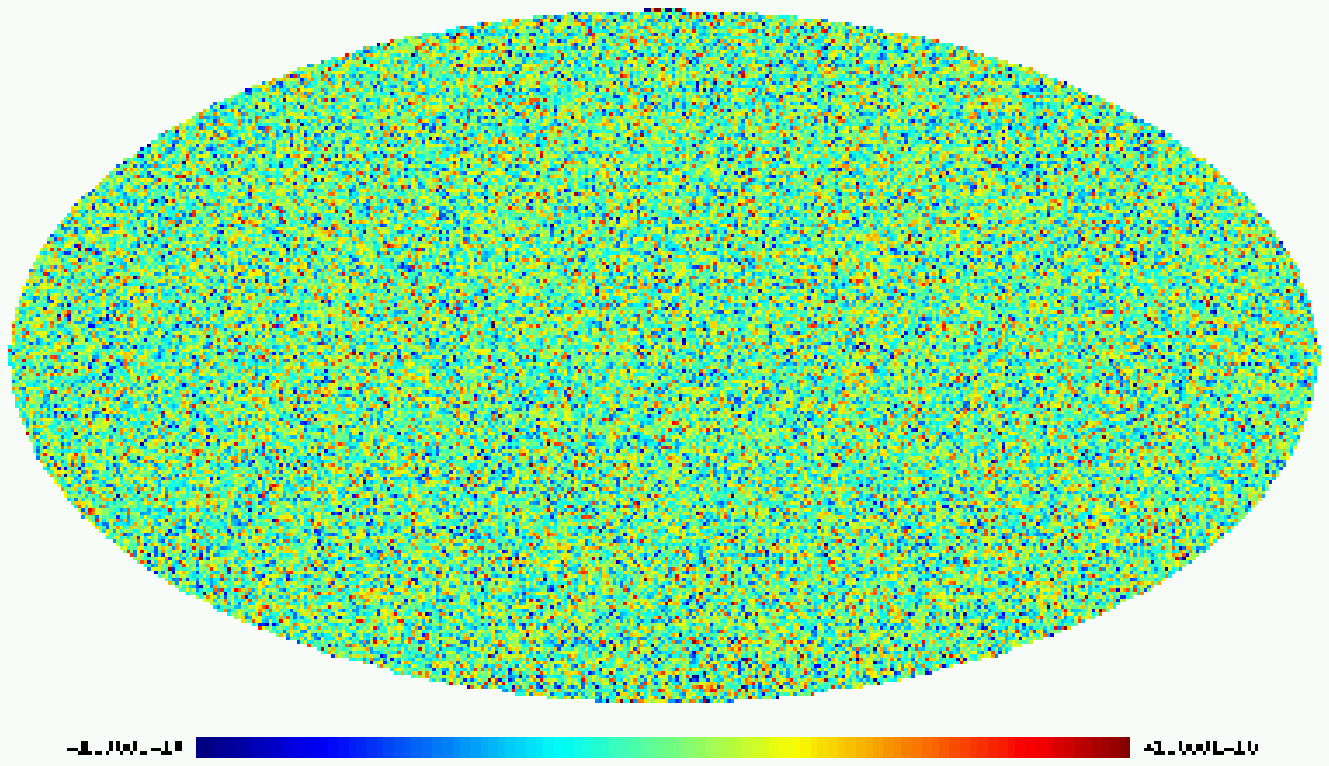}}}
\hbox{\hspace*{0.01cm}
\centerline{\includegraphics[width=0.3\linewidth]{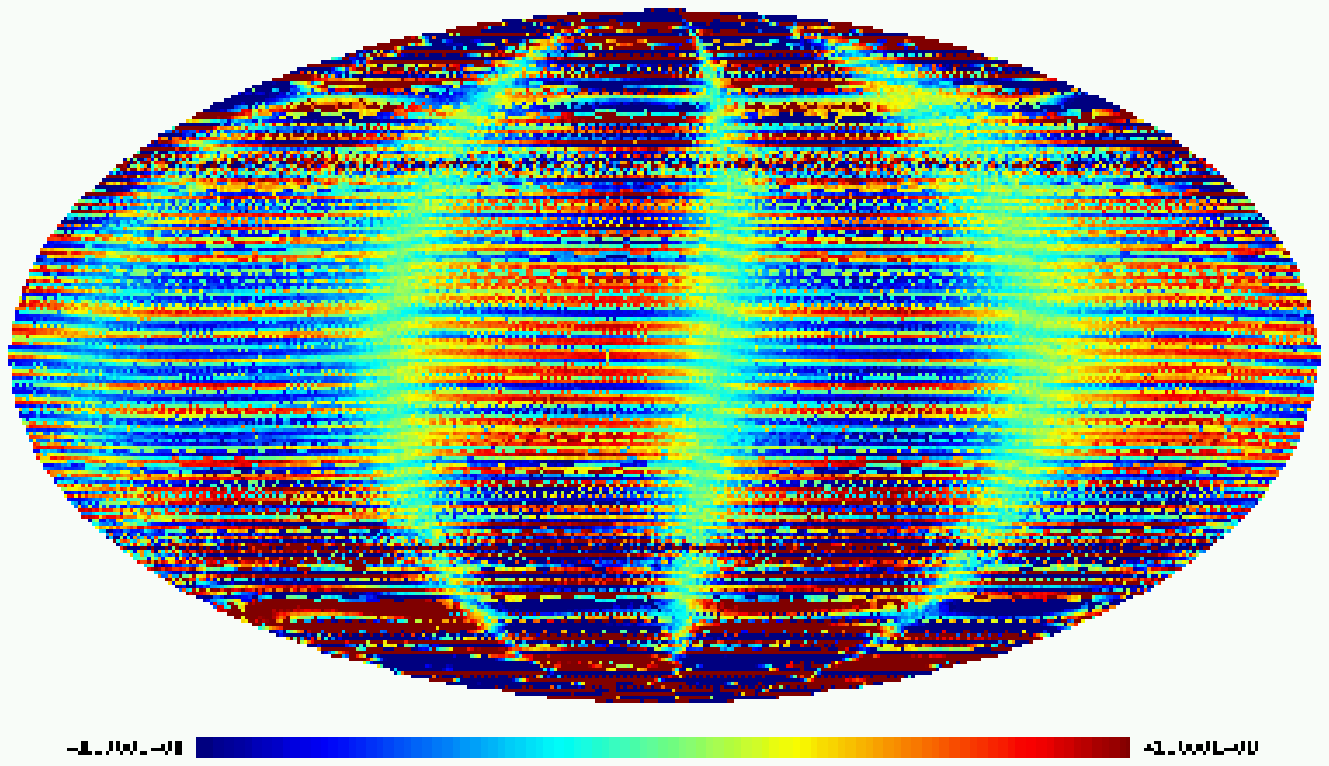}%}}
%\hbox{\hspace*{0.01cm}
\includegraphics[width=0.3\linewidth]{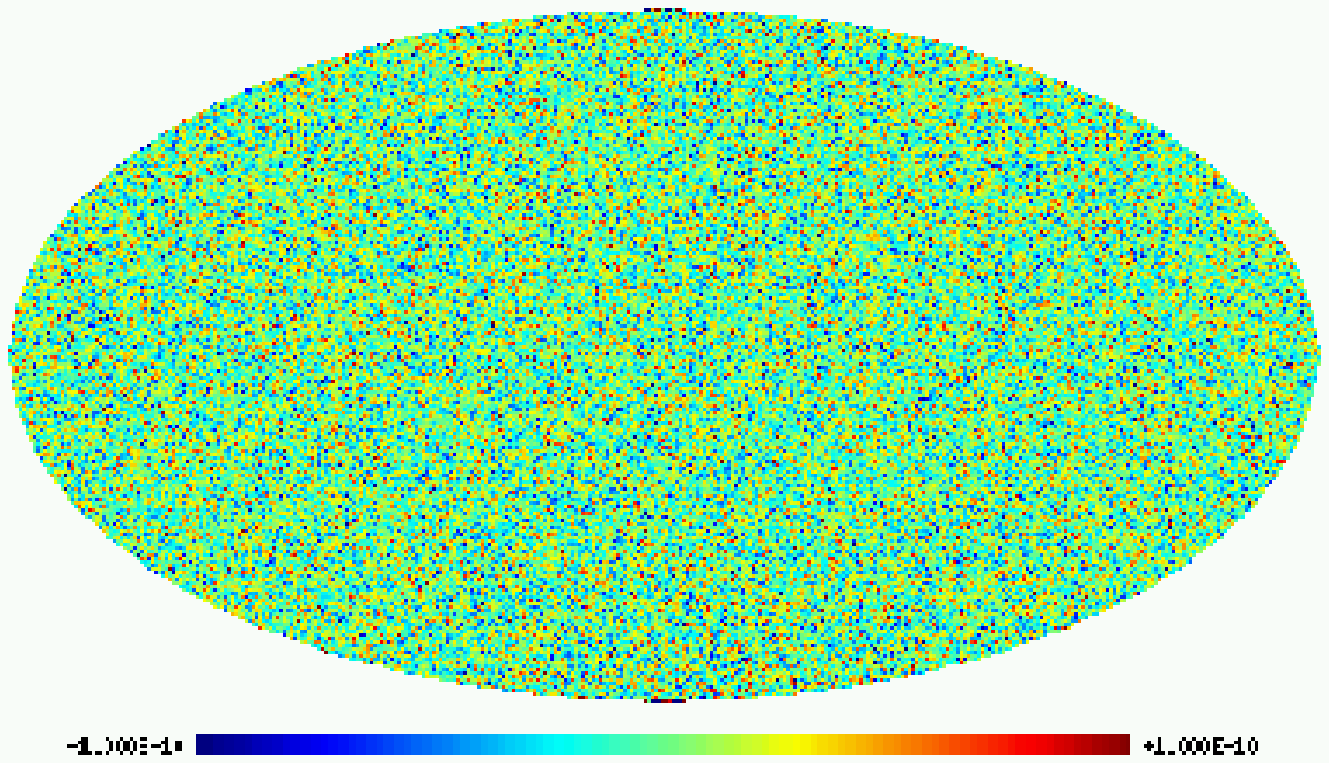}}}
\caption{ The differences of reconstructed and input maps for the
HEALPix\,2.11 for Nside=1024 (the top pair corresponds to Q components
and  the bottom pair is for U. Top left plot corresponds to the
``zero iteration'' key (the color scale is $-10^{-8},10^{-8} mK$),
top right is for 4 iterations (the color scale is $10^{-10},10^{-10}mK$).
Bottom left ---
the GLESP-pol reconstruction for the grN mode
(the color scale is $10^{-8},10^{-8}mK$).
Bottom right corresponds to the grS mode (the color scale is
$-10^{-10},10^{-10}mK$). The number of pixels for the
HEALPix\,2.11 and the GLESP-pol are practically the same.
No correction by the  window function of the pixels.
 }
\label{fig7}
\end{figure}

  As one can see from this figure, the reconstruction of Q and U
components by the grN and grS pixelization is characterized by very
high accuracy,  but the grN pixelization looks slightly better.
For the HEALPix\,2.11
the corresponding maps for differences are shown in Fig.\ref{fig7}.

\begin{figure}[!th]
\hbox{\hspace*{0.01cm}
%\begingroup
\centerline{\includegraphics[width=0.3\linewidth]{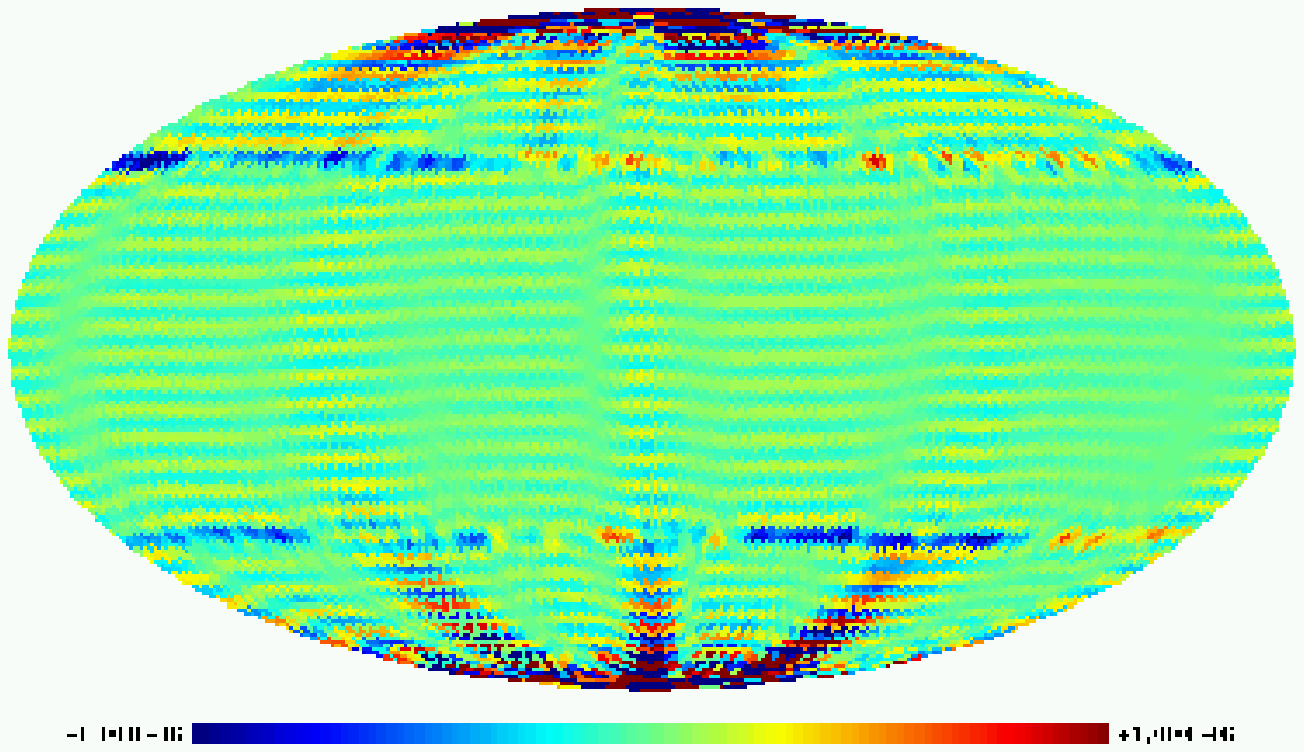}%}}
%\hbox{\hspace*{0.01cm}
\includegraphics[width=0.3\linewidth]{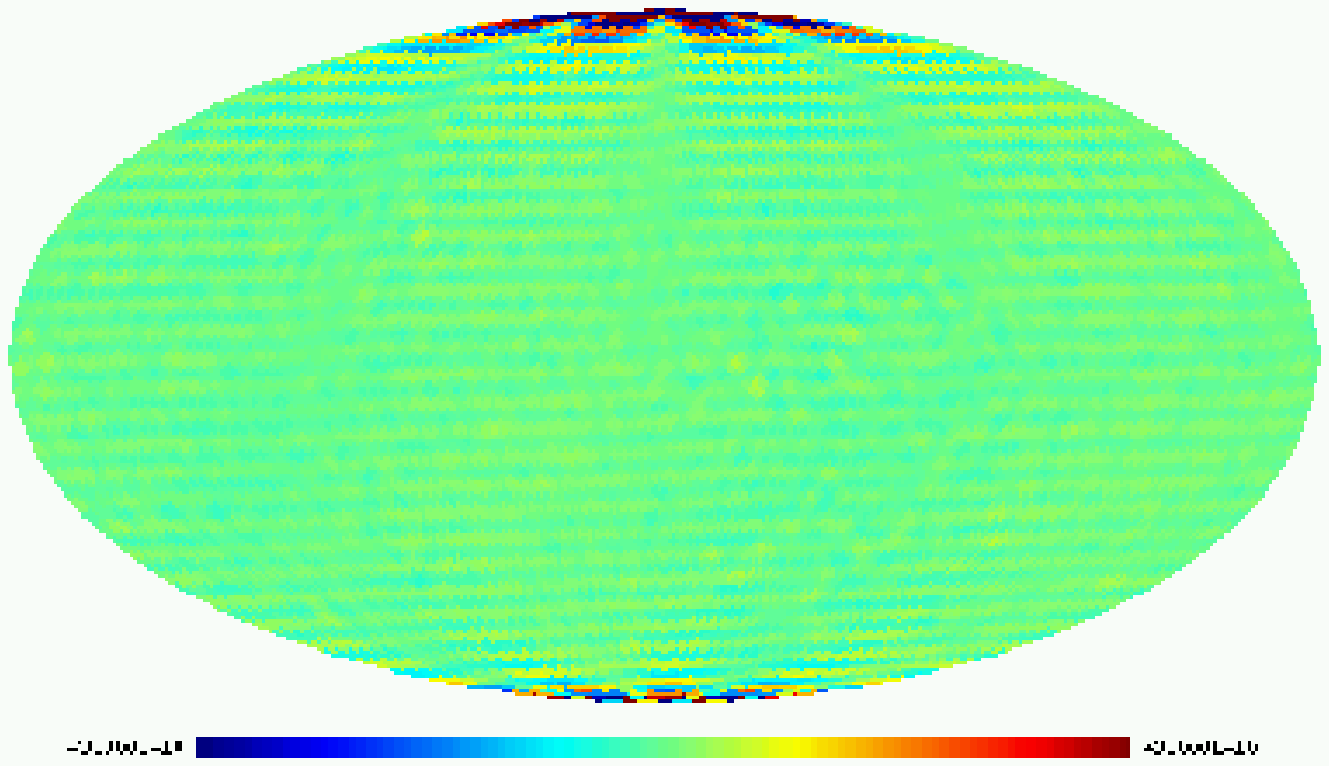}}}
\hbox{\hspace*{0.01cm}
\centerline{\includegraphics[width=0.3\linewidth]{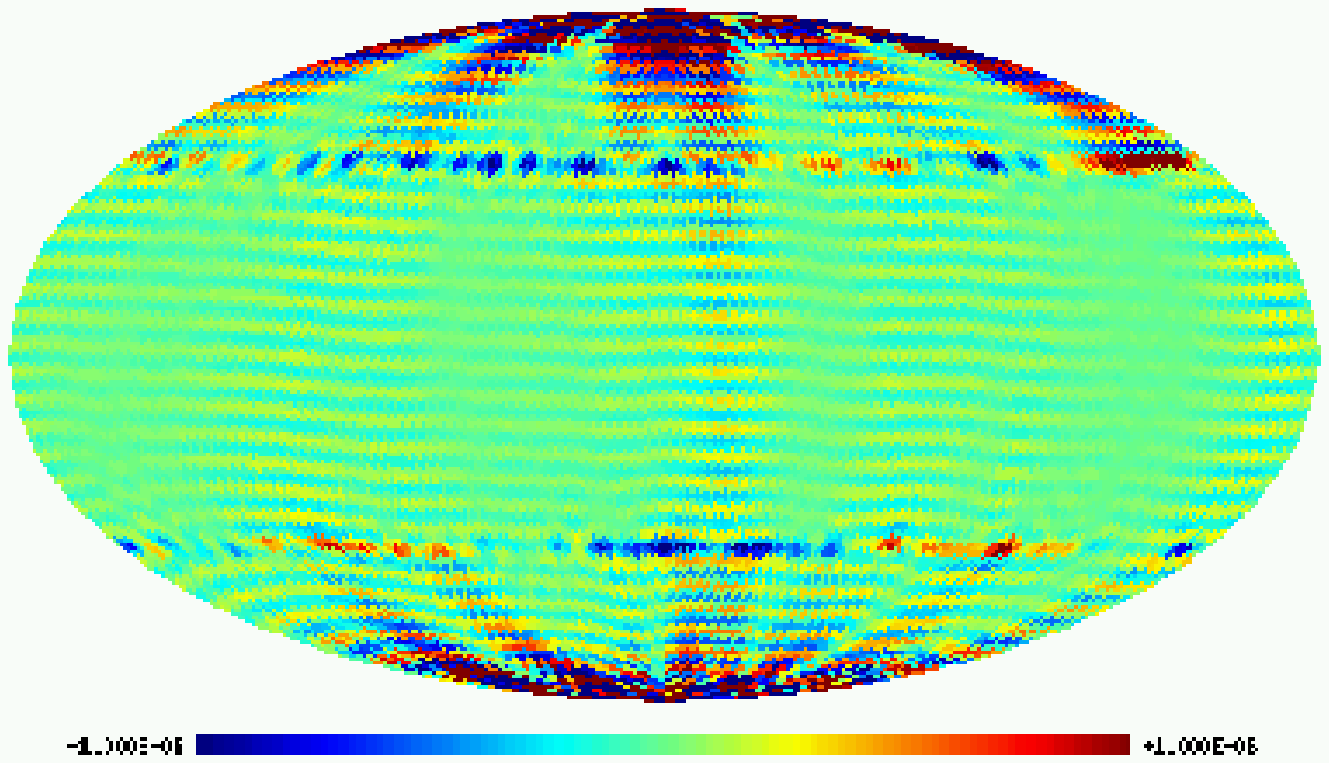}%}}
%\hbox{\hspace*{0.01cm}
\includegraphics[width=0.3\linewidth]{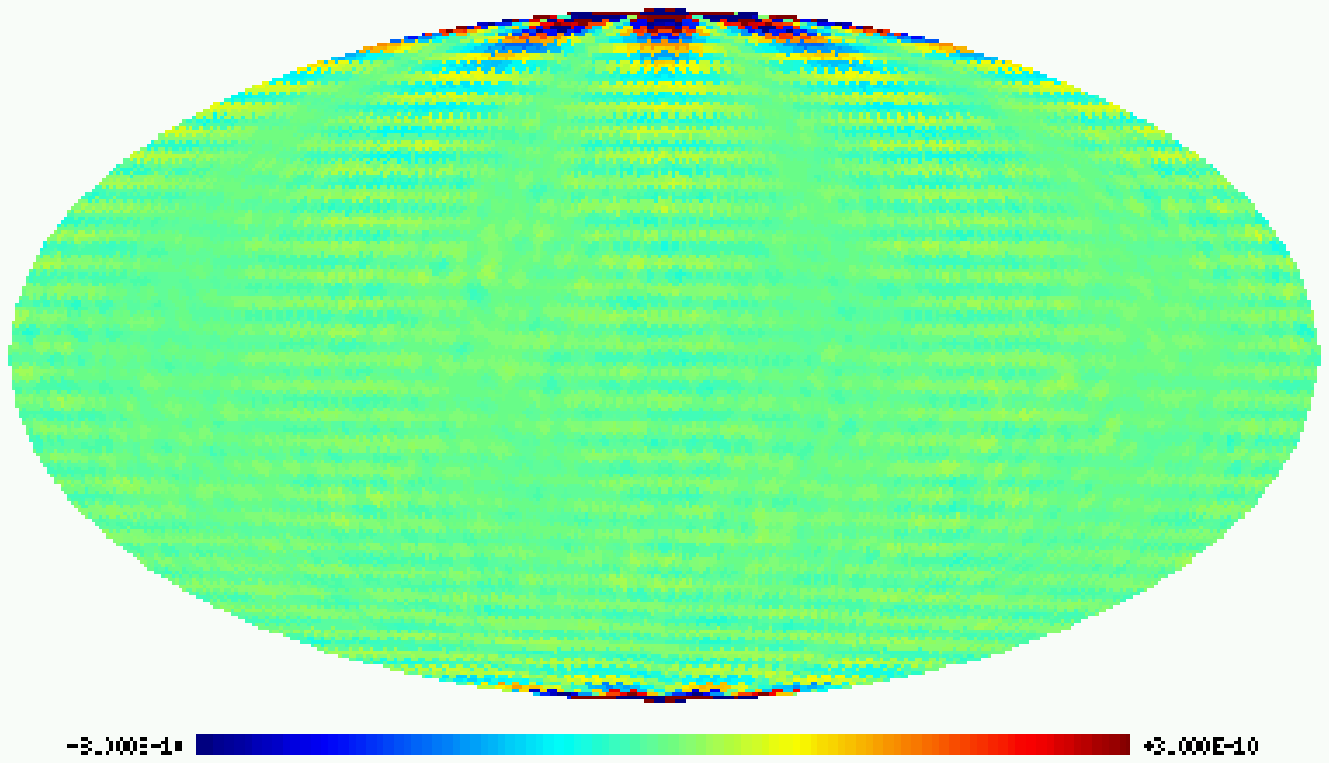}}}
\caption{ The differences of reconstructed and input maps for
the HEALPix\,2.11 for Nside=32 (the top pair corresponds to Q components
and  the bottom pair is for U. Top left plot corresponds to the
``zero iteration'' key (the color scale is $-10^{-6},10^{-6} mK$),
top right is for 4 iterations
(the color scale is $-3\cdot10^{-10},3\cdot10^{-10}mK$).
Bottom left ---
the U component for the zero iteration (the color scale is
$-10^{-6},10^{-6} mK$). Bottom right corresponds to 4 iterations
(the color scale is $-3\cdot10^{-10},3\cdot10^{-10}mK$).
 }
\label{fig8}
\end{figure}

For  low resolution pixelization with Nside=32 and $\ell_{max}=32$ the
corresponding maps are shown in
Fig.\ref{fig8}

\begin{figure}[!th]
\hbox{\hspace*{0.01cm}
%\begingroup
\centerline{\includegraphics[width=0.3\linewidth]{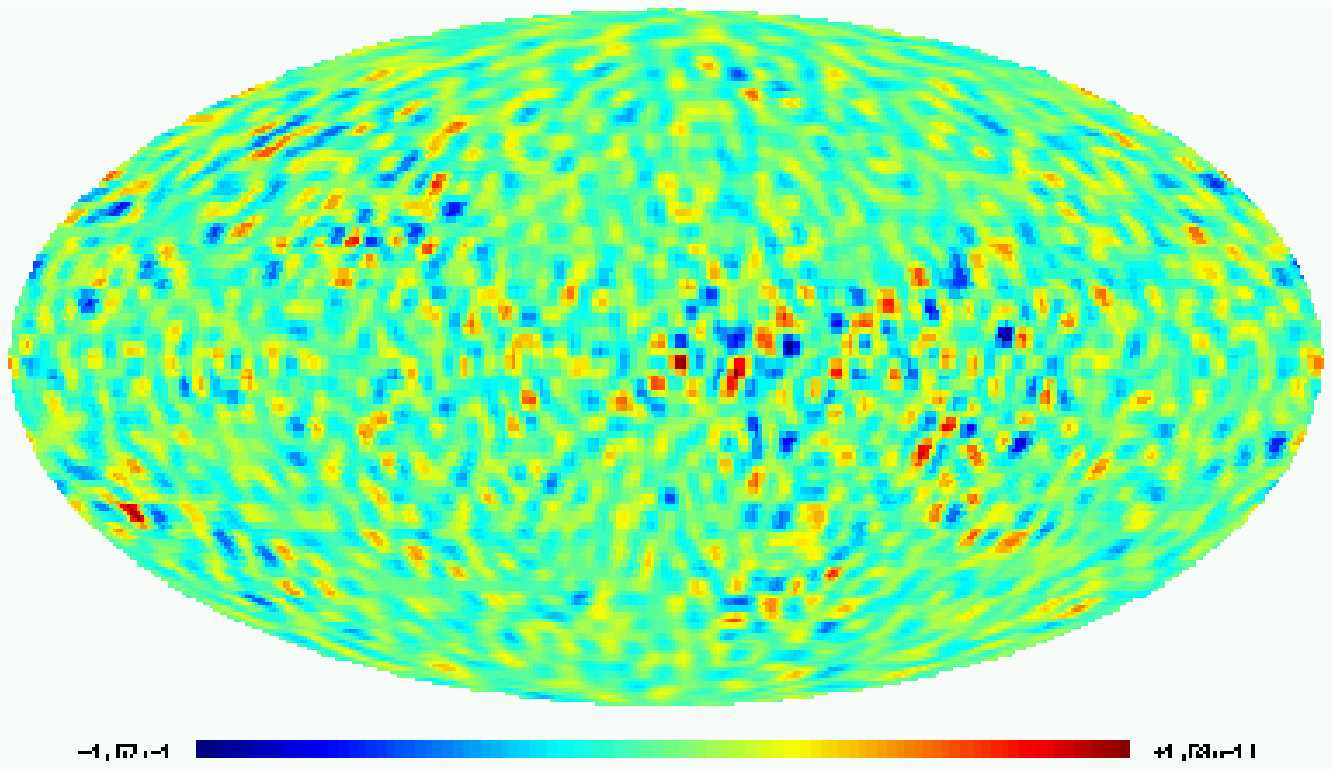}%}}
%\hbox{\hspace*{0.01cm}
\includegraphics[width=0.3\linewidth]{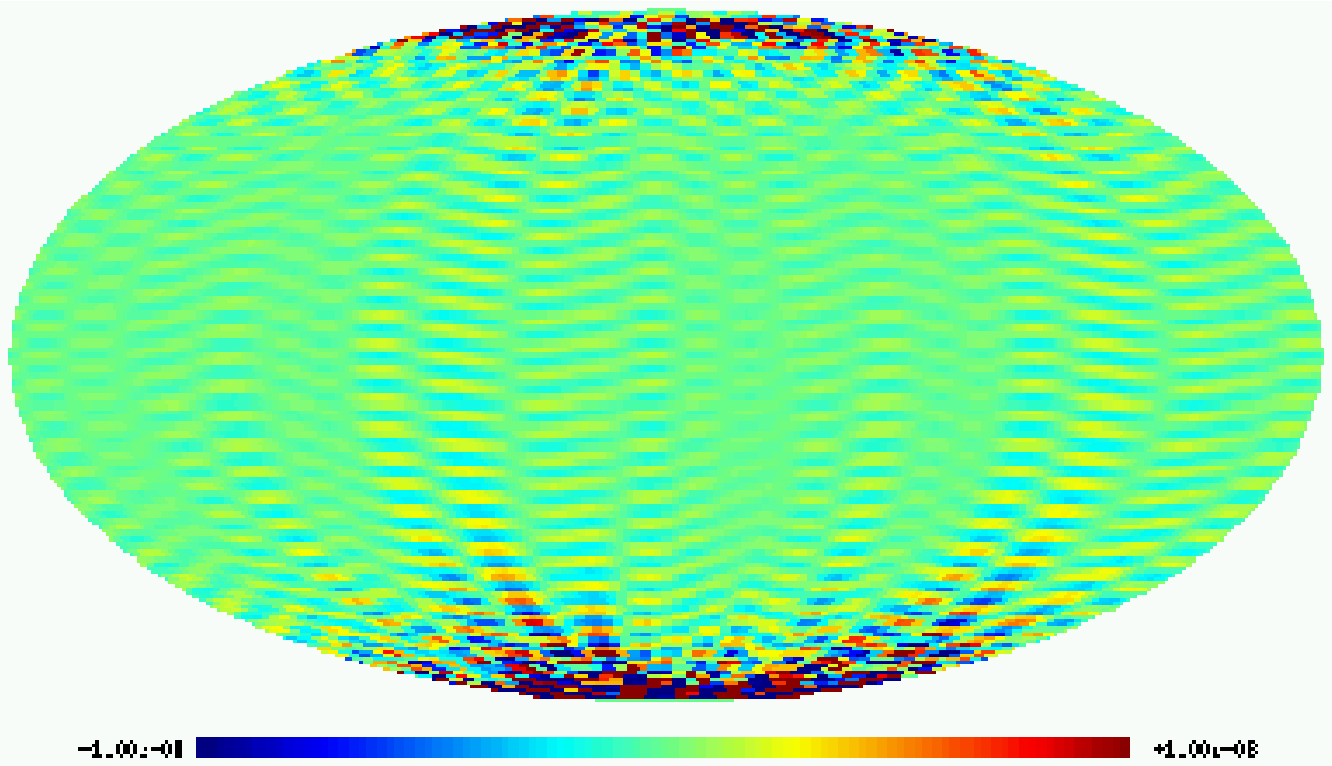}}}
\hbox{\hspace*{0.01cm}
\centerline{\includegraphics[width=0.3\linewidth]{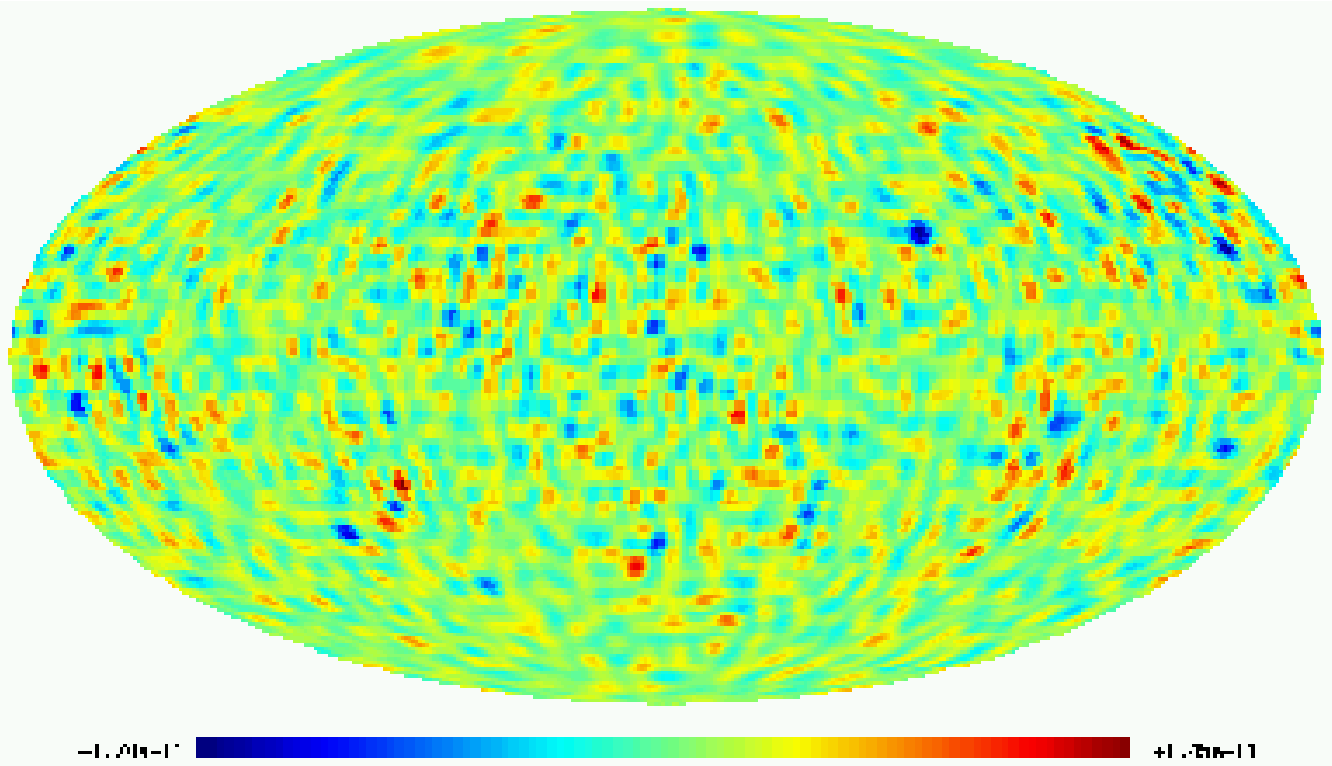}%}}
%\hbox{\hspace*{0.01cm}
\includegraphics[width=0.3\linewidth]{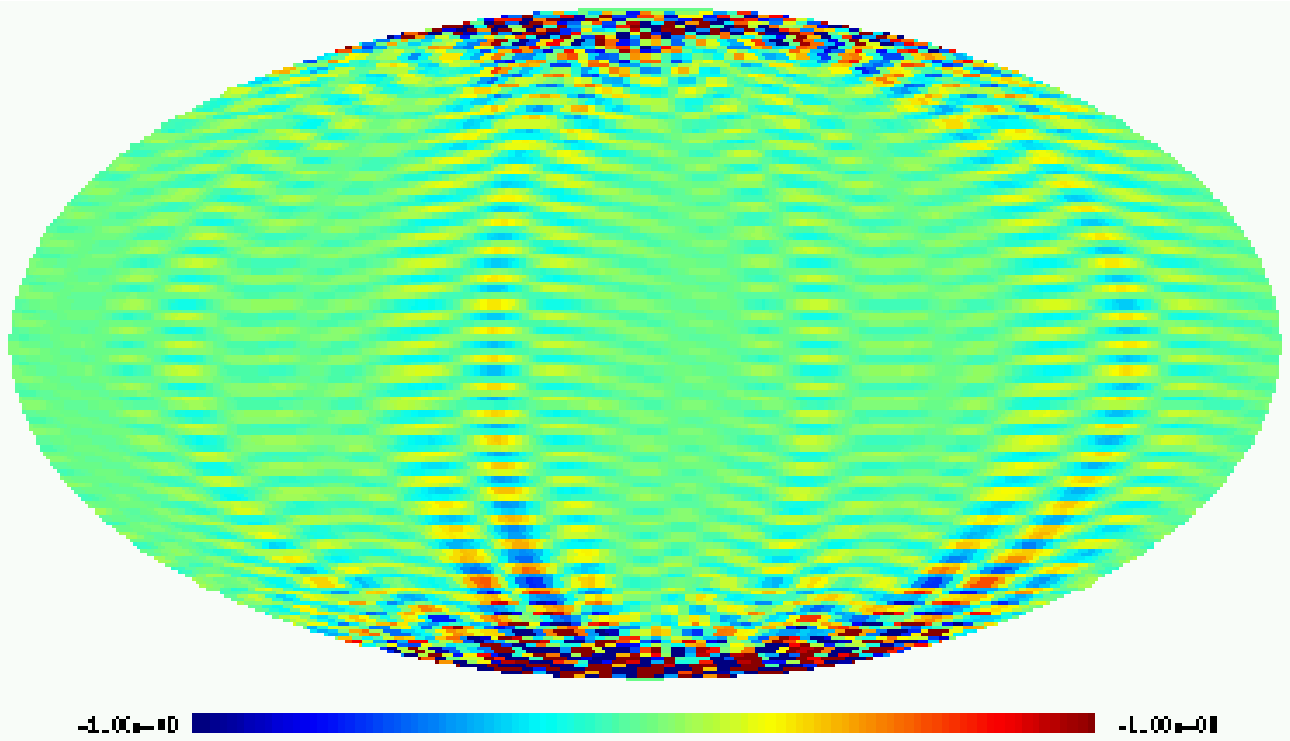}}}
\caption{ The errors of reconstructions  for  the GLESP-pol
(Q for the top pair and U for the bottom pair). Top left plot corresponds
to the grN pixelization
(the color scale is $-2\cdot10^{-11},2\cdot10^{-11}mK$),
top right is for grN (the color scale is $-10^{-8},10^{-8}mK$).
Bottom left ---
the GLESP-pol reconstruction for the grS mode (the color scale is
$-2\cdot10^{-11},2\cdot10^{-11}mK$. Bottom right corresponds to the grS
mode (the color scale is $-10^{-8},10^{-8}mK$).
 }
\label{fig9}
\end{figure}

\begin{figure}[!th]
\hbox{\hspace*{0.01cm}
%\begingroup
\centerline{\includegraphics[width=0.2\linewidth]{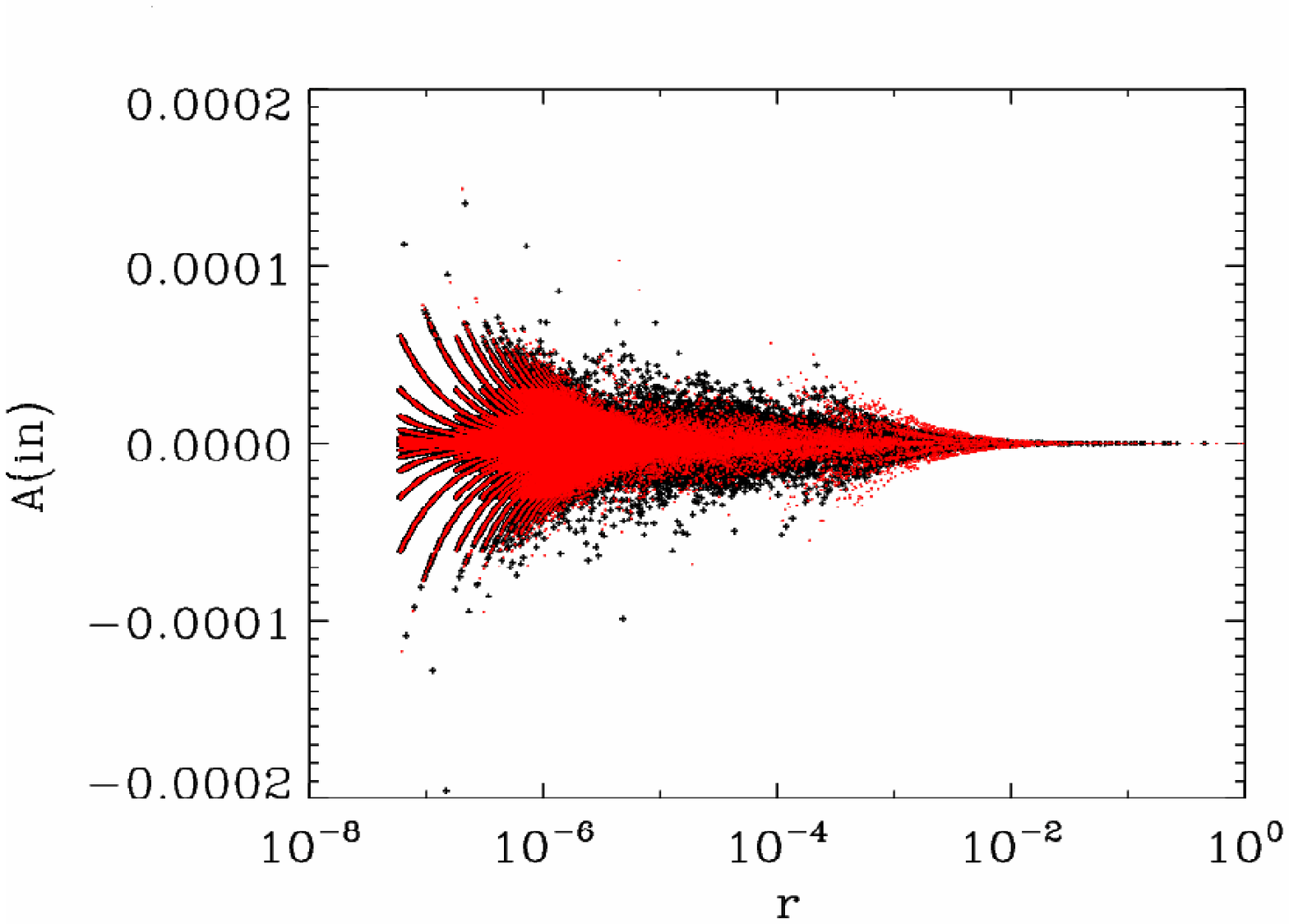}%}}
%\hbox{\hspace*{0.01cm}
\includegraphics[width=0.2\linewidth]{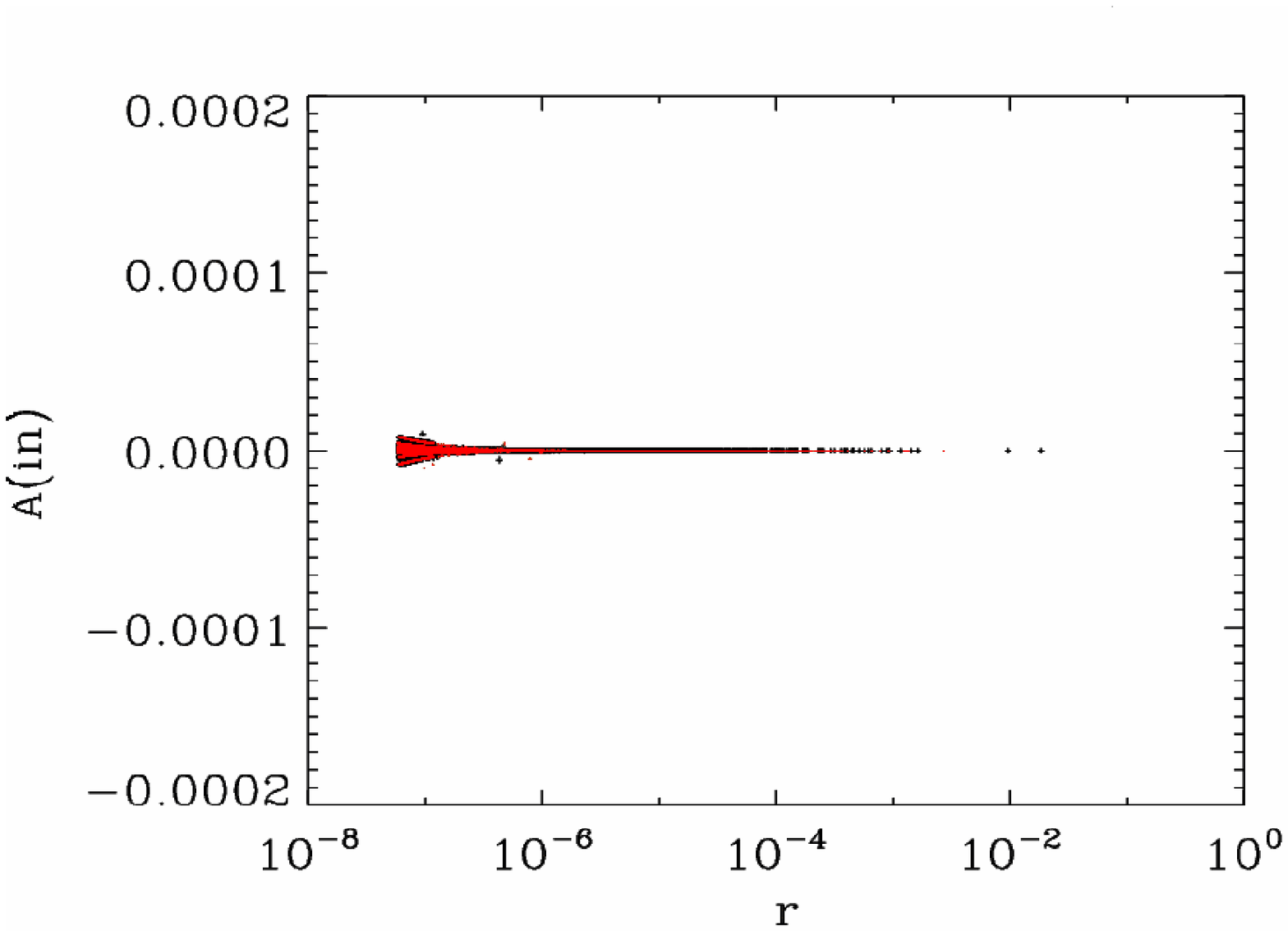}
%\hbox{\hspace*{0.01cm}
%\centerline{
\includegraphics[width=0.2\linewidth]{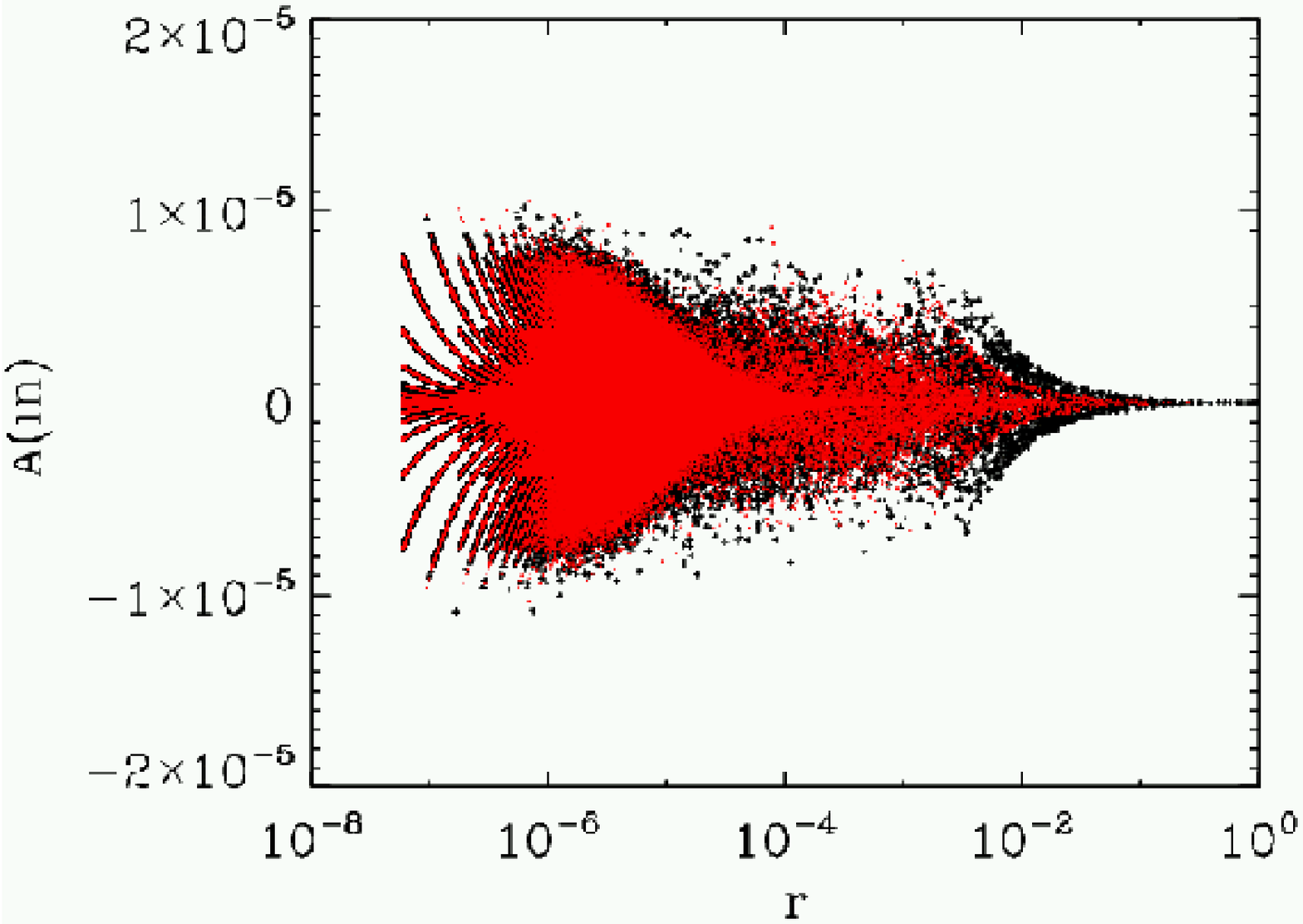}%}}
%\hbox{\hspace*{0.01cm}
\includegraphics[width=0.2\linewidth]{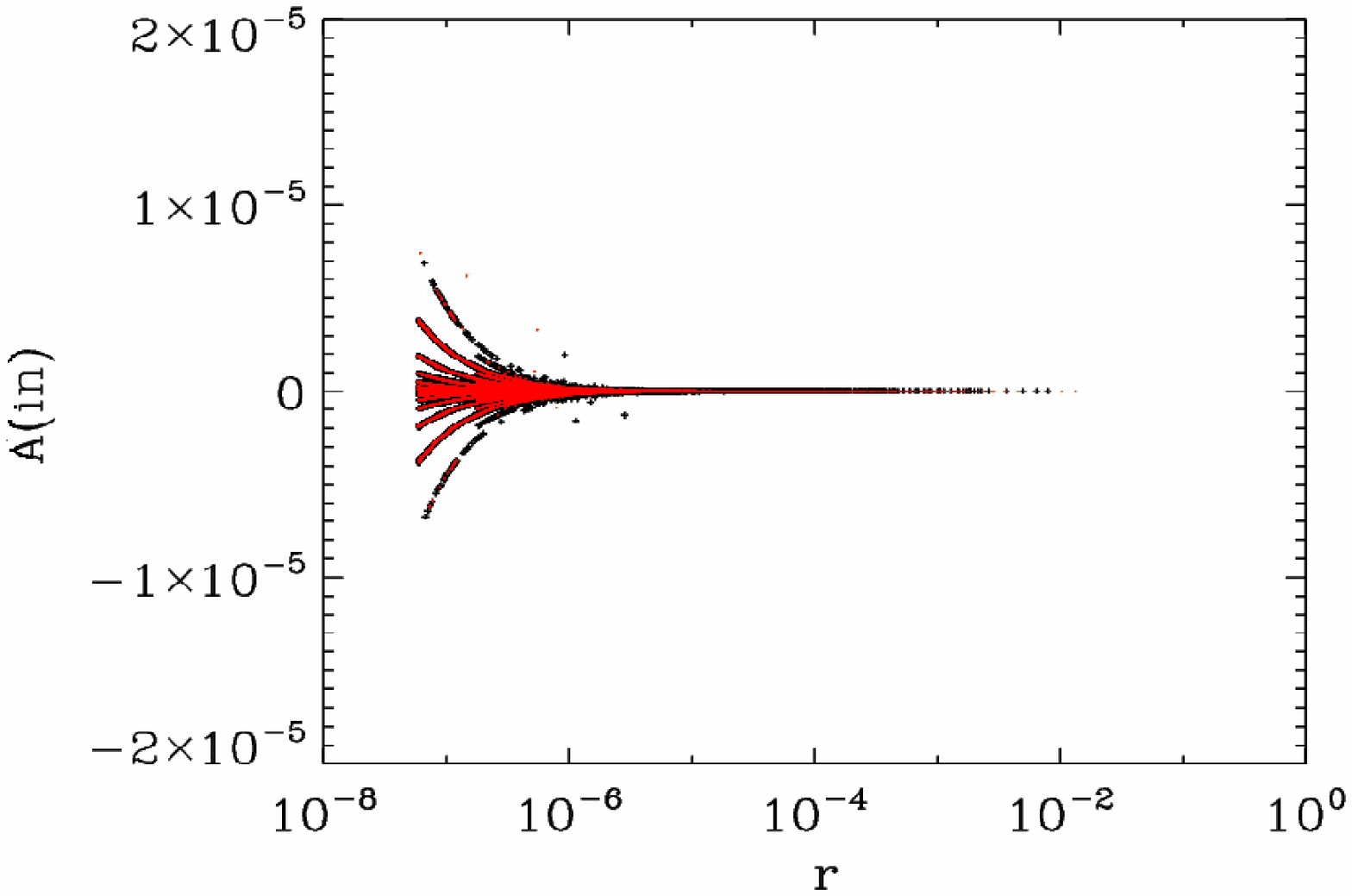}}}
\hbox{\hspace*{0.01cm}
\centerline{\includegraphics[width=0.2\linewidth]{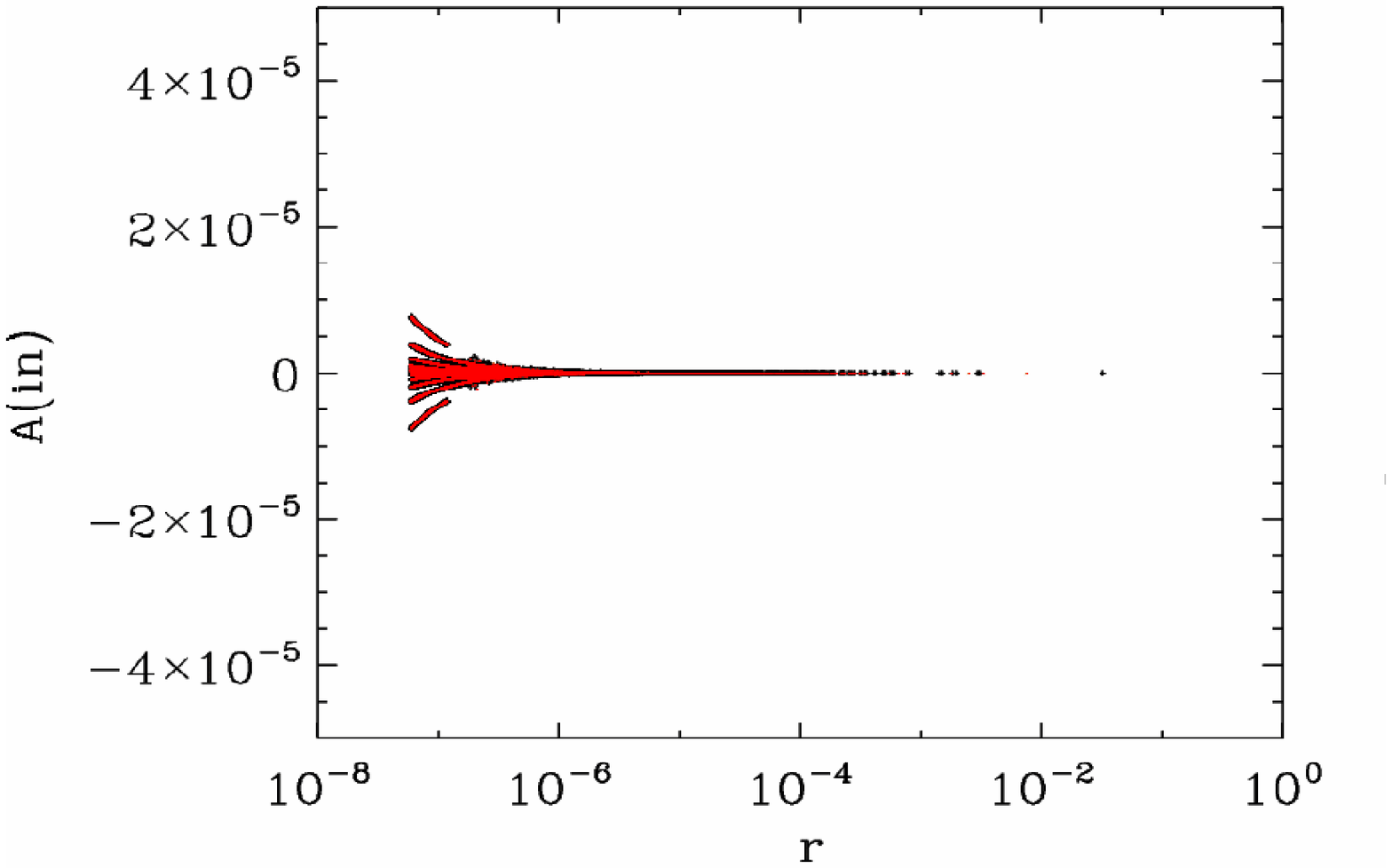}%}}
%\hbox{\hspace*{0.01cm}
\includegraphics[width=0.2\linewidth]{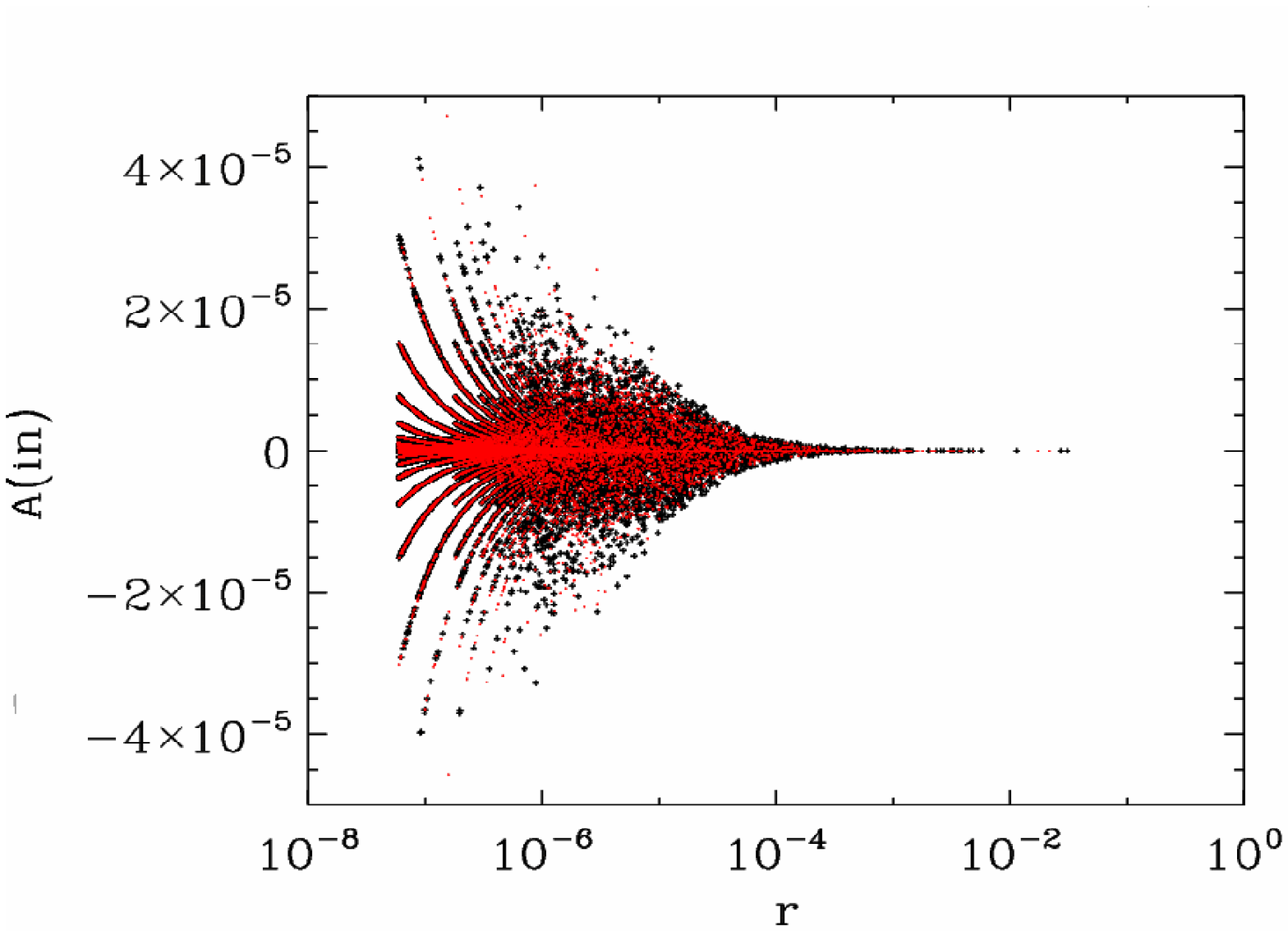}
%\hbox{\hspace*{0.01cm}
%\centerline{
\includegraphics[width=0.2\linewidth]{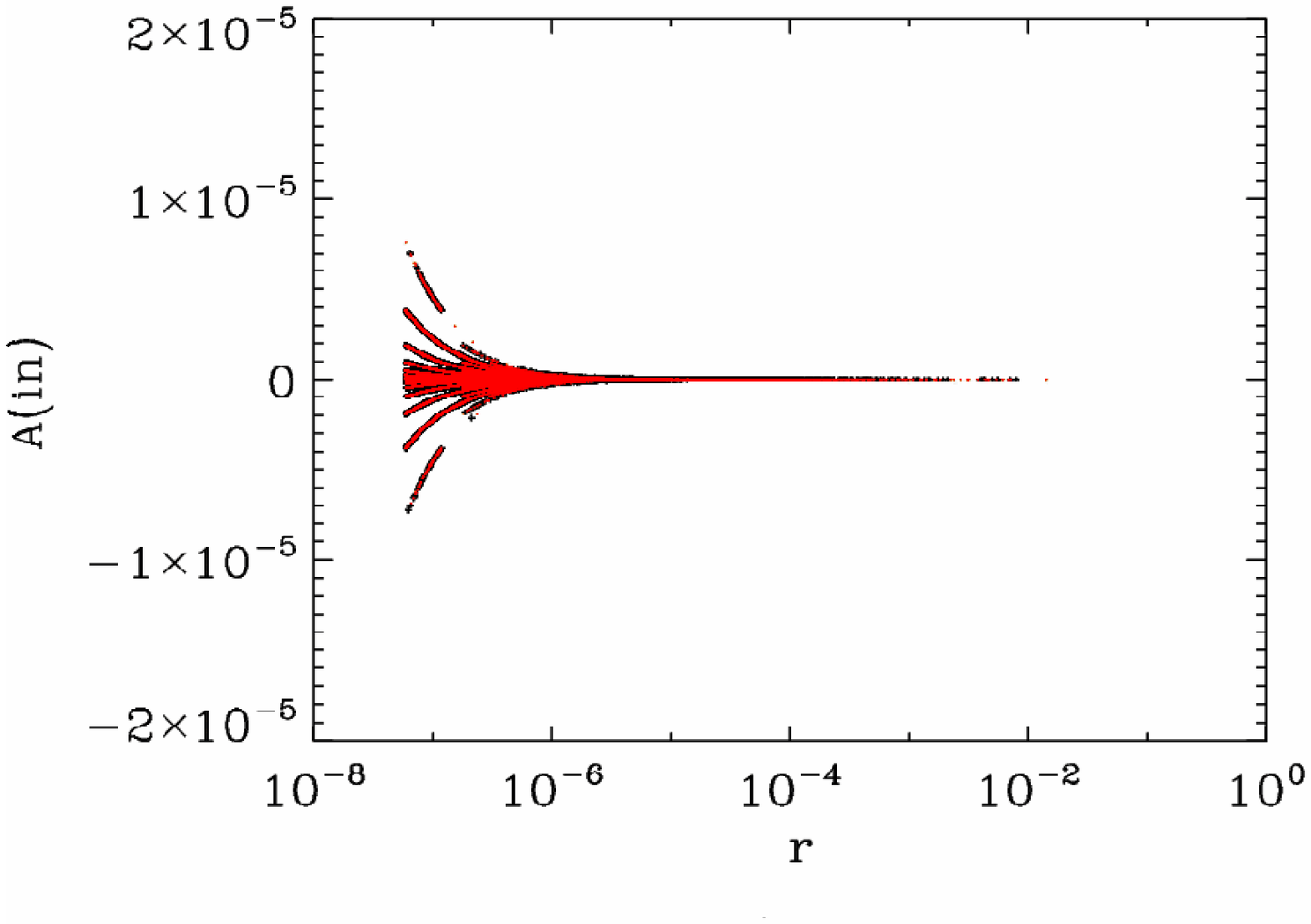}%}}
%\hbox{\hspace*{0.01cm}
\includegraphics[width=0.2\linewidth]{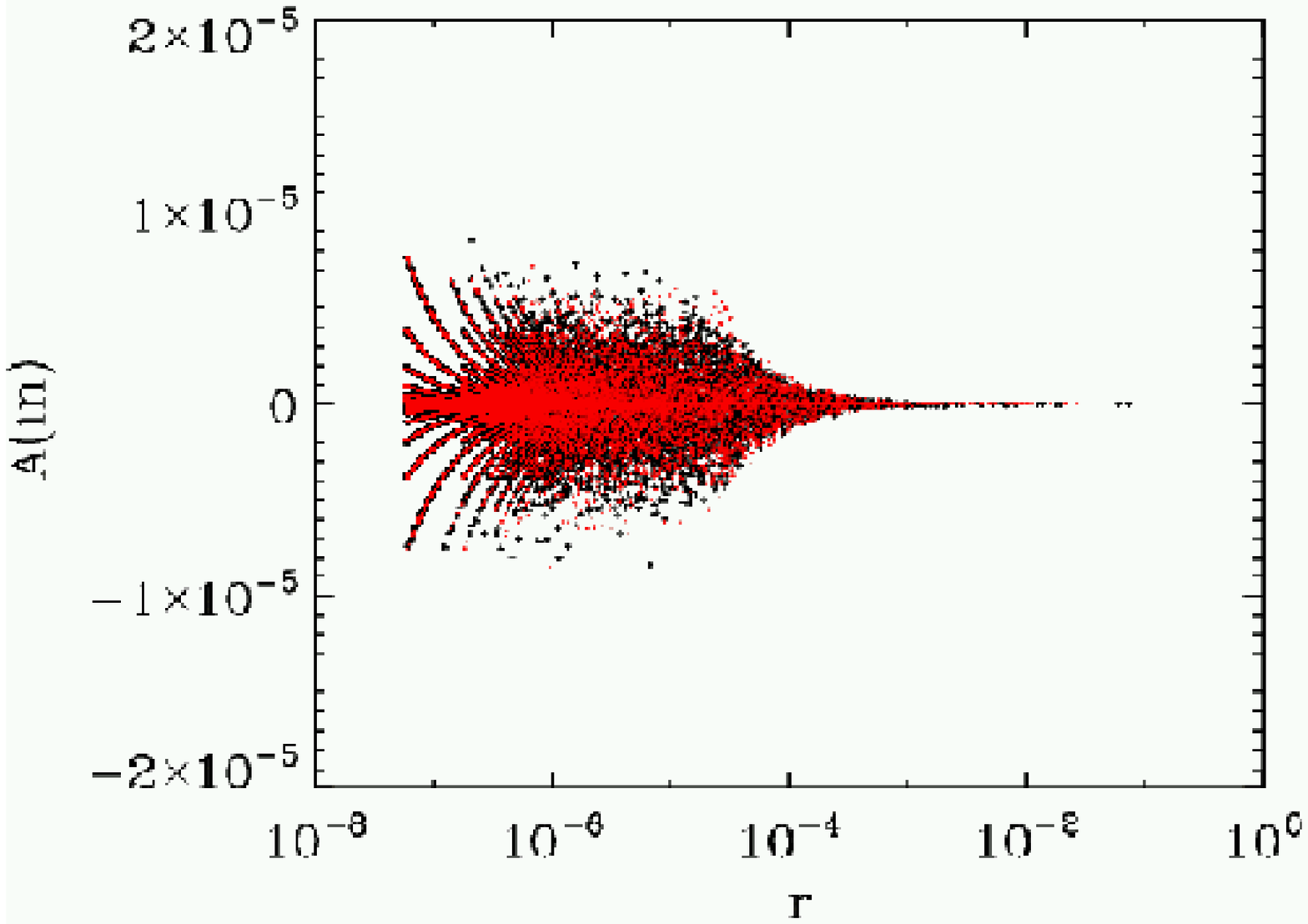}}}
\caption{ The errors of reconstructions of $\Re e a^E_{\l,m}$
and $\Im m a^E_{\l,m}$. Top row corresponds to the HEALPix\,2.11
(Nside=1024). From the left to the right are: E mode with zero
iteration, E mode with 4 iteration, B mode with zero iteration,
B mode with 4 iteration. Bottom row is the same, but for the GLESP-pol
grN and grS modes and $\ell_{max}=1500$.
 Black dots corresponds to
the real part of $\ell,m$ modes, the red dots are for imaginary part.
 }
\label{fig10}
\end{figure}

Thus, one can see that implementation of the 4 iterations for
the HEALPix package gives us practically the same result, as
from GLESP-pol even for the low resolution maps (see Fig.\ref{fig9}).
In Fig.\ref{fig10} we show the diagram, similar to the Fig.\ref{fig3},
but for E and B components of polarization.

One can see that the GLESP-pol mode grN gives us practically the same
errors, as the HEALPix\,2.11 4 iteration mode. However, the CPU time
for the GLESP-pol is  about 4 time smaller then for the HEALPix\,2.11
due to absence of iterations.

\begin{figure}[!th]
\hbox{\hspace*{0.01cm}
\centerline{
\includegraphics[width=0.32\linewidth]{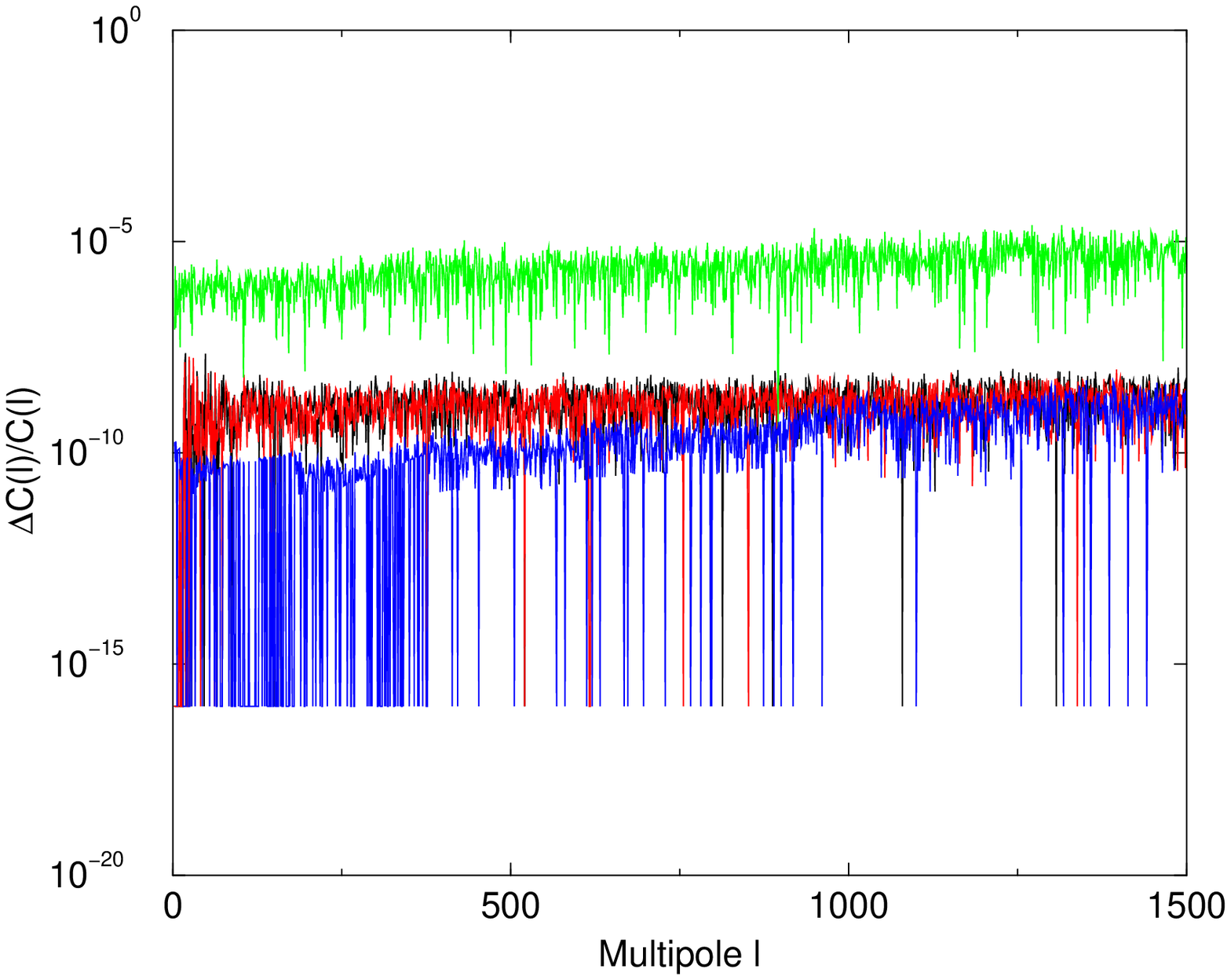}
\includegraphics[width=0.32\linewidth]{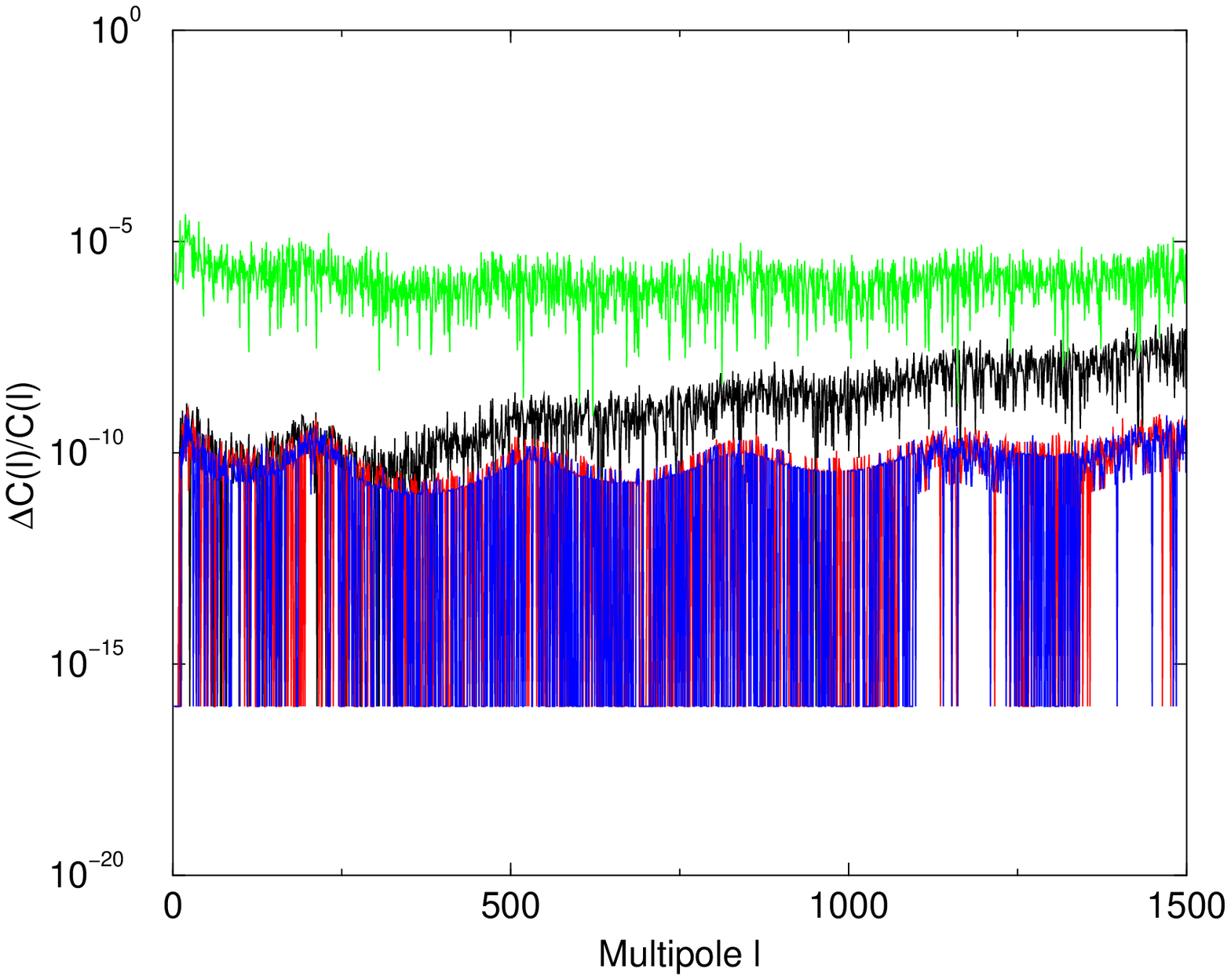}
\includegraphics[width=0.32\linewidth]{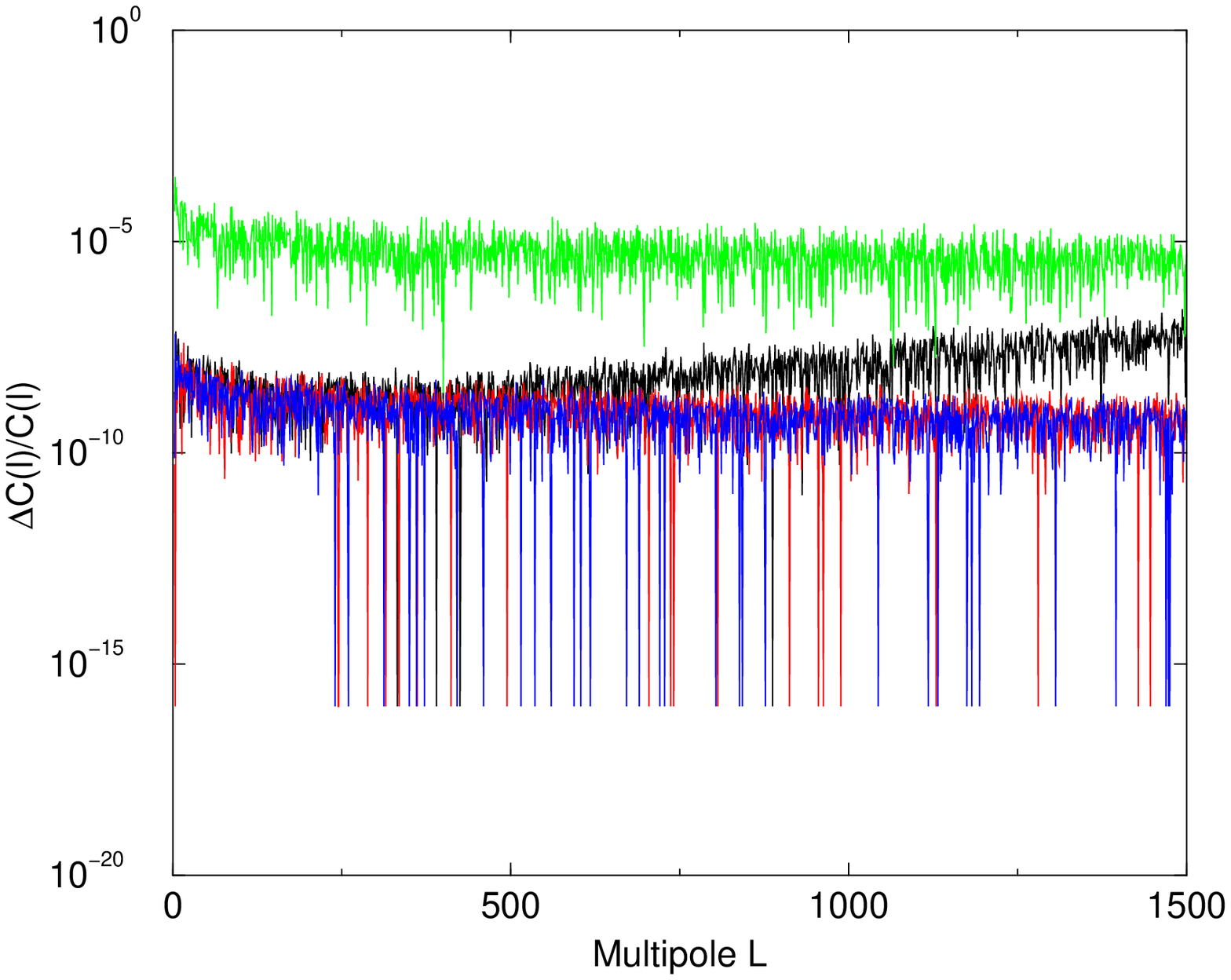}
}}
\caption{
The relative accuracy of $C(\l)$ specrum restoration for
temperature anisotropy (left), E-polarization (middle) and
B-polarization(right). The grS GLESP type is shown
with the black line, the grN type is shown with red, the HEALpix calculations
of $C(\l)$ of 0 iteration mode are plotted with green color, and
of 4 iteration with blue.
}
\label{fig11}
\end{figure}

Finally, we demonstrate absolute relative accuracy for power spectra
($\Delta C(\l)/C(\l)$)
for the GLESP\,2.0 (grS and grN) and for the HEALPix\,2.11 (0 and 4
iteration modes) shown in Fig.\,(\ref{fig11}).
As one could see the realtive accuracy for restoration of the
power spectrum of temperature anisotropy in
the GLESP is approximately the same for both types grS and grN,
and the HEALPix 4 iteration mode give better accuracy at low multipoles
and reaches the GLESP one at higher.
For polarization, we have approximately the same accuracy
for the GLESP grN and HEALPix 4 iteration modes.

\section{Conclusions}
Here we present the GLESP-pol package which incorporates
calculations of polarization on the sphere into
the CMB analysis package based on the Gauss--Legendre Sky
Pixelization. We developed corresponding software for data
processing.

According to our numerical calculations, the described scheme
for polarization preserves the same level of precision as GLESP\,1.0
can provide for the  temperature anisotropy.

The code continues to be developed now and different
definitions of polarization modes
Refs.~\refcite{kami},      % (Kamionkowski et al., 1997;  %+
\refcite{hpix}              % G\'orski et al., 2005)       %+
are planned to be included.
GLESP-pol is open for new approaches and  could
be implemented for extension of  the present code, e.g. such as
the fast spin-weighted harmonics calculation
Ref.~\refcite{wiaux}.     % (Wiaux et al., 2005).
A completely new algorithm for fast Spherical Harmonics
Transform has been proposed recently
by Mark Tygert
Ref.~\refcite{tygert}.
It is worth mentioning that this
algorithm can be applied only for GLESP pixelization of the sky and
effective as $O(A\times \l^2Log_2\l)$ operations instead of usual $O(\l^3)$.
Unfortunately a huge prefactor $A$ makes this algorithm only about 3 times
faster than existing HEALPix and GLESP for $\l=2048$.

The most important part of our investigation is that the HEALPix\,2.11
package
provides the same accuracy as the GLESP-pol only by implementation of
4 iterations.
The zero iteration  mode of the  HEALPix\,2.11 can provide significant
error for the
coefficients of expansion and be used with caution.
\section*{Acknowledgments}
Athours are very thankful
to Radek Stompor for useful remarks to the paper and
to Per Rex Christensen for very fruitful discussions and
checking GLESP procedures.
This paper was supported in part by the RFBR grants
07-02-00886, 08-02-00090, 08-02-00159, 09-02-00298, by
S.S. 2469.2008.2 and
by FNU grants
272-06-0417, 272-07-0528 and 21-04-0355.

\section{Appendix}

\subsection{Description of polarization}
\subsubsection{Basic relations for the flat space}

The polarization is characterized by the symmetric traceless
matrix $\mathbf{T}$ composed by two Stokes parameters,
$Q({\bf x})\,\&\,U({\bf x})$:
\be
\mathbf{T} =
\left( \begin{array}{cc}
Q & U \\
U & - Q
\end{array}
\right)
\label{aqu}
\ee
The functions $Q$ and $U$ depend upon the coordinate frame and
components of the tensor $T_{i}^{j}$ obey the corresponding
tensor transformation law:
\be
\tilde{T}_{i}^{j} = O_i^k O_\l^j T_{k}^{\l}
\label{aij'}
\ee
where the coordinate transformation is given by $\tilde{x}^i=
O^i_k x^k $. In particular case under rotation of the coordinate
system with the rotation angle $\phi$
\be
\mathbf{O} =
\left( \begin{array}{cc}
 \cos\phi & \sin\phi \\
-\sin\phi & \cos\phi
\end{array}
\right)
\label{trot}
\ee
the functions $Q$ and $U$ are transformed as:
\be
Q' = Q \cos 2\phi + U \sin 2\phi  \nonumber \\
U' = -Q \sin 2\phi + U \cos 2\phi
\label{q'u'}
\ee
The intensity $I^2=Q^2+U^2=\frac{1}{2}T_i^k T^i_k$ does not depend
upon the coordinate system.

\subsubsection{Basic relations on the sphere}

An arbitrary traceless symmetric tensor can be presented
in terms of scalar ``electrical'', $E$, and pseudo-scalar
``magnetic'', $B$, potentials as:
\be
T_{ij} = \left(E_{;i;j}-
	  {1\over 2}g_{ij}E_{;k}^{;k}\right)
+{1\over 2}\left( \epsilon_{ik}B_{;j}^{;k} +
\epsilon_{jk}B^{;k}_{;i}\right)\,.
\label{ageneral}
\ee
where $\epsilon_{ij}$ is the completely antisymmetric tensor
($\epsilon_{11}=\epsilon_{22}=0,\, \epsilon_{12}=
-\epsilon_{21}=-sin\theta$)
and '$;$' denotes covariant differentiation on the two
dimensional sphere with the metric
\[
dl^2=d\theta^2+\sin^2\theta d\varphi^2
\]
Here $\theta,\varphi$ are the polar and azimuthal angles.
In the case we get instead of (\ref{aqu}):
\[
T^1_1 = -T^2_2=Q\quad T^1_2 = \sin\theta U,\quad
T^2_1=U/\sin\theta\,,
\]
and again we have
\[
I^2=0.5 T_i^kT_k^i=Q^2+U^2\,.
\]
Stokes parameters $Q$ and $U$ are linked with the potentials
$E$ and $B$ as follows:
\be
Q=D_1E-D_2B,\quad  U=D_2E+D_1B\,,
\label{op1}
\ee
and operators $D_1$ and $D_2$ are:
\be
D_1={1\over 2}\left({\partial^2\over\partial\theta^2}-cot\theta
{\partial\over\partial\theta}-{1\over \sin^2\theta}
{\partial^2\over\partial\varphi^2}\right) =
{1\over 2}\left((1-x^2){\partial^2\over\partial x^2}-
{1\over 1-x^2}{\partial^2\over\partial\varphi^2}\right)
\label{op2}
\ee
\[
 D_2=\left({1\over\sin\theta}{\partial^2\over\partial\theta
\partial\varphi}-{cot\theta\over \sin\theta}{\partial\over
\partial\varphi}\right)=-\left({\partial^2\over\partial x
\partial\varphi}+{x\over 1-x^2}{\partial\over\partial\varphi}
\right)\,.
\]
 where $x=cos\theta$. These operators are identical with
expressions (2.22) and (2.23) in
Ref.~\refcite{kami}. % Kamionkowski et al. (1997).

\subsubsection{Spherical harmonics}

Following to
% Doroshkevich et al. (2005 a,b)
Refs.~\refcite{glesp},
reficite{glespa}
we use the
representation of scalar (``electrical''), $E$, and
pseudoscalar (``magnetic''), $B$, polarization potentials
identical to (\ref{eq1}):
\be
E={1\over\sqrt{2\pi}}\sum_{\l=2}^{\l_{max}}\left(e_{\l 0}
f^0_\l(x)+2\sum_{m=1}^\l f^m_\l(x)[e_{\l m}^c\cos(m\varphi)-
e_{\l m}^s\sin(m\varphi)]\right)\,,
\label{s1}
\ee
\be
B={1\over\sqrt{2\pi}}\sum_{\l=2}^{\l_{max}}\left(
b_{\l 0}f^0_\l(x)+2\sum_{m=1}^\l f^m_\l(x)[b_{\l m}^c\cos(m\varphi)-
b_{\l m}^s\sin(m\varphi)]\right)\,,
\label{p1}
\ee
\be
f_\l^m=\sqrt{(\l+0.5){(\l-m)!\over(\l+m)!}}B_\l^m,\quad
0\leq m\leq \l\leq l_{max}\,,
\label{flmm}
\ee
Here $B_\l^m(x)$ and $f_\l^m(x)$ are the associated Legendre
functions (ordinary and normalized) and $e_{\l m}$ and
$b_{\l m}$ are the coefficients of decomposition
characterizing properties of polarization.
\be
e_{\l m}^c={1\over\sqrt{2\pi}}\int_{-1}^1dx\int_0^{2\pi}
d\varphi E(x,\varphi)f_\l^m(x)\cos m\varphi,
\label{slm}
\ee
\[
e_{\l m}^s={1\over\sqrt{2\pi}}\int_{-1}^1dx\int_0^{2\pi}
d\varphi E(x,\varphi)f_\l^m(x)\sin m\varphi\,.
\]
Similar expressions for the ``magnetic'' mode can be obtained
by replacing of $e_{\l m}^c\,\&\,e_{\l m}^s$ with $b_{\l m}^c\,\&\,
b_{\l m}^s$ and $E$ with $B$.

As in the previous package the functions $f_\l^m(x)$ are found
recursively:
\be
f_\l^m(x)=x\sqrt{4l^2-1\over \l^2-m^2}f_{\l-1}^m-
\sqrt{{2\l+1\over 2l-3}{(\l-1)^2-m^2\over \l^2-m^2}}f_{\l-2}^m\,,
\label{fmm}
\ee
or
\be
f_\l^m=-{2(m-1)\over \sqrt{\l^2-m^2+\l+m}}{xf_\l^{m-1}\over
\sqrt{1-x^2}}-\sqrt{{\l+2-m\over \l+1-m}\,{\l+m-1\over \l+m}}
f_\l^{m-2},
\label{fll}
\ee
\be
f_m^m=(-1)^m\sqrt{(2m+1)!!\over 2(2m)!!}(1-x^2)^{m/2},\quad
f^m_{m+1}=x\sqrt{2m+3}f^m_m\,.
\label{fmmm}
\ee
Relation (\ref{fmm}) starts with $f_m^m$ and $f_{m+1}^m$
(\ref{fmmm}) and generates functions $f_\l^m$ for all
$\l\geq m$. Relation (\ref{fll}) starts with $f_m^m$ and
$f_\l^0$ and generates functions $f_\l^m$ for all $1\leq m\leq
l$. In the case function $f_\l^0$ must be found with relation
(\ref{fmm}).

It is easy to see that
\be
D_1E=F_\l^m[e_{\l m}^c\cos(m\varphi)-
e_{\l m}^s\sin(m\varphi)]\,,
\label{d1s}
\ee
\be
D_2E=\Phi_\l^m[e_{\l m}^c\sin(m\varphi)+
e_{\l m}^s\cos(m\varphi)]\,,
\label{d2s}
\ee
where functions $F_\l^m$ and $\Phi_\l^m$ are expressed through
the normalized Legendre functions (\ref{fmm},\ref{fll}):
\be
M_\l F_\l^m=\sqrt{{2\l+1\over 2\l-1}(\l^2-m^2)}\,\,
{xf_{\l-1}^m\over 1-x^2}+{(m^2-\l)f_\l^m\over 1-x^2}-{\l(\l-1)\over 2}f_\l^m,
\label{flmm1}
\ee
\be
M_\l\Phi_\l^m={m\over 1-x^2}\left[\sqrt{{2\l+1\over 2\l-1}(\l^2-m^2)}
f_{\l-1}^m-(\l-1)xf_\l^m\right]\,,
\label{phmm}
\ee
or by other way
\be
M_\l F_\l^m={m-1\over 1-x^2}mf_\l^m-{\l^2+\l-2m\over 2}f^m_\l-
%%\sqrt{(\l+m+1)(\l-m)\over 1-x^2}{xf_\l^{m+1}}\,,
\sqrt{\l^2-m^2+\l-m\over 1-x^2}{xf_\l^{m+1}}\,,\,\,
\label{flml}
\ee
\be
M_\l\Phi_\l^m=-m\left[{\sqrt{(\l+m+1)(\l-m)}\over\sqrt{1-x^2}}
f^{m+1}_\l+(m-1){x\over 1-x^2}f^m_\l\right]\,.
\label{phll}
\ee
Here
\be
M_\l^2=0.25(\l+2)(\l+1)\l(\l-1),
\label{norma1}
\ee
and functions $F_\l^m$ and $\Phi_\l^m$ are normalized by condition
\[
\int_{-1}^1(F_\l^m+\Phi_\l^m)^2 dx=\int_{-1}^1[(F_\l^m)^2+
(\Phi_\l^m)^2] dx=1\,.
\]
In particular,
\[
F^m_m={m-1\over 2M_m}m{1+x^2\over 1-x^2}f^m_m,\quad
F^m_{m+1}=mxf^m_m{\sqrt{2m+3}\over M_{m+1}}
\left[{m-1\over 1-x^2}-{m+1\over 2}\right],
\]
\be
\Phi_m^m=-{m(m-1)\over M_m(1-x^2)}xf^m_m,\quad
\Phi^m_{m+1}=mf^m_m{\sqrt{2m+3}\over M_{m+1}}\,\,
{1-mx^2\over 1-x^2}\,,
\label{phimm}
\ee

Combining (\ref{op1}), (\ref{op2}), (\ref{s1}) and (\ref{p1}),
we get for functions $Q,\,\&\,U$:
\be
Q^m_c(x)={1\over\sqrt{2\pi}}\int_0^{2\pi} d\varphi Q\cos m
\varphi=\sum_{\l=2}^{\l_{max}}(F_\l^m(x)e_{\l m}^c-\Phi_\l^m(x)
b_{\l m}^s)\,,
\label{qcc}
\ee
\be
Q^m_s(x)={1\over\sqrt{2\pi}}\int_0^{2\pi} d\varphi Q\sin m
\varphi=\sum_{\l=2}^{\l_{max}}(-F_\l^m(x)e_{\l m}^s-\Phi_\l^m(x)
b_{\l m}^c)\,,
\label{qcs}
\ee
\be
U^m_c(x)={1\over\sqrt{2\pi}}\int_0^{2\pi} d\varphi U\cos m
\varphi=\sum_{\l=2}^{\l_{max}}(\Phi_\l^m(x)e_{\l m}^s+F_\l^m(x)
b_{\l m}^c)\,,
\label{ucc}
\ee
\be
U^m_s(x)={1\over\sqrt{2\pi}}\int_0^{2\pi} d\varphi U\sin m
\varphi=\sum_{\l=2}^{\l_{max}}(\Phi_\l^m(x)e_{\l m}^c-F_\l^m(x)
b_{\l m}^s)\,,
\label{ucs}
\ee

\subsubsection{Spin--weight functions}

Functions $F_\l^m$ and $\Phi_\l^m$ do not form an orthogonal
basis and to perform the further decomposition of polarization
we need to use the normalized spin - weight spherical functions
which can be defined as follows:
\be
\lambda_{\l,m}^+= F_\l^m+\Phi_\l^m,\quad
\lambda_{\l,m}^-=F_\l^m-\Phi_\l^m,\,
\label{ghlm}
\ee
\be
\lambda^+_{\l,m}(x)=(-1)^{m+\l}\lambda^{-}_{\l,m}(-x),\quad
\lambda^-_{\l,m}(x)=(-1)^{m+\l}\lambda^{+}_{\l,m}(-x)\,,
\label{lbd-pm}
\ee
These functions satisfy the self-adjoint equation:
\[
{d\over dx}\left[(1-x^2){d\over dx}\lambda_{\l,m}\right]-
{4+m^2+4mx\over 1-x^2}\lambda_{\l,m}+\l(\l+1)\lambda_{\l,m}=0\,,
\]
and, so, they are orthogonal for a given $m$ (see, e.g.,
\refcite{varsh},  % Varshalovich et al. 1989;                   %+
\refcite{lewis}    % Lewis, Challinor \& Turok, 2001,            %+
and references there).
Some of this functions can be written directly:
\[
\lambda_{m,m}^+=A_{m,m}(-1)^m(1-x)^2(1-x^2)^{m-2\over 2}=
F^m_m+\Phi^m_m\,,
\]
\[
\lambda_{m+1,m}^+=A_{m+1,m}(-1)^m(1-x)^2(1-x^2)^{m-2\over 2}
[2+(m+1)x]= F^m_{m+1}+\Phi^m_{m+1}\,,
\]
\[
A_{m,m}^2={m(m-1)(2m+1)!!\over (m+2)(2m+2)!!},\quad
A_{m+1,m}^2={2m+3\over (m+3)(m-1)}A_{m,m}^2\,.
\]
and using these relations functions $\lambda_{\l,m}$ with $\l>m+1$
can be evaluated recursively:
\be
\lambda_{\l,m}=\left(x+{2m\over \l(\l-1)}\right)C_{\l,m}
\lambda_{\l-1,m}-{C_{\l,m}\over C_{\l-1,m}}\lambda_{\l-2,m}\,,
\label{recur}
\ee
\[
C_{\l,m}=\sqrt{\l^2(4\l^2-1)\over (\l^2-m^2)(\l^2-4)}\,.
\]
\[
\int_{-1}^1dx[F^m_\l(x)F_k^m(x)+\Phi^m_\l(x)\Phi_k^m(x)]=
\delta_{k\l}\,,
\]
\[
\int_{-1}^1dx[F^m_\l(x)\Phi_k^m(x)+\Phi^m_\l(x)F_k^m(x)]=0\,.
\]
This means that coefficients for the decomposition of $Q$ and
$U$ can be found from ({\ref{qcc}\,-\,\ref{ucs}}) as follows:
\be
e_{\l m}^c=\int_{-1}^1dx[\Phi^m_\l(x)U_s^m(x)+F^m_\l(x)Q_c^m(x)]\,,
\label{sclm}
\ee
\be
e^s_{\l m}=\int_{-1}^1dx[\Phi^m_\l(x)U_c^m(x)-F^m_\l(x)Q_s^m(x)]\,,
\label{sslm}
\ee
\be
b_{\l m}^c=\int_{-1}^1dx[F^m_\l(x)U_c^m(x)-\Phi^m_\l(x)Q_s^m(x)]\,,
\label{pclm}
\ee
\be
b^s_{\l m}=\int_{-1}^1dx[-F^m_\l(x)U_s^m(x)-\Phi^m_\l(x)Q_c^m(x)]\,,
\label{pslm}
\ee

For numerical analysis of the polarization maps
together with the temperature fluctuations it is convenient
to use relations (\ref{flmm} - \ref{phll}) and (\ref{sclm} -
\ref{pslm}) instead of the recursive relations (\ref{recur}).
Inverse problem that is the construction of the polarization
maps from coefficients $e_{\l m}\,\&\,b_{\l m}$ can also be solved
using relations (\ref{qcc}\,-\,\ref{ucs}).

\subsubsection{Spectra of the anisotropy and polarization}

Four power spectra can be introduced for the
temperature, electrical and magnetic Stokes parameters,
they are:
\[
C^T_\l={1\over 2\l+1}\sum_{m=-\l}^\l |a_{\l m}|^2\,,
\]
\be
C^E_\l={1\over 2\l+1}\sum_{m=0}^\l [(e_{\l m}^c)^2+(e_{\l m}^s)^2]\,,
\label{spctr}
\ee
\[
C^B_\l={1\over 2\l+1}\sum_{m=0}^\l [(b_{\l m}^c)^2+(b_{\l m}^s)^2]\,,
\]
\[
C^{TE}_\l={1\over 2\l+1}\sum_{m=0}^\l [a_{\l m}^ce_{\l m}^c+
a_{\l m}^se_{\l m}^s]\,,
\]

When averaged over the sky, the mean square temperature
anisotropy is
\be
\langle \Delta T^2\rangle = T_0^2\sum_{\l=2}^\infty {2\l+1\over
4\pi}C^T_\l\,,
\label{temp}
\ee
where $T_0$ is the temperature of the CMB. The mean square
of polarization is
\[
\langle I^2\rangle = {1\over 2}\langle Q^2+U^2\rangle =
{1\over 2}[\langle E^2_*\rangle+\langle B^2_*\rangle]
\]
where
\be
\langle E^2_*\rangle = T_0^2\sum_{\l=2}^\infty {2\l+1\over
4\pi}C^E_\l\,,
\label{}
\ee
\be
\langle B^2_*\rangle = T_0^2\sum_{\l=2}^\infty {2\l+1\over
4\pi}C^B_\l\,,
\ee
\label{temp}
More details can be found in, for
example,
Refs.~\refcite{zald1},   % Zaldarriaga \& Harari\ee (1995), %+
\refcite{zald2},         % Zaldarriaga \& Seljak (1997),    %+
\refcite{seljak},        % Seljak \& Zaldarriaga (1997),    %+
\refcite{kami}.          % Kamionkowski et al.  (1997).     %+

\subsection{Incorporating of the polarization to the GLESP code}

To realize algorithms described above, special procedures had
been created. We had developed this code in parallel in the two
algorithmic languages, GNU\,C and FORTAN-77.
The procedures have been designed both like subroutines and
package utilities implementing in the GLESP pixelization scheme.

Two commands of GLESP-pol should be used for calculation of polarization,
namely, `{\it polmap}' and `polalm'. `{\it polmap}' calculates
Q and U-polarization maps by coefficients of E-- and B--
polarization modes. `{\it Polalm}' calculates coefficients of
E-- and B--polarization modes by Q-- and U-polarization maps,
respectively. E-- and B--coefficients recorded as $a_{\l m}$--
format files, and Q-- and U--maps recorded like GLESP maps.
Both procedures can be used for temperature--spherical
harmonics transformations too. These procedures are included
to the GLESP package, version 2.0.
The programs use the fast Fourier transfom FFTW-3.2.1
Ref.~\refcite{fftw}. %  Frigo and Johnson (1997).

% The default minimal number of pixels on a polar regions of the
% map generated by the basic procedure `{\it cl2map}' is 9 for
% the polarization coefficients of spherical functions. This is
% different from the previous GLESP package where minimal number
% of pixels near the poles is 3.  As for the GLESP map of the CMB
% anisotropy, the size of pixels is determined from equator pixels
% and in the case the size of polar pixels is smaller then that
% for other pixels. But it is possible to use the size of pixels
% for polarization calculated from the size of near-pole pixels
% as well.

The program `{\it difmap}' is developed for standard
polarization transforms to calculate a polarization angle
$PA=1/2 \arctan (U/Q)$ and an intensity $I=\sqrt{Q2+U2}$.
Both values can be plotted with the GLESP drawing procedure
`{\it f2fig}'.

%\subsection{M and L versions of the code}
%
% As in the previous GLESP package two versions - M and L - are
% presented. The more rapid M - version based on the relations
% (\ref{fmm}, \ref{flmm}\,\&\,\ref{phmm}) started with functions
% $f_m^m$ which are extremely small near the poles for large $m$
% what requires special attention. The more slow L - version
% based on the relation (\ref{fll}, \ref{flml}\,\&\,\ref{phll})
% can be presumably used to check results obtained with the
% M - version.

%\begin{figure}
%\centering
%\epsfxsize=10 cm
%%\epsfbox{../precise/fig23.eps}
%\epsfbox{fig35.eps}
%\vspace{0.8cm}
%\caption{Evaluations of the accuracy restrictions for the
%temperature, and Stokes parameters for HEALPix (plus, triangles
%and rombus, correspondingly) and GLESP
%(points for temperature, stars and squares for Stokes
%parameters) for $2\leq l\leq 1500$
%}
%\label{prec}
%\end{figure}

\subsection{Accuracy restrictions}

Both temperature and polarization maps and spectra processed
by the GLESP package are determined in main by four parameters.
First of all, it is the maximal number of harmonics under
consideration, $\ell_{max}$ (\ref{s1}). The second one is the
numberof rings used for map presentation, $N_\theta$. By
definition, we must have $N_\theta\geq 2\ell_{max}$. The third
one is the number of pixels for each ring of the map,
$N_\phi^i(\theta_i)$. If we like to use the same angular
resolution in azimuthal and polar directions than, evidently,
we must take $N_\phi(\pi/2)\approx 2N_\theta$, and $N_\phi^i
(\theta_i)$ decreases progressively away from the equator.
The fourth one is the Nyquist parameter, $N_y$, which
regulates the precisions achieved by calculations.

The precision depends upon the choice $N_\theta\geq 2\ell_{max}$
and the Nyquist parameter, $N_y$. Thus, the orthogonality of
spherical harmonics $Y_\ell^m$ on the sphere is achieved only
if number of pixels for each ring is at least two times larger
than number of harmonics we use in our analysis, $N_\phi^i
(\theta_i)\geq 2m,\,\,0\leq m\leq \ell$ (Nyquist restriction,
$N_\phi^i(\theta_i)\geq m/N_y\geq 2m$, $N_y\leq 0.5$).
Since we try to keep the same pixel area for different
latitude, this condition is {\it not completely satisfied}
anymore because rings with number of pixels $N_\phi^i(\theta_i)
\leq 2m$ drop out from any calculations.
% In the GLESP
% package we set $N_\phi^i(\theta_i)\geq N_{min}=9$ what
% decreases the pixel area nearby the poles but hold its
% contribution for harmonics with $m\leq 9N_y\leq 5$.
% This
% means that the polar areas progressively drop out from
% analysis of fluctuations with $m\geq 9N_y+1$ and, therefore,
% from the analysis of the spectrum $C_\ell$ with $\ell\geq 9N_y
% +1$.

Fortunately, the influence of these restrictions on the
precision achieved is moderate because the contribution
of polar areas is quite small for almost all harmonics by the
following reason:
\begin{itemize}
\item{}Spin-weighted harmonics satisfy
$_{\pm 2}Y_{\ell m}\sim sin^{m-2}(\theta)$ for $m\,> 2$. This
means that harmonics with high $m$ are weighted negligible
everywhere apart from the equator, where number of pixels is
enough for orthogonality.
\item{}Harmonics with small $m\leq m_{min}=N_y\times N_{min}$
are orthogonal.
\item{}Therefore, it can be expected that the harmonics with
$m\sim m_{min}$ are the most dangerous. For such harmonics
rings nearby the poles have number of pixels smaller than
$2\times m$ and these parts of the sphere spoil the
picture. However, the weight of these parts of the sphere is
proportional to $sin^{m-2}(\theta)$ and tends to be zero for
$\theta\rightarrow 0,\pi$.
\end{itemize}

To accelerate the computations we remain in the calculations
only larger terms restricted by conditions $|f_{\ell}^m(x)|\geq
\varepsilon,\,$$\,|F_\l^m|, |\Phi_\l^m|\geq\varepsilon$
where $\varepsilon$ depends upon $\l$ and required precision
and, so, is determined in practice. Our tests show that the
choice $\varepsilon=\varepsilon_1$ for $\l\leq 500$ and
$\varepsilon=\varepsilon_2$ for $\ell\geq 500$ allows to
moderately decrease the
calculation time and weakly change the precision achieved.

We can conclude, that for a given $\ell_{max}$ the reasonable
precision can be achieved for the parameters :
\be
N_\theta=(2.5 - 3)\ell_{max},\quad N_\phi=2\times N_\theta,
\quad N_{min}=9 - 11,\quad N_y=0.5\,,
\label{param}
\ee
and
\[
\varepsilon_1\approx 10^{-10},\quad \varepsilon_2\approx
10^{-4}\,.
\]

Evidently, the precision achieved increases for larger
$N_\theta$ and $N_\phi$, however, is accompanied by
corresponding fast growth of the calculation time.
Perhaps, for calculations of polarization the better results
can be achieved with the repixelisation of the map in
accordance with the net of roots of the spin -- weight
functions $\lambda_{\ell,m}^{\pm}$ (\ref{ghlm}) instead of
the Legendre polynomials (\ref{root}) and corresponding choice
of the weighting coefficients (\ref{eq6}).

\subsection{Data format}

Developing the GLESP package for polarization calculation,
we should change a format of data representation.
% Considering
% possibility the transformation only polarization data (i.e.
% without temperature anisotropy),
We introduce two types of
the format describing $a_{\ell m}$--coefficients and maps.

In the first case, we can use the standard $a_{\ell m}$--
coefficients data, which contain index describing number of
$\ell$ and $m$ modes corresponding to the HEALPix, real and
imaginary parts of $a_{\ell m}$. These three parameters are
described by three--fields records of the FITS Binary Table
Ref.~\refcite{FITS}. % (Hanisch et al., 2001).   %+
Map data are described by the three--
fields Binary Table FITS format containing a vector of
$x_i=\cos \theta$ positions, a vector of numbers of pixels
per each layer $N_{\phi_i}$, and set of temperature values
in each pixel recorded by layers from the North Pole.
All these data description formats are used separately for
each type and polarization mode data, i.e. maps for temperature
anisotropy, Q and U-data are contained in singles files, and
$a_{\ell m}$ and E and B-modes coefficients are stored in
singles file too.

The second type of the data representation format is similar
to HEALPix one. In this case, data are unified in 3 extensions
containing maps with anisotropy, Q- and U-polarization data
or $a_{\ell m}$-coefficients of temperature expansions and E-
and B-mode, respectively. Files with coefficients in GLESP
and in HEALPix have the same format. Each of three FITS
extensions of GLESP files with maps contains three fields
described above.

So, using two formats of data representation a user can easily
select a path of one's data processing including or excluding
any type of polarization or anisotropy data.

\end{document}